\documentclass[12pt,a4paper,oneside]{article}

\usepackage[hypertex]{hyperref}
\usepackage{verbatim,mathrsfs,array,layout,textcomp,latexsym}
\usepackage{multirow}
\usepackage{color}

\usepackage{graphicx,amssymb,amsmath,dsfont,slashed}
\usepackage[margin=2cm]{geometry}
\begin{document}

\newcommand{\sla}{\slash \hspace{-0.2cm}}
\newcommand{\slam}{\slash \hspace{-0.25cm}}
\newcommand{\be}{\begin{equation}}
\newcommand{\ee}{\end{equation}}
\newcommand{\bea}{\begin{eqnarray}}
\newcommand{\eea}{\end{eqnarray}}
\newcommand{\beq}{\begin{equation}}
\newcommand{\eeq}{\end{equation}}
\newcommand{\beqa}{\begin{eqnarray}}
\newcommand{\eeqa}{\end{eqnarray}}
\newcommand{\no}{\nonumber}

\newcommand{\cL}{\mathcal{L}}
\newcommand{\cO}{\mathcal{O}}
\newcommand{\cA}{\mathcal{A}}
\newcommand{\cB}{\mathcal{B}}

% "Bar" fields
\newcommand{\Qbar}{\overline{Q}}
\newcommand{\Dbar}{\overline{D}}
\newcommand{\Ebar}{\overline{E}}
\newcommand{\Lbar}{\overline{L}}
\newcommand{\Bbar}{\overline{B}}
\newcommand{\Mbar}{\overline{M}}
\newcommand{\Knotbar}{\overline{K^0}}
\newcommand{\Dnotbar}{\overline{D^0}}

\newcommand{\Nkk}{{N_{\rm KK}}}
\newcommand{\Mkk}{{M_{\rm KK}}}
\newcommand{\Mkks}{{M^2_{\rm KK}}}
\newcommand{\yfd}{{y_{\rm 5D}}}
\newcommand{\yfdmin}{{y_{\rm 5D}^{\rm min}}}
\def\kpirc{k\pi r_c }
\newcommand{\rg}{{r^g_{00}}}

\newcommand{\yut}{(Y_u Y_u^\dagger)_{\slashed{\mathrm{tr}}}}
\newcommand{\ydt}{(Y_d Y_d^\dagger)_{\slashed{\mathrm{tr}}}}
\newcommand{\au}{A_{Q^u}}
\newcommand{\ad}{A_{Q^d}}
\newcommand{\had}{\hat A_{Q^d}}
\newcommand{\hadp}{\hat A'_{Q^d}}
\newcommand{\hadn}{\hat A_{Q^d}^n}
\newcommand{\hau}{\hat A_{Q^u}}
\newcommand{\haun}{\hat A_{Q^u}^n}
\newcommand{\aud}{A_{Q^u,Q^d}}
\newcommand{\haud}{\hat A_{Q^u,Q^d}}
\newcommand{\haudp}{\hat A'_{Q^u,Q^d}}
\newcommand{\hD}{\mathcal{\hat {\vec D}}}
\newcommand{\cdn}{\hat C_d^n}
\newcommand{\tr}{\mathrm{tr}}
\newcommand{\hj}{\hat J}
\newcommand{\hjd}{\hat J_d}
\newcommand{\hju}{\hat J_u}
\newcommand{\hjud}{{\hat J}_{u,d}}
\newcommand{\hjq}{\hat J_Q}
\newcommand{\ltev}{\left( \frac{\Lambda_{\rm NP}}{1\,\textrm{TeV}} \right)}
\newcommand{\dmqot}{\Delta\tilde m^2_{Q_2Q_1}}
\newcommand{\dmqoth}{\Delta\tilde m^2_{Q_3Q_1}}
\newcommand{\dmqij}{\Delta\tilde m^2_{Q_iQ_j}}
\newcommand{\lmc}{\lambda}
\newcommand{\eg} {{\it e.g.}}
\newcommand{\ie}  {{\it i.e.}}
\newcommand{\et} {{\it et. al}}

\definecolor{red}{cmyk}{0,1,1,0.4}

\def\lsim{\mathrel{\rlap{\lower4pt\hbox{\hskip1pt$\sim$}}
     \raise1pt\hbox{$<$}}}         %less than or approx. symbol
\def\gsim{\mathrel{\rlap{\lower4pt\hbox{\hskip1pt$\sim$}}
     \raise1pt\hbox{$>$}}}         %greater than or approx. symbol

%%%%%%%%%%%%%My defintions
\newcommand{\LF} {{\cal L}^{\rm F}}
\newcommand{\GSM} { {\cal G}^{\rm SM}}
\newcommand{\GSMCP} { {\cal G}^{\rm SM}_{CP}}
\newcommand{\GSME} { {\cal G}^{\rm SM}_{\rm exten}}
\newcommand{\ba}{\begin{eqnarray}}
\newcommand{\ea}{\end{eqnarray}}
\newcommand{\bi}{\begin{itemize}}
\newcommand{\ei}{\end{itemize}}
\newcommand{\nn}{\nonumber}
\newcommand{\MeV}{{\rm \, MeV}}
\newcommand{\GeV}{{\rm \, GeV}}
\newcommand{\Or}{{\cal O}}
\def\Ord#1{\Or\left(#1\right)}
\def\Eq#1{Eq.~(\ref{#1})}
\newcommand{\ddbar}{D^0-\overline{D}{}^0}
\newcommand{\diag}{{\rm diag}}
\newcommand{\VKM}{V^{\rm CKM}}
\newcommand{\delKM}{\delta^{\rm KM}}
\newcommand{\CSM}{C^{\rm SM}}
\newcommand{\gW}{{g_2^\pm}}
\newcommand{\LMFV}{{\Lambda_{\rm MFV}}}
\newcommand{\LMFVF}{{\cal L}^{\Delta F=2}_{\rm MFV}}
\newcommand{\bp}{\begin{pmatrix}}
\newcommand{\ep}{\end{pmatrix}}
\newcommand{\Obsn}{{O_{b\to s\nu\bar \nu}}}
\newcommand{\Obdn}{{O_{b\to d\nu\bar \nu}}}

%GMFV definitions

\newcommand{\yydagu}{Y_uY_u^\dagger}
\newcommand{\yydagd}{Y_dY_d^\dagger}
\newcommand{\ydagyu}{Y_u^\dagger Y_u}
\newcommand{\ydagyd}{Y_d^\dagger Y_d}
\newcommand{\hchi}{\hat\chi}
\newcommand{\tb}{${\rm tan}\beta $ }
\newcommand{\G}{{\cal G}^{\rm SM}}
\renewcommand{\H}{{\cal H}^{\rm SM}}
\renewcommand{\O}{{\cal O}}
\renewcommand{\hchi}{\hat\chi}
\newcommand{\dchih}{\Delta\hat\chi}
\newcommand{\dchi}{\Delta\chi}
\newcommand{\hpi}{\hat \Pi}
\newcommand{\eps}{\epsilon}
\newcommand{\dpi}{\delta \pi}
\newcommand{\V}{V_{\rm CKM}}
\newcommand{\3}{{\bf 3}}
\newcommand{\1}{{\bf 1}}
\newcommand{\mysubsection}[1]{\vspace{.5cm}{\it #1}\newline}
\newcommand{\Mp}{M_{\rm Pl}}
\newcommand{\tm}{{\tilde m^2}}
\renewcommand{\t}{\tilde t}
\renewcommand{\u}{\tilde u^{(2)}}
\renewcommand{\b}{\tilde b}
\renewcommand{\d}{\tilde d^{(2)}}
\newcommand{\tbar}{\overline{\tilde t}}
\newcommand{\ubar}{\overline{\tilde u}^{(2)}}
\newcommand{\ubarL}{\overline{\tilde u^{(2)}_L}}
\newcommand{\bbar}{\overline {\tilde b}}
\newcommand{\dbar}{\overline{\tilde d^{(2)}}}
\newcommand{\dbarL}{\overline{\tilde d^{(2)}_L}}
\newcommand{\dbarR}{\overline{\tilde d^{(2)}_R}}
\newcommand{\OCP}{ {\cal O}_{B_{d,s}} }
\newcommand{\Lgmfv}{\Lambda_{\rm MFV}}
\newcommand{\chd}{\marginpar{\bf changed} }

\renewcommand{\Im}{\mathrm{Im}}

%\Title{Flavor Physics Within \& Beyond the Standard Model}
%\Author{Gilad Perez}
%
%\title{\huge\vspace*{-.1cm}2009 TASI Lecture --  Flavor Physics\vspace*{.5cm}}
%
%\author{{\Large Gilad Perez\vspace*{.1cm} }}
%\email{gilad.perez@weizmann.ac.il  }
%
%\author{{\Large Oram Gedalia\vspace*{.4cm} }}
%\email{oram.gedalia@weizmann.ac.il  }
%
%\affiliation{\large Weizmann Institute of Science \vspace*{1cm}}
%%\vspace*{2cm}
%%\begin{center}
%%{\large  \vspace*{.5cm} Weizmann Institute of Science \vspace*{-.3cm}}\end{center}

\title{\bf TASI 2009 Lectures~-- Flavor Physics}
\author{ Oram Gedalia and Gilad Perez 
\vspace{6pt} \\ \fontsize{12}{16}\selectfont\textit{Department
of Particle Physics and Astrophysics, Weizmann Institute of
Science,} \\ \fontsize{12}{16}\selectfont\textit{Rehovot 76100,
Israel}}
\date{}
\maketitle

\begin{abstract}
The standard model picture of flavor and CP violation is now
experimentally verified, hence strong bounds on the flavor
structure of new physics follow. We begin by discussing in
detail the unique way that flavor conversion and CP violation
arise in the standard model. The description provided is based
on a spurion, symmetry oriented, analysis, and a covariant
basis for describing flavor transition processes is introduced,
in order to make the discussion transparent for non-experts. We
show how to derive model independent bounds on generic new
physics models. Furthermore, we demonstrate, using the
covariant basis, how recent data and LHC projections can be
applied to constrain models with an arbitrary mechanism of
alignment. Next, we discuss the various limits of the minimal
flavor violation framework and their phenomenological aspects,
as well as the implications to the underlying microscopic
origin of the framework. We also briefly discuss aspects of
supersymmetry and warped extra dimension flavor violation.
Finally we speculate on the possible role of flavor physics in
the LHC era.
\end{abstract}
\newpage

\tableofcontents
\newpage

%%%%%%%%%%%%%%%%%%%%%%%%%%%%%%%%%%%%%%%%%%%%%%%%%%%%
%%%%%%%%%%%%%%%%%%%%%%%%%%%%%%%%%%%%%%%%%%%%%%%%%%%%
\section{Introduction}
%%%%%%%%%%%%%%%%%%%%%%%%%%%%%%%%%%%%%%%%%%%%%%%%%%%%
%%%%%%%%%%%%%%%%%%%%%%%%%%%%%%%%%%%%%%%%%%%%%%%%%%%%

Flavors are replications of states with identical quantum
numbers. The standard model (SM) consists of three such
replications of the five fermionic representations of the SM
gauge group. Flavor physics describes the non-trivial spectrum
and interactions of the flavor sector. What makes this field
particularly interesting is that the SM flavor sector is rather
unique, and its special characteristics make it testable and
predictive. \footnote{This set of lectures discusses the quark
sector only. Many of the concepts that are explained here can
be directly applied to the lepton sector.} Let us list few of
the SM unique flavor predictions: \bi
\item It Contains a single CP violating parameter.\footnote{The
    SM contains an additional flavor diagonal CP violating
    parameter, namely the strong CP phase. However, experimental data
    constrains it to be smaller than
    $\Or\left(10^{-10}\right)$, hence negligibly small.}
\item Flavor conversion is driven by three mixing angles.
\item To leading order, flavor conversion proceeds through
    weak charged current interactions.
\item To leading order, flavor conversion involves left handed
    (LH) currents.
\item CP violating processes must involve all three
    generations.
\item The dominant flavor breaking is due to the top Yukawa
    coupling, hence the SM possesses a large approximate global
    flavor symmetry (as shown below, technically it is given by
    $U(2)_Q\times U(2)_U\times U(1)_t \times U(3)_D$).
\ei In the last four decades or so, a huge effort was invested
towards testing the SM predictions related to its flavor
sector. Recently, due to the success of the B factories, the
field of flavor physics has made a dramatic progress,
culminated in Kobayashi and Maskawa winning the Nobel prize. It
is now established that the SM contributions drive the observed
flavor and CP violation (CPV) in nature, via the
Cabibbo-Kobayashi-Maskawa
(CKM)~\cite{Cabibbo:1963yz,Kobayashi:1973fv} description. To
verify that this is indeed the case, one can allow new physics
(NP) to contribute to various clean observables, which can be
calculated precisely within the SM. Analyses of the data before
and after the B factories data have
matured~\cite{Ligeti,NMFV1,NMFV2,Buras:2009us}, demonstrating
that the NP contributions to these clean processes cannot be
bigger than $\Ord{30\%}$ of the SM
contributions~\cite{UTFit,CKMFitter}.

Very recently, the SM passed another non-trivial test. The
neutral $D$ meson system (for formalism
see~\eg~\cite{DDbarform1,DDbarform2,DDbarform3, DDbarform4,
DDbarform5} and refs.~therein) bears two unique aspects among
the four neutral meson system ($K,D,B,B_s$): (i) The long
distance contributions to the mixing are orders of magnitude
above the SM short distance ones~\cite{Dlong1, Dlong2}, thus
making it difficult to theoretically predict the width and mass
splitting. (ii) The SM contribution to the CP violation in the
mixing amplitude is expected to be below the permil
level~\cite{DCPV}, hence $\ddbar$ mixing can unambiguously
signal new physics if CPV is observed. Present
data~\cite{Ciuchini:2007cw,Gedalia:2009kh,otherD1,otherD2,
otherD3,otherD4,combine,indirect} implies that generic CPV
contributions can be only of $\Ord{20\%}$ of the total
(un-calculable) contributions to the mixing amplitudes, again
consistent with the SM null prediction.

We have just given rather solid arguments for the validity of
the SM flavor description. What else is there to say then?
Could this be the end of the story? We have several important
reasons to think that flavor physics will continue to play a
significant role in our understanding of microscopical physics
at and beyond the reach of current colliders. Let us first
mention a few examples that demonstrate the role that flavor
precision tests played in the past: \bi
\item The smallness of $\Gamma(K_L\to\mu^+\mu^-)/
    \Gamma(K^+\to\mu^+\nu)$ led to predicting a fourth quark
    (the charm) via the discovery of the GIM
    mechanism~\cite{GIM}.
\item The size of the mass difference in the neutral Kaon
    system, $\Delta m_K$, led to a successful prediction of the
    charm mass~\cite{Charm}.
\item The size of $\Delta m_B$ led to a successful prediction of the
  top mass (for a review see~\cite{Franzini:1988fs} and refs.~therein).
\ei This partial list demonstrates the power of flavor
precision tests in terms of being sensitive to short distance
dynamics. Even in view of the SM successful flavor story, it is
likely that there are missing experimental and theoretical
ingredients, as follows: \bi
\item Within the SM, as mentioned, there is a single CP
    violating parameter. We shall see that the unique structure
    of the SM flavor sector implies that CP violating phenomena
    are highly suppressed. Baryogenesis, which requires a sizable
    CP violating source, therefore cannot be accounted for by the SM
    CKM phase. Measurements of CPV in flavor changing
  processes might provide evidence for additional sources coming
  from short distance physics.
\item The SM flavor parameters are hierarchical, and most of
    them are small (excluding the top Yukawa and the CKM
    phase), which is denoted as the flavor puzzle. This peculiarity
    might stem from unknown flavor dynamics. Though it might be
    related to very short distance physics, we can still get
    indirect information about its nature via combinations of
    flavor precision and high $p_T$ measurements.
\item The SM fine tuning problem, which is related to the
    quadratic divergence of the Higgs mass, generically
    requires new physics at, or below, the
  TeV scale. If such new physics has a generic flavor structure, it
  would contribute to flavor changing neutral current (FCNC) processes
  orders of magnitude above the observed rates. Putting it differently,
  the flavor scale at which NP is allowed to have a generic flavor structure
  is required to be larger than $\Ord{10^5}\,$TeV, in order to be consistent with flavor
  precision tests. Since this is well
  above the electroweak symmetry breaking scale, it implies an ``intermediate''
  hierarchy puzzle ({\it cf.}~the little hierarchy~\cite{LHierarchy1,LHierarchy2} problem). We
use the term ``puzzle'' and not ``problem'' since in general,
the smallness of the flavor parameters, even within NP models,
implies the presence of approximate symmetries. One can
imagine, for instance, a situation where the suppression of the
NP contributions to FCNC processes is linked with the SM small
mixing angles and small light quark Yukawas~\cite{NMFV1,NMFV2}.
In such a case, this intermediate hierarchy is resolved in a
technically natural way, or radiatively stable manner, and no
fine tuning is required.\footnote{Unlike, say, the case of the
$S$ electroweak parameter, where in general one cannot
associate an approximate symmetry with the limit of small NP
contributions to $S$.} \ei

%%%%%%%%%%%%%%%%%%%%%%%%%%%%%%%%%%%%%%%%%%%%%%%%%%%%
%%%%%%%%%%%%%%%%%%%%%%%%%%%%%%%%%%%%%%%%%%%%%%%%%%%%
\section{The standard model flavor sector}
%%%%%%%%%%%%%%%%%%%%%%%%%%%%%%%%%%%%%%%%%%%%%%%%%%%%
%%%%%%%%%%%%%%%%%%%%%%%%%%%%%%%%%%%%%%%%%%%%%%%%%%%%
The SM quarks furnish three representations of the SM gauge
group, $SU(3)\times SU(2)\times U(1)$: $Q(3,2)_{\frac16}\times
U(3,1)_{\frac23}\times D(3,1)_{-{\frac13}}$, where $Q,U,D$
stand for $SU(2)$ weak doublet, up type and down type singlet
quarks, respectively. Flavor physics is related to the fact
that the SM consists of three replications/generations/flavors
of these three representations. The flavor sector of the SM is
described via the following part of the SM Lagrangian
\begin{equation} \label{Lflavor}
\LF=\overline{q^i} \slashed{D}\ q^j \delta_{ij}+(Y_U)_{ij}\overline{ Q^i} U^j H_U+
(Y_D)_{ij}\overline{ Q^i} D^j H_D+{\rm h.c.}\,,
\end{equation}
where $\slashed{D}\equiv D_\mu\gamma^\mu$ with $D_\mu$ being a
covariant derivative, $q=Q,U,D$, within the SM with a single
Higgs $H_U=i\sigma_2 H_D^*$ (however, the reader should keep in
mind that at present, the nature and content of the SM Higgs
sector is unknown) and $i,j=1,2,3$ are flavor indices.

If we switch off the Yukawa interactions, the SM would possess
a large global flavor symmetry, $\GSM$,\footnote{At the quantum
level, a linear combination of the diagonal $U(1)$'s inside the
$U(3)$'s, which corresponds to the axial current, is
anomalous.}
\begin{eqnarray}
  \label{GSM}
  \GSM= U(3)_Q\times U(3)_U\times U(3)_D\,.
\end{eqnarray}
Inspecting Eq.~\eqref{Lflavor} shows that the only non-trivial
flavor dependence in the Lagrangian is in the form of the
Yukawa interactions. It is encoded in a pair of $3\times 3$
complex matrices, $Y_{U,D}$.

%%%%%%%%%%%%%%%%%%%%%%%%%%%%%%%%%%%%%%%%%%%%%
\subsection{The SM quark flavor parameters}
%%%%%%%%%%%%%%%%%%%%%%%%%%%%%%%%%%%%%%%%%%%%%
Naively one might think that the number of the SM flavor
parameters is given by $2\times 9=18$ real numbers and $2\times
9=18$ imaginary ones, the elements of $Y_{U,D}$. However, some
of the parameters which appear in the Yukawa matrices are
unphysical. A simple way to see that (see
\eg~\cite{Nirrev1,Nirrev2,Nirrev3} and refs.~therein) is to use
the fact that a flavor basis transformation,
\begin{eqnarray}
  \label{Qtrans}
  Q \to V_Q Q\,, \qquad U\to V_U U\,, \qquad  D \to V_D  D\,,
\end{eqnarray}
leaves the SM Lagrangian invariant, apart from redefinition of
the Yukawas,
\begin{eqnarray} \label{Ytrans}
Y_U \to V_Q Y_U V_U^\dagger\,, \qquad  Y_D \to V_Q Y_D V_D^\dagger\,,
\end{eqnarray}
where $V_i$ is a $3\times 3$ unitary rotation matrix. Each of
the three rotation matrices $V_{Q,U,D}$ contains three real
parameters and six imaginary ones (the former ones correspond
to the three generators of the $SO(3)$ group, and the latter
correspond to the remaining six generators of the $U(3)$
group). We know, however, that physical observables do not
depend on our choice of basis. Hence, we can use these
rotations to eliminate unphysical flavor parameters from
$Y_{U,D}$. Out of the 18 real parameters, we can remove 9
($3\times 3$) ones. Out of the 18 imaginary parameters, we can
remove 17 (3$\times6-1$) ones. We cannot remove all the
imaginary parameters, due to the fact that the SM Lagrangian
conserves a $U(1)_B$ symmetry.\footnote{More precisely, only
the combination $U(1)_{B-L}$ is non-anomalous.} Thus, there is
a linear combination of the diagonal generators of $\GSM$ which
is unbroken even in the presence of the Yukawa matrices, and
hence cannot be used in order to remove the extra imaginary
parameter.

An explicit calculation shows that the 9 real parameters
correspond to 6 masses and 3 CKM mixing angles, while the
imaginary parameter corresponds to the CKM celebrated CPV
phase. To see that, we can define a mass basis where $Y_{U,D}$
are both diagonal. This can be achieved by applying a
bi-unitary transformation on each of the Yukawas:
\begin{eqnarray} \label{Qmtrans}
Q^{u,d} \to V_{Q^{u,d}} Q^{u,d}\,, \qquad U\to V_U U\,, \qquad  D \to V_D  D\,,
\end{eqnarray}
which leaves the SM Lagrangian invariant, apart from
redefinition of the Yukawas,
\begin{eqnarray} \label{Ymtrans}
Y_U \to V_{Q^u} Y_U V_U^\dagger\,, \qquad  Y_D \to V_{Q^d} Y_D V_D^\dagger\,.
\end{eqnarray}
The difference between the transformations used in
Eqs.~\eqref{Qtrans} and~\eqref{Ytrans} and the ones
above~(\ref{Qmtrans},\ref{Ymtrans}), is in the fact that each
component of the $SU(2)$ weak doublets (denoted as $Q^u\equiv
U_L$ and $Q^d\equiv D_L$) transforms independently. This
manifestly breaks the $SU(2)$ gauge invariance, hence such a
transformation makes sense only for a theory in which the
electroweak symmetry is broken. This is precisely the case for
the SM, where the masses are induced by spontaneous electroweak
symmetry breaking via the Higgs mechanism. Applying the above
transformation amounts to ``moving'' to the mass basis. The SM
flavor Lagrangian, in the mass basis, is given by (in a unitary
gauge),
\begin{equation}
\begin{split}
\LF_m=&\left(\overline{q^i} D
\hspace*{-.25cm\slash}\ q^j \delta_{ij}\right)_{\rm NC}
+\begin{pmatrix} \overline{u_L} \,\overline{c_L}\,
\overline{t_L}\end{pmatrix}
\begin{pmatrix}y_u&0&0\cr 0&y_c&0\cr 0&0&y_t\end{pmatrix}
\begin{pmatrix} u_R \cr c_R \cr t_R \end{pmatrix}
\left( v+h\right)+(u, c, t)\leftrightarrow(d,s,
b)\\ &+ \frac{g_2}{\sqrt2}{\overline {u_{Li}}}\gamma^\mu
\VKM_{ij}d_{Lj} W_\mu^++{\rm h.c.} ,\,\label{Lflavormass}
\end{split}
\end{equation}
where the subscript NC stands for neutral current interaction
for the gluons, the photon and the $Z$ gauge bosons, $W^\pm$
stands for the charged electroweak gauge bosons, $h$ is the
physical Higgs field, $v\sim 176\,$GeV, $m_i=y_i v$ and $\VKM$
is the CKM matrix \ba\label{VCKM}
\VKM=V_{Q^u}V_{Q^d}^\dagger\,. \ea In general, the CKM is a
$3\times3$ unitary matrix, with 6 imaginary parameters.
However, as evident from Eq. (\ref{Lflavormass}), the charged
current interactions are the only terms which are not invariant
under individual quark vectorial $U(1)^6$ field redefinitions,
\ba u_i,d_j\to e^{i\theta_{u_i,d_j}}\,. \ea The diagonal part
of this transformation corresponds to the classically conserved
baryon current, while the non-diagonal, $U(1)^5$, part of the
transformation can be used to remove 5 out of the 6 phases,
leaving the CKM matrix with a single physical phase. Notice
also that a possible permutation ambiguity for ordering the CKM
entries is removed, given that we have ordered the fields in
\Eq{Lflavormass} according to their masses, light fields first.
This exercise of explicitly identifying the mass basis rotation
is quite instructive, and we have already learned several
important issues regarding how flavor is broken within the SM
(we shall derive the same conclusions using a spurion analysis
in a symmetry oriented manner in Sec.~\ref{spurion}): \bi
\item Flavor conversions only proceed via the three CKM mixing angles.
\item Flavor conversion is mediated via the charged current electroweak interactions.
\item The charge current interactions only involve LH fields.
\ei

Even after removing all the unphysical parameters, there are
various possible forms for the CKM matrix. For example, a
parameterization used by the particle data group~\cite{PDG}, is
given by \ba\label{stapar} \VKM=\begin{pmatrix}
c_{12}c_{13}&s_{12}c_{13}& s_{13}e^{-i\delKM}\cr
-s_{12}c_{23}-c_{12}s_{23}s_{13}e^{i\delKM}&
c_{12}c_{23}-s_{12}s_{23}s_{13}e^{i\delKM}&s_{23}c_{13}\cr
s_{12}s_{23}-c_{12}c_{23}s_{13}e^{i\delKM}&
-c_{12}s_{23}-s_{12}c_{23}s_{13}e^{i\delKM}&c_{23}c_{13}\cr
\end{pmatrix}, \ea where $c_{ij}\equiv\cos\theta_{ij}$ and
$s_{ij}\equiv\sin\theta_{ij}$. The three $\sin\theta_{ij}$ are
the three real mixing parameters, while $\delKM$ is the
Kobayashi-Maskawa phase.

%%%%%%%%%%%%%%%%%%%%%%%%%%%%%%%%%%%%%%%
\subsection{CP violation}
%%%%%%%%%%%%%%%%%%%%%%%%%%%%%%%%%%%%%%%
The SM predictive power picks up once CPV is considered. We
have already proven that the SM flavor sector contains a single
CP violating parameter. Once presented with a SM Lagrangian
where the Yukawa matrices are given in a generic basis, it is
not trivial to determine whether CP is violated or not. This is
even more challenging when discussing beyond the SM dynamics,
where new CP violating sources might be present. A brute force
way to establish that CP is violated would be to show that no
field redefinitions would render a Lagrangian real. For
example, consider a Lagrangian with a single Yukawa matrix,
\ba\label{Yukpairs} {\cal
L}^Y=Y_{ij}\overline{\psi^i_{L}}\phi\psi^j_{R}
+Y_{ij}^*\overline{\psi^j_{R}}\phi^\dagger\psi^i_{L}, \ea where
$\phi$ is a scalar and $\psi^i_X$ is a fermion field. A CP
transformation exchanges the operators \ba\label{CPoper}
\overline{\psi^i_{L}}\phi\psi^j_{R}\leftrightarrow
\overline{\psi^j_{R}}\phi^\dagger\psi^i_{L}, \ea but leaves
their coefficients, $Y_{ij}$ and $Y_{ij}^*$, unchanged, since
CP is a linear unitary non-anomalous transformation. This means
that CP is conserved if \ba Y_{ij}=Y_{ij}^*\,. \ea This is,
however, not a basis independent statement. Since physical
observables do no depend on a specific basis choice, it is
enough to find a basis in which the above relation
holds.\footnote{It is easy to show that in this example, in
fact, CP is not violated for any number of generations.}

Sometimes the brute force way is tedious and might be rather
complicated. A more systematic approach would be to identify a
phase reparameterization invariant or basis independent
quantity, that vanishes in the CP conserving limit. As
discovered in~\cite{Jarlskog1,Jarlskog2}, for the SM case one
can define the following quantity
\beq\label{JarCon}
\CSM =\det[Y_D Y_D^\dagger,Y_U Y_U^\dagger]\,,
\eeq
and the SM is CP violating if and only if
\beq
\Im \left(\CSM\right)\neq0.
\eeq
It is trivial to prove that only if the number of generations
is three or more, then CP is violated. Hence, within the SM,
where CP is broken explicitly in the flavor sector, any CP
violating process must involve all three generations. This is
an important condition, which implies strong predictive power.
Furthermore, all the CPV observables are correlated, since they
are all proportional to a single CP violating parameter,
$\delKM$. Finally, it is worth mentioning that CPV observables
are related to interference between different processes, and
hence are measurements of amplitude ratios. Thus, in various
known cases, they turn out to be cleaner and easier to
interpret theoretically.

%%%%%%%%%%%%%%%%%%%%%%%%%%%%%%%%%%%%%%%
\subsection{The flavor puzzle}
%%%%%%%%%%%%%%%%%%%%%%%%%%%%%%%%%%%%%%%

Now that we have precisely identified the SM physical flavor
parameters, it is interesting to ask what is their experimental
value (using $\rm \overline{MS}$)~\cite{PDG}:
\begin{equation}\label{flavorpara}
\begin{split}
m_u&= 1.5..3.3 \MeV \,, \ m_d=3.5..6.0 \MeV\,, \ m_s = 150^{+30}_{-40}\MeV\,, \nn \\
m_c&=1.3\GeV\,, \ m_b=4.2 \GeV\,, \ m_t = 170\GeV\,, \nn \\
\left|\VKM_{ud}\right|&=0.97\,, \ \left|\VKM_{us}\right|= 0.23\,, \ \left|\VKM_{ub}\right| = 3.9\times 10^{-3}\,, \nn \\
\left|\VKM_{cd}\right|&=0.23 \,, \ \left|\VKM_{cs}\right|=1.0 \,, \ \left|\VKM_{cb}\right| = 41\times 10^{-3}\,, \nn \\
\left|\VKM_{td}\right|&=8.1\times 10^{-3}\,, \
\left|\VKM_{ts}\right|=39\times 10^{-3} \,, \
\left|\VKM_{tb}\right| = 1\,, \ \delKM=77^o\,,
\end{split}
\end{equation}
where $\VKM_{ij}$ corresponds to the magnitude of the $ij$
entry in the CKM matrix, $\delKM$ is the CKM phase, only
uncertainties bigger than~10\% are shown, numbers are shown to
a 2-digit precision and the $\VKM_{ti}$ entries involve
indirect information (a detailed description and refs.~can be
found in~\cite{PDG}).

Inspecting the actual numerical values for the flavor
parameters given in Eq.~\eqref{flavorpara}, shows a peculiar
structure. Most of the parameters, apart from the top mass and
the CKM phase, are small and hierarchical. The amount of
hierarchy can be characterized by looking at two different
classes of observables: \bi
\item Hierarchies between the masses, which are not related to
    flavor converting processes~-- as a measure of these
    hierarchies, we can just estimate what is the size of the
    product of the Yukawa coupling square differences (in the mass basis) \ba
    \frac{\left(m_t^2-m_c^2\right) \left(m_t^2-m_u^2\right)
    \left(m_c^2-m_u^2\right) \left(m_b^2-m_s^2\right)
    \left(m_b^2-m_d^2\right)   \left(m_s^2-m_d^2\right)}{
    v^{12}}=\Ord{10^{-17}}\,. \ea
\item Hierarchies in the mixing which mediate flavor
    conversion~-- this is related to the tiny misalignment between the up and
    down Yukawas; one can quantify this effect in a basis
    independent fashion as follows.
 A CP violating quantity, associated with $\VKM$, that is independent
of parametrization \cite{Jarlskog1,Jarlskog2}, $J_{\rm KM}$, is
defined through
\begin{equation}\label{defJ}
\begin{split}
\Im&\big[\VKM_{ij}\VKM_{kl}
\big(\VKM_{il}\big)^*\big(\VKM_{kj}\big)^*\big]= J_{\rm
KM}\sum_{m,n=1}^3\epsilon_{ikm}\epsilon_{jln}
=\\&=c_{12}c_{23}c_{13}^2s_{12}s_{23}s_{13}\sin\delKM\simeq
\lambda^6 A^2\eta=\Ord{10^{-5}},
\end{split}
\end{equation}
where $i,j,k,l=1,2,3$. We see that even though $\delKM$ is of
order unity, the resulting CP violating parameter is small, as
it is ``screened'' by small mixing angles. If any of the mixing
angles is a multiple of $\pi/2$, then the SM Lagrangian becomes
real. Another explicit way to see that $Y_U$ and $Y_D$ are
quasi aligned is via the Wolfenstein parametrization of the CKM
matrix, where the four mixing parameters are
$(\lambda,A,\rho,\eta)$, with $\lambda=|V_{us}|=0.23$ playing
the role of an expansion parameter~\cite{Wolfenstein}: \ba
\label{WCKM} \VKM=\begin{pmatrix}
1-\frac{\lambda^2}{2}&\lambda&A\lambda^3(\rho-i\eta)\cr
-\lambda&1-\frac{\lambda^2}{2}&A\lambda^2\cr
A\lambda^3(1-\rho-i\eta)&-A\lambda^2&1\cr\end{pmatrix}+{\cal
O}(\lambda^4). \ea Basically, to zeroth order, the CKM matrix
is just a unit matrix\,! \ei

As we shall discuss further below, both kinds of hierarchies
described in the bullets lead to suppression of CPV. Thus, a
nice way to quantify the amount of hierarchies, both in masses
and mixing angles, is to compute the value of the
reparameterization invariant measure of CPV introduced in
Eq.~(\ref{JarCon}) \ba\label{JarConSM} \CSM =J_{\rm KM} \,
\frac{\left(m_t^2-m_c^2\right) \left(m_t^2-m_u^2\right)
\left(m_c^2-m_u^2\right) \left(m_b^2-m_s^2\right)
\left(m_b^2-m_d^2\right) \left(m_s^2-m_d^2\right)}{
v^{12}}={\cal O}\big(10^{-22}\big). \ea This tiny value of
$\CSM$ that characterizes the flavor hierarchy in nature would
be of order 10\% in theories where $Y_{U,D}$ are generic order
one complex matrices. The smallness of $\CSM$ is something that
many flavor models beyond the SM try to address. Furthermore,
SM extensions that have new sources of CPV tend not to have the
SM built-in CP screening mechanism. As a result, they give too
large contributions to the various observables that are
sensitive to CP breaking. Therefore, these models are usually
excluded by the data, which is, as mentioned, consistent with
the SM predictions.

%%%%%%%%%%%%%%%%%%%%%%%%%%%%%%%%%%%%%%%%%%%%%%%%
%%%%%%%%%%%%%%%%%%%%%%%%%%%%%%%%%%%%%%%%%%%%%%%%
\section{Spurion analysis of the SM flavor sector }\label{spurion}
%%%%%%%%%%%%%%%%%%%%%%%%%%%%%%%%%%%%%%%%%%%%%%%%
%%%%%%%%%%%%%%%%%%%%%%%%%%%%%%%%%%%%%%%%%%%%%%%%
In this part we shall try to be more systematic in
understanding the way flavor is broken within the SM. We shall
develop a spurion, symmetry-oriented description for the SM
flavor structure, and also generalize it to NP models with
similar flavor structure, that goes under the name minimal
flavor violation (MFV).

%%%%%%%%%%%%%%%%%%%%%%%%%%%%%%%%%%%%%%%%%%%%%%%%
\subsection{Understanding the SM flavor breaking}\label{spurionsub}
\begin{figure}[htb]
  \begin{center}
    \includegraphics[width=0.4\textwidth]{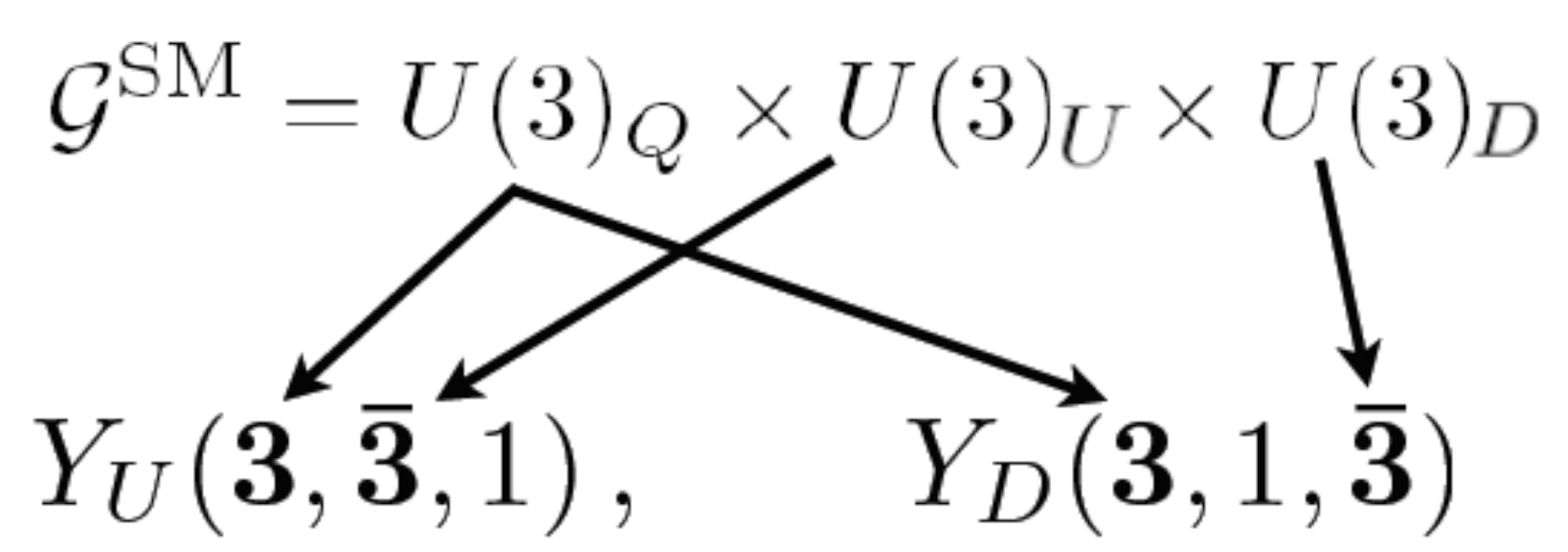}
    \caption{The SM flavor symmetry breaking by the Yukawa matrices.}
    \label{MFVbreaking}
  \end{center}
\end{figure}

\begin{figure}[htb]
  \begin{center}
    \includegraphics[width=0.57\textwidth]{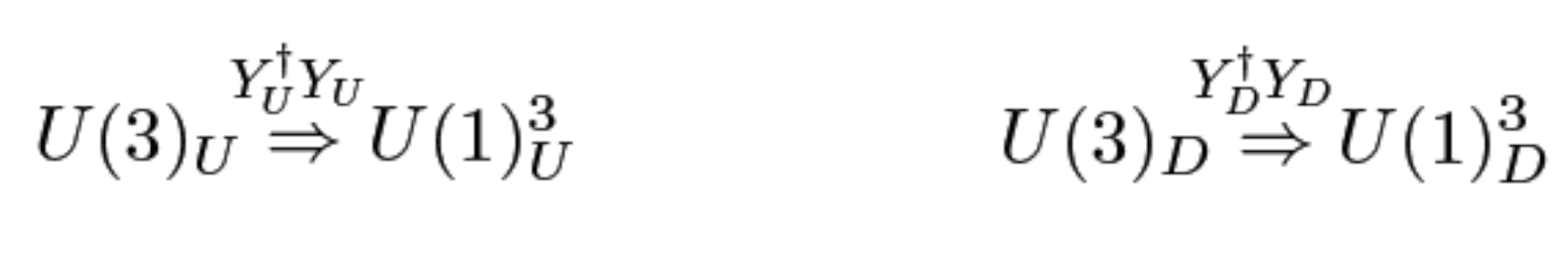}
    \caption{Breaking of the $U(3)_{U,D}$ groups by the Yukawa matrices,
    which form an appropriate LH (RH) flavor group singlet (adjoint+singlet).}
    \label{RHbreaking}
  \end{center}
\end{figure}

%%%%%%%%%%%%%%%%%%%%%%%%%%%%%%%%%%%%%%%%%%%%%%%%
It is clear that if we set the Yukawa couplings of the SM to
zero, we restore the full global flavor group, $ \GSM=
U(3)_Q\times U(3)_U\times U(3)_D\,.$ In order to be able to
better understand the nature of flavor and CPV within the SM,
in the presence of the Yukawa terms, we can use a spurion
analysis as follows. Let us formally promote the Yukawa
matrices to spurion fields, which transform under $\GSM$ in a
manner that makes the SM invariant under the full flavor group
(see \eg~\cite{MFVspurions1} and refs.~therein). From the
flavor transformation given in
Eqs.~(\ref{Qtrans},\ref{Ytrans}), we can read the
representation of the various fields under $\GSM$ (see
illustration in Fig.~\ref{MFVbreaking})
\begin{equation}
\begin{split}
{\rm Fields:}&\ \ Q(\mathbf{3},1,1), \ U(1,\mathbf{3},1), \ D(1,1,\mathbf{ 3})\,;\\
{\rm Spurions:}&\ \ Y_U(\mathbf{3},\mathbf{\bar 3},1), \
Y_D(\mathbf{3},1,\mathbf{\bar 3})\,.
\end{split}
\end{equation}
The flavor group is broken by the ``background'' value of the
spurions $Y_{U,D}\,$, which are bi-fundamentals of $\GSM$. It
is instructive to consider the breaking of the different flavor
groups separately (since $Y_{U,D}$ are bi-fundamentals, the
breaking of quark doublet and singlet flavor groups are linked
together, so this analysis only gives partial information to be
completed below). Consider the quark singlet flavor group,
$U(3)_U\times U(3)_D$, first. We can construct a polynomial of
the Yukawas with simple transformation properties under the
flavor group. For instance, consider the objects \ba
\label{AUD} A_{U,D}\equiv Y_{U,D}^\dagger
Y_{U,D}-\frac{1}{3}\tr\left(Y_{U,D}^\dagger
Y_{U,D}\right)\mathds{1}_3\,. \ea Under the flavor group
$A_{U,D}$ transform as \ba A_{U,D}\to V_{U,D} A_{U,D}
V_{U,D}^\dagger\,. \ea Thus, $A_{U,D}$ are adjoints of
$U(3)_{U,D}$ and singlets of the rest of the flavor group
[while $\tr(Y_{U,D}^\dagger Y_{U,D})$ are flavor singlets]. Via
similarity transformation, we can bring $A_{U,D}$ to a diagonal
form, simultaneously. Thus, we learn that the background value
of each of the Yukawa matrices separately breaks the
$U(3)_{U,D}$ down to a residual $U(1)^3_{U,D}$ group, as
illustrated in Fig.~\ref{RHbreaking}.

Let us now discuss the breaking of the LH flavor group. We can,
in principle, apply the same analysis for the LH flavor group,
$U(3)_Q$, via defining the adjoints (in this case we have two
independent ones),
\beq \label{AQ}
A_{Q^u,Q^d}\equiv Y_{U,D} Y_{U,D}^\dagger- \frac{1}{3}
\tr\left(Y_{U,D} Y_{U,D}^\dagger\right)\mathds{1}_3\,.
\eeq
However, in this case the breaking is more involved, since
$A_{Q^{u,d}}$ are adjoints of the same flavor group. This is a
direct consequence of the $SU(2)$ weak gauge interaction, which
relates the two components of the $SU(2)$ doublets. This
actually motivates one to extend the global flavor group as
follows. If we switch off the electroweak interactions, the SM
global flavor group is actually enlarged to~\cite{Perez:2009xj}
\beq \label{GSMweakless}
\GSM_{\rm weakless}= U(6)_Q\times U(3)_U\times U(3)_D\,,
\eeq
since now each $SU(2)$ doublet, $Q_i\,$, can be split into two
independent flavors, $Q_i^{u,d}\,$, with identical $SU(3)\times
U(1)$ gauge quantum numbers~\cite{weakless}. This limit,
however, is not very illuminating, since it does not allow for
flavor violation at all. To make a progress, it is instructive
to distinguish the $W^3$ neutral current interactions from the
$W^\pm$ charged current ones, as follows: The $W^3$ couplings
are flavor universal, which, however, couple up and down quarks
separately. The $W^\pm$ couplings, $\gW$, link between the up
and down LH quarks. In the presence of only $W^3$ couplings,
the residual flavor group is given by\footnote{To get to this
limit formally, one can think of a model where the Higgs field
is an adjoint of $SU(2)$ and a singlet of color and
hypercharge. In this case the Higgs vacuum expectation value
(VEV) preserves a $U(1)$ gauge symmetry, and the $W^3$ would
therefore remain massless. However, the $W^\pm$ will acquire
masses of the order of the Higgs VEV, and therefore charged
current interactions would be suppressed.}
\beq \label{GSMW3}
\GSME= U(3)_{Q^u}\times  U(3)_{Q^d}\times U(3)_U\times
U(3)_D\,.
\eeq
In this limit, even in the presence of the Yukawa matrices,
flavor conversion is forbidden. We have already seen explicitly
that only the charged currents link between different flavors
(see \Eq{Lflavormass}). It is thus evident that to formally
characterize flavor violation, we can extend the flavor group
from $\GSM\to \GSME$, where now we break the quark doublets to
their isospin components, $U_L,D_L$, and add another spurion,
$\gW$
\begin{equation}\label{flavextent}
\begin{split}
{\rm Fields:}&\ \ U_L(\mathbf{3},1,1,1), \ D_L(1,\mathbf{3},1,1),
\ U(1,1,\mathbf{3},1), \ D(1,1,1,\mathbf{ 3})\\
{\rm Spurions:}&\ \ \gW(\mathbf{3},\mathbf{\bar 3},1,1), \
Y_U(\mathbf{3},1,\mathbf{\bar 3},1), \ Y_D(1,\mathbf{3},1,\mathbf{\bar 3}) \,.
\end{split}
\end{equation}
Flavor breaking within the SM occurs only when $\GSME$ is fully
broken via the Yukawa background values, but also due to the
fact that $\gW$ has a background value. Unlike $Y_{U,D}\,$,
$\gW$ is a special spurion in the sense that its eigenvalues
are degenerate, as required by the weak gauge symmetry. Hence,
it breaks the $U(3)_{Q^u}\times U(3)_{Q^d}$ down to a diagonal
group, which is nothing but $U(3)_Q$. We can identify two bases
where $\gW$ has an interesting background value: The weak
interaction basis, in which the background value of $\gW$ is
simply a unit matrix\footnote{Note that the interaction basis
is not unique, given that $\gW$ is invariant under a flavor
transformation where $Q^u$ and $Q^d$ are rotated by the same
amount~-- see more in the following.} \ba\label{gWint}
\left(\gW\right)_{\rm int}\propto \mathds{1}_3\,, \ea and the
mass basis, where (after removing all unphysical parameters)
the background value of $\gW$ is the CKM matrix \ba
\left(\gW\right)_{\rm mass}\propto \VKM\,. \ea

\begin{figure}[h]
  \begin{center}
    \includegraphics[width=.7\textwidth]{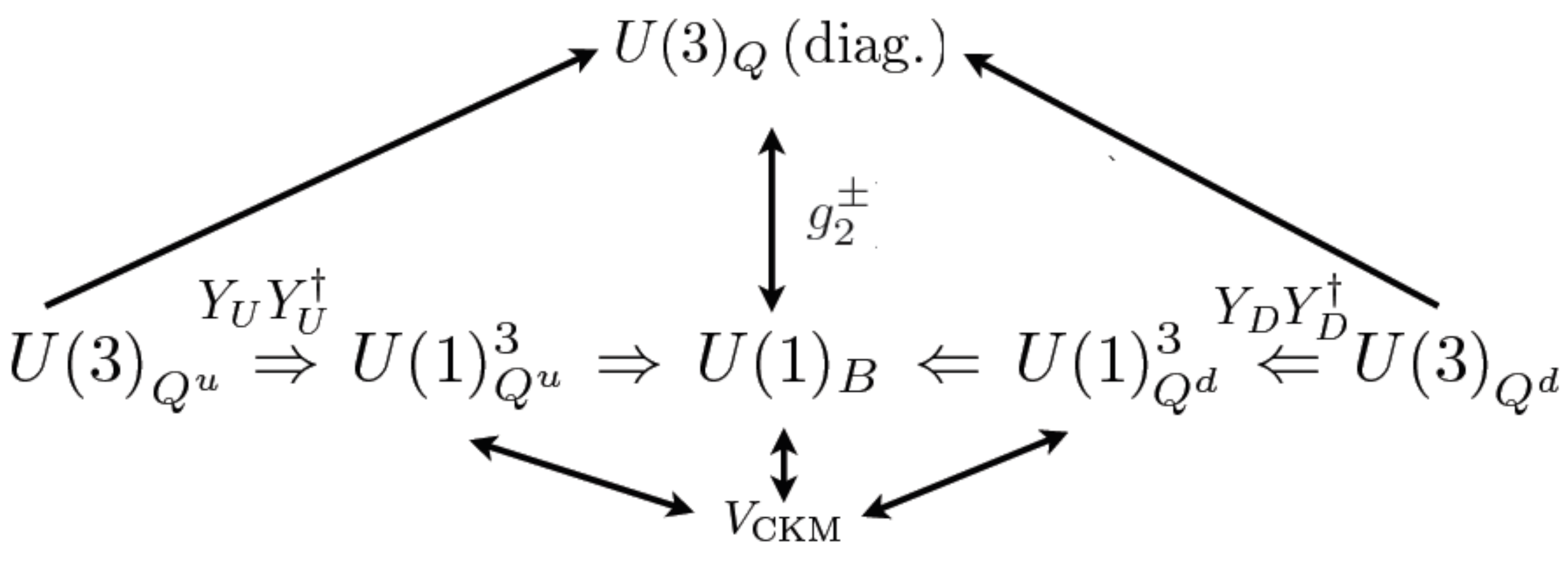}
    \caption{$U(3)_{Q^u,Q^d}$ breaking by $A_{Q^u,Q^d}$ and $\gW$.}
    \label{LHbreaking}
  \end{center}
\end{figure}

\begin{figure}[h]
  \begin{center}
    \includegraphics[width=.55\textwidth]{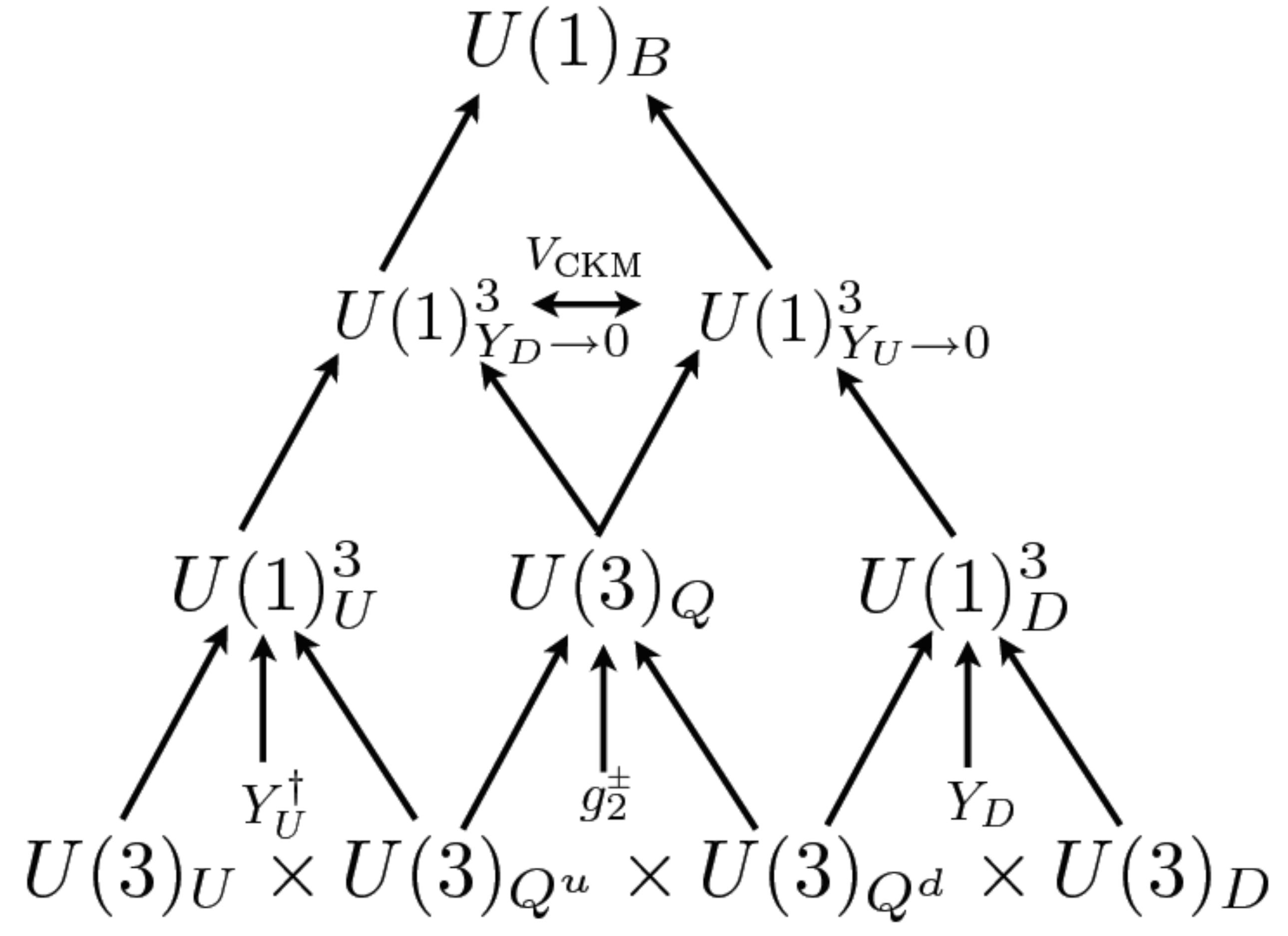}
    \caption{The schematic structure of the various ingredients that mediate flavor breaking within the SM.}
    \label{SMflavorbreaking}
  \end{center}
\end{figure}

Now we are in a position to understand the way flavor
conversion is obtained in the SM. Three spurions must
participate in the breaking: $Y_{U,D}$ and $\gW$. Since $\gW$
is involved, it is clear that generation transitions must
involve LH charged current interactions. These transitions are
mediated by the spurion backgrounds, $A_{Q^u,Q^d}$ (see
Eq.~\eqref{AQ}), which characterize the breaking of the
individual LH flavor symmetries, \ba U(3)_{Q^u}\times
U(3)_{Q^d}\to U(1)^3_{Q^u}\times U(1)^3_{Q^d}\,.\ea Flavor
conversion occurs because of the fact that in general we cannot
diagonalize simultaneously $A_{Q^u,Q^d}$ and $\gW$, where the
misalignment between $A_{Q^u}$ and $A_{Q^d}$ is precisely
characterized by the CKM matrix. This is illustrated in Fig.
\ref{LHbreaking}, where it is shown that the flavor breaking
within the SM goes through collective breaking~\cite{GMFV}~-- a
term often used in the context of little Higgs models (see
\eg~\cite{LHRev} and refs.~therein). We can now combine the LH
and RH quark flavor symmetry breaking to obtain the complete
picture of how flavor is broken within the SM. As we saw, the
breaking of the quark singlet groups is rather trivial. It is,
however, linked to the more involved LH flavor breaking, since
the Yukawa matrices are bi-fundamentals~-- the LH and RH flavor
breaking are tied together. The full breaking is illustrated in
Fig.~\ref{SMflavorbreaking}.

%%%%%%%%%%%%%%%%%%%%%%%%%%%%%%%
\subsection{A comment on description of flavor conversion in physical processes}
%%%%%%%%%%%%%%%%%%%%%%%%%%%%%%%
The above spurion structure allows us to describe SM flavor
converting processes. However, the reader might be confused,
since we have argued above that flavor converting processes
must involve the three spurions, $A_{Q^{u,d}}$ and $\gW$. It is
well known that the rates for charge current processes, which
are described via conversion of down quark to an up one (and
vise a versa), say like beta decay or $b\to u$ transitions, are
only suppressed by the corresponding CKM entry, or $\gW$. What
happened to the dependence on $A_{Q^{u,d}}$? The key point here
is that in a typical flavor precision measurement, the
experimentalists produce mass eigenstate (for example a neutron
or a $B$ meson), and thus the fields involved are chosen to be
in the mass basis. For example, a $b\to c$ process is
characterized by producing a $B$ meson which decays into a
charmed one. Hence, both $A_{Q^{u}}$ and $A_{Q^{d}}$
participate, being forced to be diagonal, but in a nonlinear
way. Physically, we can characterize it by writing an operator
\beq
{\cal O}_{b\to c}=\bar c_{\rm mass} \left(\gW\right)^{cb}_{\rm
mass} b_{\rm mass}\,,
\eeq
where both the $b_{\rm mass}$ and $c_{\rm mass}$ quarks are
mass eigenstate. Note that this is consistent with the
transformation rules for the extended gauge group, $\GSME\,$,
given in Eqs.~\eqref{GSMW3} and~\eqref{flavextent}, where the
fields involved belong to different representations of the
extended flavor group.

The situation is different when FCNC processes are considered.
In such a case, a typical measurement involves mass eigenstate
quarks belonging to the same representation of $\GSME$. For
example, processes that mediate $B^0_d-\overline B^0_d$
oscillation due to the tiny mass difference $\Delta m_{B_d}$
between the two mass eigenstates (which was first measured by
the ARGUS experiment~\cite{ARGUS}), are described via the
following operator, omitting the spurion structure for
simplicity,
\beq
{\cal O}_{\Delta m_{B_d}}=\left(\bar b_{\rm mass}\, d_{\rm
mass}\right)^2\,.
\eeq
Obviously, this operator cannot be generated by SM processes,
as it is violates the $\GSME$ symmetry explicitly. Since it
involves flavor conversion (it violates $b$ number by two
units, hence denoted as $\Delta b=2$ and belongs to $\Delta
F=2$ class of FCNC processes), it must have some power of
$\gW$. A single power of $\gW$ connects a LH down quark to a LH
up one, so the leading contribution should go like $\bar D_L^i
\left(\gW\right)^{ik} \left(\gW^*\right)^{kj} D_L^j$
($i,k,j=1,2,3$). Hence, as expected, this process is mediated
at least via one loop. This would not work as well, since we
can always rotate the down quark fields into the mass basis,
and simultaneously rotate also the up type quarks (away from
their mass basis) so that  $\gW\propto \mathds{1}_3$. These
manipulations define the interaction basis, which is not unique
(see Eq.~\eqref{gWint}). Therefore, the leading flavor
invariant spurion that mediates FCNC transition would have to
involve the up type Yukawa spurion as well. A naive guess would
be
\begin{equation} \label{dmd}
\begin{split}
{\cal O}_{\Delta m_{B_d}}&\propto\left[\bar b_{\rm mass} \left(\gW
\right)_{\rm mass} ^{bk} \left(A_{Q^u}\right)_{kl} \left(\gW^*\right)_{\rm mass} ^{ld} d_{\rm mass}\right]^2 \\
&\sim \left\{\bar b_{\rm mass}  \left[m_t^2 \VKM_{tb}
\left(\VKM_{td}\right)^*+m_c^2 \VKM_{cb} \left(\VKM_{cd}\right)^*\right] d_{\rm mass}\right\}^2\,,
\end{split}
\end{equation}
where it is understood that $\left(A_{Q^u}\right)_{kl}$ is
evaluated in the down quark mass basis (tiny corrections of
order  $m_u^2$ are neglected in the above). This expression
captures the right flavor structure, and is correct for a
sizeable class of SM extensions. However, it is actually
incorrect in the SM case. The reason is that within the SM, the
flavor symmetries are strongly broken by the large top quark
mass~\cite{GMFV}. The SM corresponding amplitude consists of a
rather non-trivial and non-linear function of $A_{Q^u}\,$,
instead of the above naive expression~(see \eg~\cite{BBL} and
refs.~therein), which assumes only the simplest polynomial
dependence of the spurions. The SM amplitude for $\Delta
m_{B_d}$ is described via a box diagram, and two out of the
four powers of masses are canceled, since they appear in the
propagators.

%%%%%%%%%%%%%%%%%%%%%%%%%%%%%%%
\subsection{The SM approximate symmetry structure}
\label{sec:appsym}
%%%%%%%%%%%%%%%%%%%%%%%%%%%%%%%

\begin{figure}[htb]
\centering
\includegraphics[width=5In]{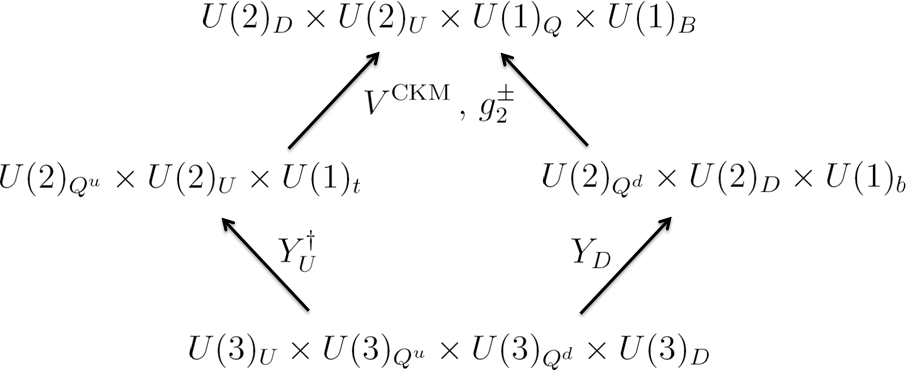}
\caption{The approximate flavor symmetry breaking pattern.
Note that there is also a residual $U(1)_Q$ symmetry, as explained
in Sec.~\ref{sec:3g}.}
\label{fig:approxbreak}
\end{figure}

In the above we have considered the most general breaking
pattern. However, as discussed, the essence of the flavor
puzzle is the large hierarchies in the quark masses, the
eigenvalues of $Y_{U,D}$ and their approximate alignment. Going
back to the spurions that mediate the SM flavor conversions
defined in Eqs.~\eqref{AUD} and~\eqref{AQ}, we can write them
as
\begin{equation}\label{approx}
\begin{split}
A_{U,D}&=\diag\left(0,0,y_{t,b}^2\right)-\frac{y_{t,b}^2}{3}\mathds{1}_3+\Ord{\frac{m_{c,s}^2}{
m_{t,b}^2}}\,, \\
\aud&=\diag\left(0,0,y_{t,b}^2\right)-\frac{y_{t,b}^2}{3}\mathds{1}_3+\Ord{\frac{m_{c,s}^2}{
m_{t,b}^2}}+\Ord{\lambda^2}\,,
\end{split}
\end{equation}
where in the above we took advantage of the fact that
${m_{c,s}^2/m_{t,b}^2},\lambda^2=\Ord{10^{-5,-4,-2}}$ are
small. The hierarchies in the quark masses are translated to an
approximate residual RH $U(2)_U\times U(2)_D$ flavor group (see
Fig.~\ref{fig:approxbreak}), implying that RH currents which
involve light quarks are very small.

We have so far only briefly discussed the role of FCNCs. In the
above we have argued, both based on an explicit calculation and
in terms of a spurion analysis, that at tree level there are no
flavor violating neutral currents, since they must be mediated
through the $W^\pm$ couplings or $\gW$. In fact, this
situation, which is nothing but the celebrated GIM
mechanism~\cite{GIM}, goes beyond the SM to all models in which
all LH quarks are $SU(2)$ doublets and all RH ones are
singlets. The $Z$ boson might have flavor changing couplings in
models where this is not the case.

Can we guess what is the leading spurion structure that induces
FCNC within the SM, say which mediates the $b\to d\nu\bar \nu$
decay process via an operator $\Obdn$? The process changes $b$
quark number by one unit (belongs to $\Delta F=1$ class of FCNC
transitions). It clearly has to contain down type LH quark
fields (let us ignore the lepton current, which is
flavor-trivial; for effects related to neutrino masses and
lepton number breaking in this class of models see
\eg~\cite{KLn1,KLn2,KLn3, KLn4,KLn5,KLn6, KLn7,KLn8,KLn9,
KLn10,KLn11}). Therefore, using the argument presented when
discussing $\Delta m_{B_d}$ (see \Eq{dmd}), the leading flavor
invariant spurion that mediates FCNC would have to involve the
up type Yukawa spurion as well
\beq
\Obdn \propto \bar D_L^i \gW_{ik} \left(A_{Q^u}\right)_{kl}
\gW^*_{lj} D_L^j \times\bar \nu\nu\,.
\eeq

The above considerations demonstrate how the GIM mechanism
removes the SM divergencies from various one loop FCNC
processes, which are naively expected to be log divergent. The
reason is that the insertion of $A_{Q^u}$ is translated to
quark mass difference insertion. It means that the relevant one
loop diagram has to be proportional to $m_i^2-m_j^2$ ($i\neq
j$). Thus, the superficial degree of divergency is lowered by
two units, which renders the amplitude finite.\footnote{For
simplicity, we only consider cases with hard GIM, in which the
dependence on mass differences is polynomial. There is a large
class of amplitudes, for example processes that are mediated
via penguin diagrams with gluon or photon lines, where the
quark mass dependence is more complicated, and may involve
logarithms. The suppression of the corresponding amplitudes
goes under the name soft GIM~\cite{BBL}.} Furthermore, as
explained above (see also~\Eq{tdom}), we can use the fact that
the top contribution dominates the flavor violation to simplify
the form of $\Obdn$
\beq \label{estimate}
\Obdn \sim \frac{g_2^4}{16\pi^2 M_W^2}\, \bar b_L \VKM_{tb}
\left(\VKM_{td}\right)^* d_L\times \bar \nu \nu\,,
\eeq
where we have added a one loop suppression factor and an
expected weak scale suppression. This rough estimation actually
reproduces the SM result up to a factor of about 1.5 (see
\eg~\cite{BBL,BurasMFV1,BurasMFV2, BurasMFV3}).

We thus find that down quark FCNC amplitudes are expected to be
highly suppressed due to the smallness of the top off-diagonal
entries of the CKM matrix. Parameterically, we find the
following suppression factor for transition between the $i$th
and $j$th generations:
\beq\label{fsup}
\begin{split}
b\to s &\propto \left|\VKM_{tb} \VKM_{ts}\right| \sim \lambda^2\,,\\
b\to d &\propto  \left|\VKM_{tb} \VKM_{td}\right| \sim \lambda^3\,,\\
s\to d &\propto \left|\VKM_{td} \VKM_{ts}\right| \sim
\lambda^5\,,
\end{split}
\eeq
where for the $\Delta F=2$ case one needs to simply square the
parametric suppression factors. This simple exercise
illustrates how powerful is the SM FCNC suppression mechanism.
The gist of it is that the rate of SM FCNC processes is small,
since they occur at one loop, and more importantly due to the
fact that they are suppressed by the top CKM off-diagonal
entries, which are very small. Furthermore, since
\beq \label{tdom}
\left|\VKM_{ts,td}\right|\gg \frac{m_{c,u}^2}{m_{t}^2}\,,
\eeq
in most cases the dominant flavor conversion effects are
expected to be mediated via the top Yukawa
coupling.\footnote{This is definitely correct for CP violating
processes, or any ones which involve the third generation
quarks. It also generically holds for new physics MFV models.
Within the SM, for CP conserving processes which involve only
the first two generations, one can find exceptions, for
instance when considering the Kaon and $D$ meson mass
differences, $\Delta m_{D,K}$.}

We can now understand how the SM uniqueness related to
suppression of flavor converting processes arises: \bi
\item RH currents for light quarks are suppressed due to their
    small Yukawa couplings (them being light).
\item Flavor transition occurs to leading order only via LH
    charged current interactions.
\item To leading order, flavor conversion is only due to the
    large top Yukawa coupling. \ei

%%%%%%%%%%%%%%%%%%%%%%%%%%%%%%%%%%%%%%%%%%%%%%%%%%%%%%%
%%%%%%%%%%%%%%%%%%%%%%%%%%%%%%%%%%%%%%%%%%%%%%%%%%%%%%%
\section{Covariant description of flavor violation} \label{covdes}
%%%%%%%%%%%%%%%%%%%%%%%%%%%%%%%%%%%%%%%%%%%%%%%%%%%%%%%
%%%%%%%%%%%%%%%%%%%%%%%%%%%%%%%%%%%%%%%%%%%%%%%%%%%%%%%

The spurion language discussed in the previous section is
useful in understanding the flavor structure of the SM. In the
current section we present a covariant formalism, based on this
language, that enables to express physical observables in an
explicitly basis independent form. This formalism, introduced
in~\cite{Gedalia:2010zs,Gedalia:2010mf}, can be later used to
analyze NP contributions to such observables, and obtain model
independent bounds based on experimental data. We focus only on
the LH sector.

%%%%%%%%%%%%%%%%%%%%%%%%%%%%%%%%%%%%%%%%%%%%%%%%%%%%
%--------------------------------------------------------------
\subsection{Two generations} \label{sec:2g}
%--------------------------------------------------------------
%%%%%%%%%%%%%%%%%%%%%%%%%%%%%%%%%%%%%%%%%%%%%%%%%%%%

We start with the simpler two generation case, which is
actually very useful in constraining new physics, as a result
of the richer experimental precision data. Any hermitian
traceless $2 \times 2$ matrix can be expressed as a linear
combination of the Pauli matrices $\sigma_i$. This combination
can be naturally interpreted as a vector in three dimensional
real space, which applies to $\ad$ and $\au$. We can then
define a length of such a vector, a scalar product, a cross
product and an angle between two vectors, all of which are
basis independent\footnote{The factor of $-i/2$ in the cross
product is required in order to have the standard geometrical
interpretation $\left| \vec{A} \times \vec{B} \right|
=|\vec{A}||\vec{B}|\sin \theta_{AB}$, with $\theta_{AB}$
defined through the scalar product as in
Eq.~\eqref{definitions}.}:
\beq \label{definitions}
\begin{split}
|&\vec{A}| \equiv \sqrt{\frac{1}{2} \tr(A^2)} \, , \quad
\vec{A} \cdot \vec{B} \equiv \frac{1}{2} \tr(A \, B) \, , \quad
\vec{A}
\times \vec{B} \equiv -\frac{i}{2} \left[ A,B \right] \, , \\
&\cos (\theta_{AB}) \equiv \frac{\vec{A} \cdot
\vec{B}}{|\vec{A}| |\vec{B}|}= \frac{\tr(A \, B)}{\sqrt{\tr
(A^2) \tr (B^2)}} \, .
\end{split}
\eeq

\begin{figure}[hbt]
\centering
\includegraphics[width=2.5In]{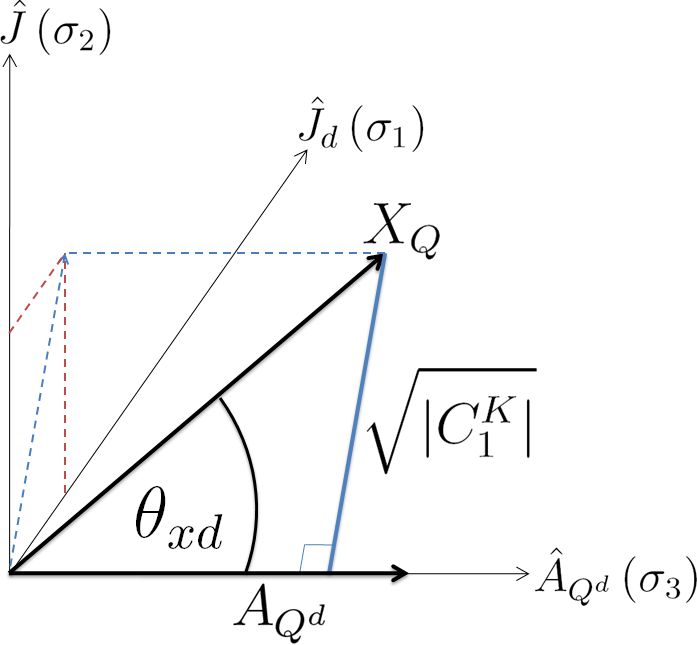}
\caption{The contribution of $X_Q$ to $K^0-\overline{K^0}$ mixing, $\Delta m_K$, given by
the solid blue line. In the down mass basis, $\had$ corresponds to $\sigma_3$,
$\hj$ is $\sigma_2$ and $\hjd$ is $\sigma_1$. The figure is taken from~\cite{Gedalia:2010mf}.}
\label{fig:2g_fv}
\end{figure}

These definitions allow for an intuitive understanding of the
flavor and CP violation induced by a new physics source, based
on simple geometric terms. Consider a dimension six
$SU(2)_L$-invariant operator, involving only quark doublets,
\beq \label{o1}
\frac{C_1}{\Lambda_{\rm NP}^2} O_1=\frac{1}{\Lambda_{\rm NP}^2}
\left[ \overline{Q}_{i} (X_Q)_{ij} \gamma_\mu Q_{j} \right]
\left[ \overline{Q}_{i} (X_Q)_{ij} \gamma^\mu Q_{j} \right] \,
,
\eeq
where $\Lambda_{\rm NP}$ is some high energy
scale.\footnote{This use of effective field theory to describe
NP contributions will be explained in detail in the next
section. Note also that we employ here a slightly different
notation, more suitable for the current needs, than in the next
section.} $X_Q$ is a traceless hermitian matrix, transforming
as an adjoint of $SU(3)_Q$ (or $SU(2)_Q$ for two generations),
so it ``lives'' in the same space as $\ad$ and $\au$. In the
down sector for example, the operator above is relevant for
flavor violation through $K-\overline{K}$ mixing. To analyze
its contribution, we define a covariant orthonormal basis for
each sector, with the following unit vectors
\beq \label{2g_basis}
\haud \equiv \frac{\aud}{\left| \aud \right|} \, , \quad \hj
\equiv \frac{\ad \times \au}{\left|\ad \times \au\right|} \, ,
\quad \hjud \equiv \haud \times \hj \, .
\eeq
Then the contribution of the operator in Eq.~\eqref{o1} to
$\Delta c,s=2$ processes is given by the misalignment between
$X_Q$ and $\aud$, which is equal to
\beq \label{2g_fv}
\left| C_1^{D,K} \right|=\left| { X_Q} \times {\haud} \right|^2
\, .
\eeq
This result is manifestly invariant under a change of basis.
The meaning of Eq.~\eqref{2g_fv} can be understood as follows:
We can choose an explicit basis, for example the down mass
basis, where $\ad$ is proportional to $\sigma_3$. $\Delta s=2$
transitions are induced by the off-diagonal element of $X_Q$,
so that $\left| C_1^K \right|=\left| (X_Q)_{12} \right|^2$.
Furthermore, $\left| (X_Q)_{12} \right|$ is simply the combined
size of the $\sigma_1$ and $\sigma_2$ components of $X_Q$. Its
size is given by the length of $X_Q$ times the sine of the
angle between $X_Q$ and $\ad$ (see Fig.~\ref{fig:2g_fv}). This
is exactly what Eq.~\eqref{2g_fv} describes.

\begin{figure}[htb]
\centering
\includegraphics[width=3.05In]{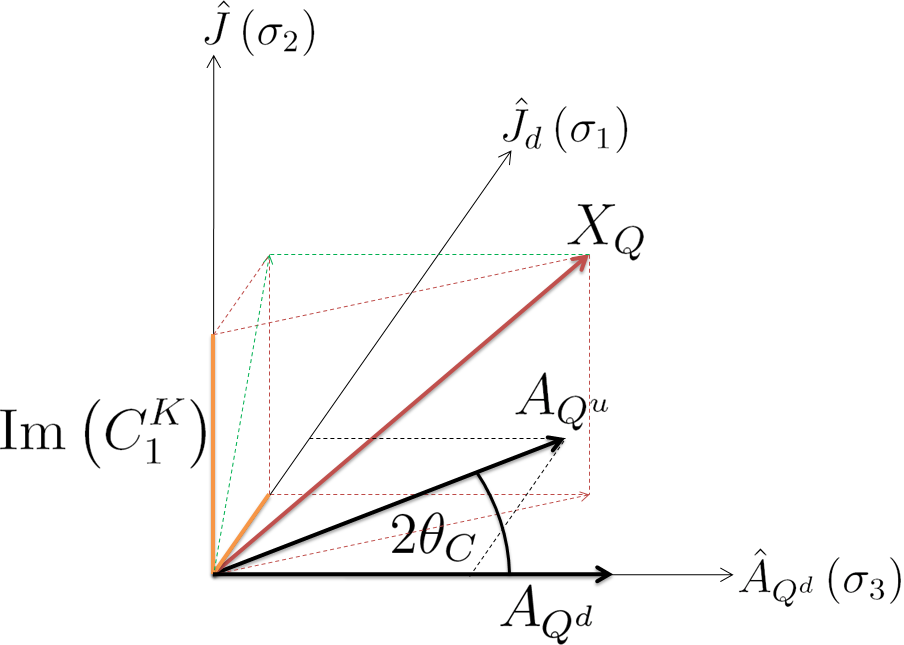}
\caption{CP violation in the Kaon system induced by $X_Q$. $\mathrm{Im}(C_1^K)$ is twice
the product of the two solid orange lines. Note that the angle between $\ad$ and $\au$
 is twice the Cabibbo angle, $\theta_C$.  The figure is taken from~\cite{Gedalia:2010mf}.}
\label{fig:2g_cp}
\end{figure}

Next we discuss CPV, which is given by
\beq \label{cpv_2g}
\mathrm{Im}\left(C_1^{K,D}\right)=2\left(X_Q \cdot \hat
J\right)\left( X_Q \cdot \hat J_{u,d} \right)\,.
\eeq
The above expression is easy to understand in the down basis,
for instance. In addition to diagonalizing $\ad$, we can also
choose $\au$ to reside in the $\sigma_1-\sigma_3$ plane
(Fig.~\ref{fig:2g_cp}) without loss of generality, since there
is no CPV in the SM for two generations. As a result, all of
the potential CPV originates from $X_Q$ in this basis. $C_1^K$
is the square of the off-diagonal element in $X_Q$,
$(X_Q)_{12}$, thus Im$\left(C_1^K\right)$ is simply twice the
real part ($\sigma_1$ component) times the imaginary part
($\sigma_2$ component). In this basis we have $\hj\propto
\sigma_1$ and $\hjd\propto \sigma_2$, this proves the validity
of Eq.~\eqref{cpv_2g}.

An interesting conclusion can be inferred from the analysis
above: In addition to the known necessary condition for CPV in
two generation~\cite{combine}
\beq
X^J \propto \tr \left( X_Q \left[ \ad,\au \right] \right) \neq
0 \,,
\eeq
we identify a second necessary condition, exclusive for $\Delta
F=2$ processes:
\beq
X^{J_{u,d}} \propto\tr \left( X_Q \left[ \aud , \left[ \ad,\au
\right] \right] \right) \neq 0 \, ,
\eeq
These conditions are physically transparent and involve only
observables.

%%%%%%%%%%%%%%%%%%%%%%%%%%%%%%%%%%%%%%%%%%%%%%%%%%%%
%--------------------------------------------------------------
\subsection{Three generations} \label{sec:3g}
%--------------------------------------------------------------
%%%%%%%%%%%%%%%%%%%%%%%%%%%%%%%%%%%%%%%%%%%%%%%%%%%%%

%%%%%%%%%%%%%%%%%%%%%%%%%%%%%%%%%%%%%%%%%%%%%%%%%
\subsubsection{Approximate $U(2)_Q$ limit of massless light quarks}
\label{sec:3g_u2}
%%%%%%%%%%%%%%%%%%%%%%%%%%%%%%%%%%%%%%%%%%%%%%%%%%

For three generations, a simple 3D geometric interpretation
does not naturally emerge anymore, as the relevant space is
characterized by the eight Gell-Mann matrices\footnote{We
denote the Gell-Mann matrices by $\Lambda_i$, where
$\tr(\Lambda_i \Lambda_j)=2\delta_{ij}$. Choosing this
convention allows us to keep the definitions of
Eq.~\eqref{definitions}.}. A useful approximation appropriate
for third generation flavor violation is to neglect the masses
of the first two generation quarks, where the breaking of the
flavor symmetry is characterized by
$[U(3)/U(2)]^2$~\cite{GMFV}. This description is especially
suitable for the LHC, where it would be difficult to
distinguish between light quark jets of different flavor. In
this limit, the 1-2 rotation and the phase of the CKM matrix
become unphysical, and we can, for instance, further apply a
$U(2)$ rotation to the first two generations to ``undo'' the
1-3 rotation. Therefore, the CKM matrix is effectively reduced
to a real matrix with a single rotation angle, $\theta$,
between an active light flavor (say, the 2nd one) and the 3rd
generation,
\beq \label{theta}
\theta\cong \sqrt{\theta_{13}^2+\theta_{23}^2}\,,
\eeq
where $\theta_{13}$ and $\theta_{23}$ are the corresponding CKM
mixing angles. The other generation (the first one) decouples,
and is protected by a residual $U(1)_Q$ symmetry. This can be
easily seen when writing $\ad$ and $\au$ in, say, the down mass
basis
\beq \label{3g_yukawas}
\ad = \frac{y_b^2}{3} \begin{pmatrix} -1 & 0 & 0\\ 0 & -1 & 0\\
0 & 0 & 2 \end{pmatrix} \, , \qquad \au= y_t^2
\begin{pmatrix} \spadesuit & 0 & 0 \\ 0 & \spadesuit & \spadesuit \\
0 & \spadesuit & \spadesuit \end{pmatrix} \, ,
\eeq
where $\spadesuit$ stands for a non-zero \emph{real} entry. The
resulting flavor symmetry breaking scheme is depicted in
Fig.~\ref{fig:approxbreak}, where we now focus only on the LH
sector.

An interesting consequence of this approximation is that a
complete basis cannot be defined covariantly, since $\aud$ in
Eq.~\eqref{3g_yukawas} clearly span only a part of the eight
dimensional space. More concretely, we can identify four
directions in this space: $\hj$ and $\hjud$ from
Eq.~\eqref{2g_basis} and either one of the two orthogonal pairs
\beq \label{3g_basis1}
\haud \quad \mathrm{and} \quad \hat C_{u,d} \equiv 2 \hj \times
\hjud-\sqrt{3} \haud \, ,
\eeq
or
\beq \label{3g_basis2}
\haudp \equiv \hj \times \hjud \quad \mathrm{and} \quad \hjq
\equiv \sqrt3 \hj \times \hjud-2 \haud \, .
\eeq
Note that $\hjq$ corresponds to the conserved $U(1)_Q$
generator, so it commutes with both $\ad$ and $\au$, and takes
the same form in both bases\footnote{The meaning of these basis
elements can be understood from the following: In the down mass
basis we have $\had=-\Lambda_8$, $\hj=\Lambda_7$,
$\hjd=\Lambda_6$ and $\hat C_d=\Lambda_3$. The alternative
diagonal generators from Eq.~\eqref{3g_basis2} are
$\hadp=(\Lambda_3-\sqrt3 \Lambda_8)/2=\mathrm{diag}(0,-1,1)$
and $\hjq=(\sqrt3 \Lambda_3+\Lambda_8)/2=
\mathrm{diag}(2,-1,-1)/\sqrt3$. It is then easy to see that
$\hjq$ commutes with the effective CKM matrix, which is just a
2-3 rotation, and that it corresponds to the $U(1)_Q$
generator, $\mathrm{diag}(1,0,0)$, after trace subtraction and
proper normalization.}. There are four additional directions,
collectively denoted as $\hD$, which transform as a doublet
under the CKM (2-3) rotation, and do not mix with the other
generators. The fact that these cannot be written as
combinations of $\aud$ stems from the approximation introduced
above of neglecting light quark masses. Without this
assumption, it is possible to span the entire space using the
Yukawa matrices~\cite{CPdiag1,CPdiag4,Ellis:2009di}. Despite
the fact that this can be done in several ways, in the next
subsection we focus on a realization for which the basis
elements have a clear physical meaning.

\begin{figure}[htb]
\centering
\includegraphics[width=2.5In]{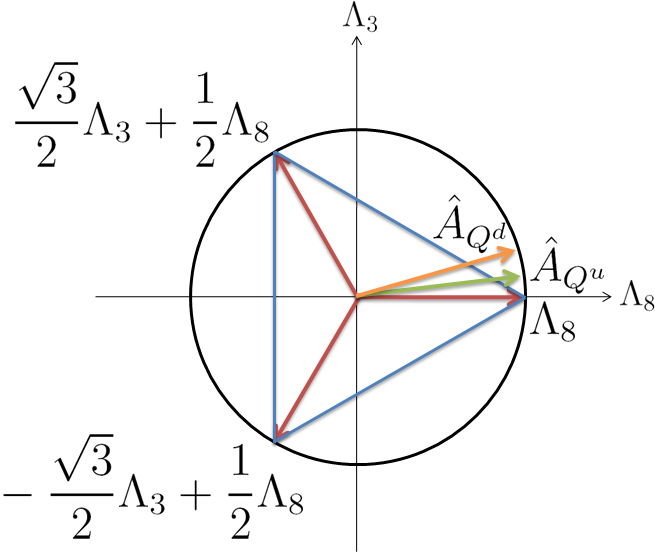}
\caption{The three unit-length diagonal traceless matrices with an inherent
$SU(2)$ symmetry. $\had$ and $\hau$ were schematically added
(their angle to the $\Lambda_8$ axis is actually much smaller than what appears in the plot).
 The figure is taken from~\cite{Gedalia:2010mf}.}
\label{fig:u2}
\end{figure}

It is interesting to notice that a given traceless adjoint
object $X$ in three generations flavor space has an inherent
$SU(2)$ symmetry (that is, two identical eigenvalues) if and
only if it satisfies
\beq
\left[\tr \left( X^2\right)\right]^{3/2}=\sqrt6\, \tr \left(
X^3 \right) \, .
\eeq
In this case it must be a unitary rotation of either
$\Lambda_8$ or its permutations $(\Lambda_8 \pm \sqrt3
\Lambda_3)/2$, which form an equilateral triangle in the
$\Lambda_3-\Lambda_8$ plane (see Fig.~\eqref{fig:u2}).

As before, we wish to characterize the flavor violation induced
by $X_Q$ in a basis independent form. The simplest observable
we can construct is the overall flavor violation of the third
generation quark, that is, its decay to any quark of the first
two generations. This can be written as
\beq \label{inclusive_decay}
\frac{2}{\sqrt{3}} \left| X_Q \times \haud \right| \, ,
\eeq
which extracts $\sqrt{\left| (X_Q)_{13}\right|^2+\left|
(X_Q)_{23}\right|^2}$ in each basis.

%%%%%%%%%%%%%%%%%%%%%%%%%%%%%%%%%%%%%%%%%%%%%%%%%%
\subsubsection{No $U(2)_Q$ limit~-- complete covariant basis}
\label{sec:3g_full}
%%%%%%%%%%%%%%%%%%%%%%%%%%%%%%%%%%%%%%%%%%%%%%%%%%

It is sufficient to restore the masses of the second generation
quarks in order to describe the full flavor space. A
simplifying step to accomplish this is to define the following
object: We take the $n$-th power of $\left( Y_D Y_D^\dagger
\right)$, remove the trace, normalize and take the limit $n \to
\infty$. This is denoted by $\hadn$:
\beq
\hadn \equiv \lim_{n \to \infty} \left\{ \frac{\left( Y_D
Y_D^\dagger\right)^n-\mathds{1}_3\tr \left[ \left( Y_D
Y_D^\dagger\right)^n \right]/3}{\left| \left( Y_D
Y_D^\dagger\right)^n-\mathds{1}_3\tr \left[ \left( Y_D
Y_D^\dagger\right)^n \right]/3 \right|}\right\} \,,
\eeq
and we similarly define $\haun$. Once we take the limit $n\to
\infty$, the small eigenvalues of $\haud$ go to zero, and the
approximation assumed before is formally reproduced. As before,
we compose the following basis elements:
\beq
\hj^n \equiv \frac{\hadn \times \haun}{\left| \hadn \times
\haun \right|} \, , \quad \hjd^n \equiv \frac{\hadn \times
\hj^n}{\left| \hadn \times \hj^n \right|} \, , \quad \cdn
\equiv 2 \hj^n \times \hjd^n -\sqrt{3} \had^n \, ,
\eeq
which are again identical to the previous case. The important
observation for this case is that the $U(1)_Q$ symmetry is now
broken. Consequently, the $U(1)_Q$ generator, $J_Q$, does not
commute with $\ad$ and $\au$ anymore (nor does $\cdn$, which is
different from $J_Q$ only by normalization and a shift by
$\ad$, see Eqs.~\eqref{3g_basis1} and~\eqref{3g_basis2}). It is
thus expected that the commutation relation $[ \ad,\cdn]$
(where $\ad$ now contains also the strange quark mass) would
point to a new direction, which could not be obtained in the
approximation used before. Further commutations with the
existing basis elements should complete the description of the
flavor space.

We thus define
\beq
\hat D_2 \equiv \frac{\had \times \cdn}{\left| \had \times \cdn
\right|} \, .
\eeq
In order to understand the physical interpretation, note that
$\hat D_2$ does not commute with $\ad$, so it must induce
flavor violation, yet it does commute with $\hadn$. The latter
can be identified as a generator of a $U(1)$ symmetry for the
bottom quark (it is proportional to diag(0,0,1) in its diagonal
form, without removing the trace), so this fact means that
$\hat D_2$ preserves this symmetry. Therefore, it must
represent a transition between the first two generations of the
down sector.

We further define
\beq
\hat D_1 \equiv \frac{\had \times \hat D_2}{\left| \had \times
\hat D_2 \right|} \, , \quad \hat D_4 \equiv \frac{\hjd^n
\times \hat D_2}{\left| \hjd^n \times \hat D_2 \right|} \, ,
\quad \hat D_5 \equiv \frac{\hj^n \times \hat D_2}{\left| \hj^n
\times \hat D_2 \right|} \, ,
\eeq
which complete the basis. All of these do not commute with
$\ad$, thus producing down flavor violation. $\hat D_1$
commutes with $\hadn$, so it is of the same status as $\hat
D_2$. The last two elements, $\hat D_{4,5}$, are responsible
for third generation decays, similarly to $\hj^n$ and $\hjd^n$.
More concretely, the latter two involve transitions between the
third generation and what was previously referred to as the
``active'' generation (a linear combination of the first two),
while $\hat D_{4,5}$ mediate transitions to the orthogonal
combination. It is of course possible to define linear
combinations of these four basis elements, such that the decays
to the strange and the down mass eigenstates are separated, but
we do not proceed with this derivation. It is also important to
note that this basis is not completely orthogonal.

In order to give a sense of the physical interpretation of the
different basis elements, it is helpful to see their
decomposition in terms of Gell-Mann matrices, in the down mass
basis (writing only the dependence of the leading terms on
$\lmc$ and $\eta$, and omitting for simplicity $\mathcal{O}(1)$
factors such as the Wolfenstein parameter $A$). This is given
by
\beq
\begin{split}
\hat D_1 \sim& \left\{-1,\eta,0,0,0,0,0,0\right\} \, , \\ \hat
D_2 \sim& \left\{-\eta,-1,0,0,0,0,0,0\right\} \, ,\\ \cdn \sim&
\left\{2\lmc,-2\eta\lmc,1,0,0,0,0,0\right\} \, ,\\
\hat D_4 \sim& \left\{0,0,0,-1,\eta,-\lmc,-\eta\lmc^3,0\right\} \, ,\\
\hat D_5 \sim& \left\{0,0,0,-\eta,-1,\eta\lmc^3,-\lmc,0\right\} \, ,\\
\hjd^n \sim& \left\{0,0,0,-\lmc,\eta\lmc,1,\eta\lmc^2,0\right\} \, ,\\
\hj^n \sim& \left\{0,0,0,-\eta\lmc,-\lmc,-\eta\lmc^2,1,0\right\} \,, \\
\hadn =& \left\{0,0,0,0,0,0,0,-1\right\} \, ,
\end{split}
\eeq
where the values in each set of curly brackets stand for the
$\Lambda_1, \ldots ,\Lambda_8$ components. This shows which
part of an object each basis element extracts under a dot
product, relative to the down sector. For instance, the leading
term in $\hat D_1$ is $\Lambda_1$, therefore it represents the
real part of a $2 \to 1$ transition.

Similarly, it is also useful to see the leading term
decomposition of $\au$ in the down mass basis,
\beq
\au \sim \left\{ -\lmc y_c^2- \lmc^5 yt^2, \eta\lmc^5 y_t^2,
-(y_c^2+\lmc^4 y_t^2)/2, \lmc^3 y_t^2,-\eta\lmc^3 y_t^2,-\lmc^2
y_t^2, -\eta \lmc^4 y_t^2,-y_t^2/\sqrt3 \right\} \, ,
\eeq
neglecting the mass of the up quark.

Finally, an instructive exercise is to decompose $\au$ in this
covariant ``down'' basis, since $\au$ is a flavor violating
source within the SM. Focusing again only on leading terms, we
have
\beq
\begin{split}
\au& \cdot \left\{\hat D_1,\hat D_2,\cdn,\hat D_4,\hat
D_5,\hjd^n,\hj^n,\hadn\right\} \sim \\ &\left\{\lambda y_c^2+
\lambda^5 y_t^2,\lambda y_c^2,(y_c^2+ \lambda^4
y_t^2)/2,\lambda^3 y_c^2,\lambda^3 y_c^2,\lambda^2
y_t^2,0,y_t^2/\sqrt3 \right\} \,.
\end{split}
\eeq
This shows the different types of flavor violation in the down
sector within the SM. It should be mentioned that the $\hat
D_2$ and $\hat D_5$ projections of $\au$ vanish when the CKM
phase is taken to zero, and also when either of the CKM mixing
angles is zero or $\pi/2$. Therefore these basis elements can
be interpreted as CP violating, together with $\hj^n$.

In order to derive model independent bounds in the next
section, we use the simpler description based on the
approximate $U(2)_Q$ symmetry, rather than the full basis.

%%%%%%%%%%%%%%%%%%%%%%%%%%%%%%%%%%%%%%%%%%%%%%%%%%
%%%%%%%%%%%%%%%%%%%%%%%%%%%%%%%%%%%%%%%%%%%%%%%%%%
\section{Model independent bounds}\label{sec:indep}
%%%%%%%%%%%%%%%%%%%%%%%%%%%%%%%%%%%%%%%%%%%%%%%%%%%
%%%%%%%%%%%%%%%%%%%%%%%%%%%%%%%%%%%%%%%%%%%%%%%%%%%

In order to describe NP effects in flavor physics, we can
follow two main strategies: (i) build an explicit ultraviolet
completion of the model, and specify which are the new fields
beyond the SM, or (ii) analyze the NP effects using a generic
effective theory approach, by integrating out the new heavy
fields. The first approach is more predictive, but also more
model dependent. We follow this approach in
Secs.~\ref{sec:susy} and~\ref{sec:exdim} in two well-motivated
SM extensions. In this and the next section we adopt the second
strategy, which is less predictive but also more general.

Assuming the new degrees of freedom to be heavier than SM
fields, we can integrate them out and describe NP effects by
means of a generalization of the Fermi Theory.  The SM
Lagrangian becomes the renormalizable part of a more general
local Lagrangian. This Lagrangian includes an infinite tower of
operators with dimension $d>4$, constructed in terms of SM
fields and suppressed by inverse powers of an effective scale
$\Lambda > M_W$:
\beq
\cL_{\rm eff} = \cL_{\rm SM}  + \sum ~
\frac{C_{i}^{(d)}}{\Lambda^{(d-4)}} ~ O_i^{(d)}({\rm
SM~fields}). \label{eq:effL}
\eeq
This general bottom-up approach allows us to analyze all
realistic extensions of the SM in terms of a limited number of
parameters (the coefficients of the higher dimensional
operators).  The drawback of this method is the impossibility
to establish correlations of NP effects at low and high
energies~-- the scale $\Lambda$ defines the cutoff of the
effective theory.  However, correlations among different low
energy processes can still be established implementing specific
symmetry properties, such as the MFV hypothesis
(Sec.~\ref{MFV}). The experimental tests of such correlations
allow us to test/establish general features of the new theory,
which hold independently of the dynamical details of the model.
In particular, $B$, $D$ and $K$ decays are extremely useful in
determining the flavor symmetry breaking pattern of the NP
model.

%%%%%%%%%%%%%%%%%%%%%%%%%%%%%%%%%%%%%%%%%%%%%%%%%%%%%%%%%%%%%%%%
\subsection{$\Delta F=2$ transitions}
%%%%%%%%%%%%%%%%%%%%%%%%%%%%%%%%%%%%%%%%%%%%%%%%%%%%%%%%%%%%%%%%
The starting point for this analysis is the observation that in
several realistic NP models, we can neglect non-standard
effects in all cases where the corresponding effective operator
is generated at tree level within the SM. This general
assumption implies that the experimental determination of the
CKM matrix via tree level processes is free from the
contamination of NP contributions. Using this determination, we
can unambiguously predict meson-antimeson mixing and FCNC
amplitudes within the SM and compare it with data, constraining
the couplings of the $\Delta F=2$ operators in
Eq.~\eqref{eq:effL}.

%%%%%%%%%%%%%%%%%%%%%%%%%%%%%%%%%%%%%%%%%%%%%%%%%%%%
\subsubsection{From short distance physics to observables}
%%%%%%%%%%%%%%%%%%%%%%%%%%%%%%%%%%%%%%%%%%%%%%%%%%%%
In order to derive bounds on the microscopic dynamics, one
needs to take into account the fact that the experimental input
is usually given at the energy scale in which the measurement
is performed, while the bound is presented at some other scale
(say 1 TeV). Moreover, the contributing higher dimension
operators mix, in general. Finally, all such processes include
long distance contributions (that is, interactions at the
hadronic level) in actual experiments. Therefore, a careful
treatment of all these effects is required. For completeness,
we include here all the necessary information needed in order
to take the above into account.

A complete set of four quark operators relevant for $\Delta
F=2$ transitions is given by
\beq \label{4q_operators}
\begin{split}
Q_1^{q_i q_j}&= \bar q^\alpha_{jL} \gamma_\mu q^\alpha_{iL}
\bar q^\beta_{jL} \gamma_\mu q^\beta_{iL} \,, \\ Q_2^{q_i
q_j}&= \bar q^\alpha_{jR} q^\alpha_{iL} \bar q^\beta_{jR}
q^\beta_{iL} \,, \\ Q_3^{q_i q_j}&= \bar q^\alpha_{jR}
q^\beta_{iL} \bar q^\beta_{jR} q^\alpha_{iL} \,, \\ Q_4^{q_i
q_j}&= \bar q^\alpha_{jR} q^\alpha_{iL} \bar q^\beta_{jL}
q^\beta_{iR} \,, \\ Q_5^{q_i q_j}&= \bar q^\alpha_{jR}
q^\beta_{iL} \bar q^\beta_{jL} q^\alpha_{iR} \,,
\end{split}
\eeq
where $i,j$ are generation indices and $\alpha,\beta$ are color
indices\footnote{note that the operator $Q_1$ has actually
already been defined in Eq.~\eqref{o1} in the previous section,
using a slightly different notation.}. There are also operators
$\tilde Q^{q_i q_j}_{1,2,3}$, which are obtained from $Q^{q_i
q_j}_{1,2,3}$ by the exchange $L \leftrightarrow R$, and the
results given for the latter apply to the former as well.

The Wilson coefficients of the above operators, $C_i(\Lambda)$,
are obtained in principle by integrating out all new particles
at the NP scale\footnote{When a bound is written in terms of an
energy scale, the running should start from this scale, which
is not known {\it a priori}. This is done in an iterative
process, which converges quickly due to the very slow running
of $\alpha_s$ at high scales.}. Then they have to be evolved
down to the hadronic scales $\mu_b=m_b=4.6\,$GeV for bottom
mesons, $\mu_D=2.8\,$GeV for charmed mesons and $\mu_K=2\,$GeV
for Kaons. We denote the Wilson coefficients at the relevant
hadronic scale, which are the measured observables, as $\langle
\overline M \vert \cL_{\rm eff} \vert M \rangle_i$, where $M$
represents a meson (note that $\langle \overline M \vert
\cL_{\rm eff} \vert M \rangle$ has dimension of [mass]). These
should be functions of the Wilson coefficients at the NP scale,
$C_i(\Lambda)$, the running of $\alpha_s$ between the NP and
the hadronic scales and the hadronic matrix elements of the
meson, $\langle \overline M \vert Q_r^{q_i q_j}\vert M \rangle$
(here $q_i q_j$ stand for the quarks that compose the meson
$M$). For bottom and charmed mesons, the analytic formula that
describes this relation is given by~\cite{UTFit,
Becirevic:2001jj}
\beq
\langle \overline M \vert \cL_{\rm eff} \vert M \rangle_i=
\sum_{j=1}^5 \sum_{r=1}^5 \left(b^{(r,i)}_j + \eta\,
c^{(r,i)}_j\right) \eta^{a_j} \, \frac{C_i(\Lambda)}{\Lambda^2}
\, \langle \overline M \vert Q_r^{q_i q_j}\vert M \rangle \,,
\eeq
where $\eta \equiv \alpha_s(\Lambda)/\alpha_s(m_t)$ and $a_j$,
$b^{(r,i)}_j$ and $c^{(r,i)}_j$ are called ``magic numbers''.

For both types of bottom mesons, the non-vanishing magic
numbers are given by
\begin{equation} \label{bmagic}
\begin{array}{l l}
a_i=(0.286, -0.692, 0.787, -1.143, 0.143),& \\
& \\
b^{(11)}_i=(0.865, 0, 0, 0, 0),&
c^{(11)}_i=(-0.017,0,0,0,0),\\
b^{(22)}_i=(0,1.879,0.012,0,0),&
c^{(22)}_i=(0,-0.18,-0.003,0,0),\\
b^{(23)}_i=(0,-0.493,0.18,0,0),&
c^{(23)}_i=(0,-0.014,0.008,0,0),\\
b^{(32)}_i=(0,-0.044,0.035,0,0),&
c^{(32)}_i=(0,0.005,-0.012,0,0),\\
b^{(33)}_i=(0,0.011,0.54,0,0),&
c^{(33)}_i=(0,0,0.028,0,0),\\
b^{(44)}_i=(0,0,0,2.87,0),&
c^{(44)}_i=(0,0,0,-0.48,0.005),\\
b^{(45)}_i=(0,0,0,0.961,-0.22),&
c^{(45)}_i=(0,0,0,-0.25,-0.006),\\
b^{(54)}_i=(0,0,0,0.09,0),&
c^{(54)}_i=(0,0,0,-0.013,-0.016),\\
b^{(55)}_i=(0,0,0,0.029,0.863),&
c^{(55)}_i=(0,0,0,-0.007,0.019).\\
\end{array}
\end{equation}
The hadronic matrix elements are
\beq
\begin{split}
\langle \overline B_q \vert Q_{1}^{bq}  \vert B_q \rangle &=
\frac{1}{3} m_{B_q} f_{B_q}^{2}  B_1^B\,,\\ \langle \overline
B_q \vert Q_{2}^{bq} \vert B_q \rangle &= -\frac{5}{24} \left(
\frac{ m_{B_q} }{ m_{b} + m_q}\right)^{2} m_{B_q} f_{B_q}^{2}
B_{2}^B \,, \\ \langle \overline B_q \vert Q_{3}^{bq} \vert B_q
\rangle &= \frac{1}{24} \left( \frac{ m_{B_q} }{ m_{b} +
m_q }\right)^{2} m_{B_q} f_{B_q}^{2} B_{3}^B \,, \\
\langle \overline B_q \vert Q_{4}^{bq} \vert B_q\rangle &=
\frac{1}{4} \left( \frac{ m_{B_q} }{ m_{b} + m_q }\right)^{2}
m_{B_q} f_{B_q}^{2} B_{4}^B \,, \\ \langle \overline B_q \vert
Q_{5}^{bq} \vert B_q \rangle &= \frac{1}{12} \left(
\frac{m_{B_q}}{ m_{b} + m_q}\right)^{2} m_{B_q} f_{B_q}^{2}
B_{5}^B \,,
\end{split}
\eeq
where $q=d,s$, and the other inputs needed here
are~\cite{UTFit,PDG}
\beq
\begin{split}
m_{B_d}&=5.279 \textrm{ GeV} \,, \ f_{B_d}=0.2 \textrm{ GeV}
\,, \ m_{B_s}=5.366 \textrm{ GeV} \,, \  f_{B_s}=0.262
\textrm{ GeV} \,, \ m_b=4.237 \textrm{ GeV} \,, \\
B_1^B&=0.88 \,, \quad B_2^B=0.82 \,, \quad B_3^B=1.02 \,, \quad
B_4^B=1.15 \,, \quad B_5^B=1.99 \,.
\end{split}
\eeq

For the $D$ meson, the $a_i$ magic numbers are as in
Eq.~\eqref{bmagic}, while the others are given by~\cite{UTFit}
\begin{equation}
\begin{array}{l l}
b^{(11)}_i=(0.837, 0, 0, 0, 0),&
c^{(11)}_i=(-0.016,0,0,0,0),\\
b^{(22)}_i=(0,2.163,0.012,0,0),&
c^{(22)}_i=(0,-0.20,-0.002,0,0),\\
b^{(23)}_i=(0,-0.567,0.176,0,0),&
c^{(23)}_i=(0,-0.016,0.006,0,0),\\
b^{(32)}_i=(0,-0.032,0.031,0,0),&
c^{(32)}_i=(0,0.004,-0.010,0,0),\\
b^{(33)}_i=(0,0.008,0.474,0,0),&
c^{(33)}_i=(0,0,0.025,0,0),\\
b^{(44)}_i=(0,0,0,3.63,0),&
c^{(44)}_i=(0,0,0,-0.56,0.006),\\
b^{(45)}_i=(0,0,0,1.21,-0.19),&
c^{(45)}_i=(0,0,0,-0.29,-0.006),\\
b^{(54)}_i=(0,0,0,0.14,0),&
c^{(54)}_i=(0,0,0,-0.019,-0.016),\\
b^{(55)}_i=(0,0,0,0.045,0.839),&
c^{(55)}_i=(0,0,0,-0.009,0.018).\\
\end{array}
\end{equation}
The $D$ hadronic matrix elements are
\beq
\begin{split}
\langle \overline D \vert Q_{1}^{cu}  \vert D \rangle &=
\frac{1}{3} m_{D} f_{D}^{2}  B_1^D\,,\\ \langle \overline D
\vert Q_{2}^{cu} \vert D \rangle &= -\frac{5}{24} \left( \frac{
m_{D} }{ m_c + m_u}\right)^{2} m_{D} f_D^{2} B_{2}^D \,,
\\ \langle \overline D \vert Q_{3}^{cu} \vert D \rangle &=
\frac{1}{24} \left( \frac{ m_{D} }{ m_c + m_u }\right)^{2} m_D
f_D^{2} B_{3}^D \,, \\ \langle \overline D \vert Q_{4}^{cu}
\vert D \rangle &= \frac{1}{4} \left( \frac{ m_{D} }{ m_c + m_u
}\right)^{2} m_D f_D^{2} B_{4}^D \,, \\ \langle \overline D
\vert Q_{5}^{cu} \vert D \rangle &= \frac{1}{12} \left(
\frac{m_D}{ m_c + m_u}\right)^{2} m_D f_D^{2} B_{5}^D \,,
\end{split}
\eeq
and we also need to use
\beq
\begin{split}
m_D&=1.864 \textrm{ GeV} \,, \quad f_D=0.2 \textrm{ GeV} \,,
\quad m_c=1.3 \textrm{ GeV} \,, \\
B_1^D&=0.865 \,, \quad B_2^D=0.82 \,, \quad B_3^D=1.07 \,,
\quad B_4^D=1.08 \,, \quad B_5^D=1.455 \,.
\end{split}
\eeq

Finally, for Kaons we use a slightly different
formula~\cite{UTFit}
\beq
\langle \overline K \vert \cL_{\rm eff} \vert K \rangle_i=
\sum_{j=1}^5 \sum_{r=1}^5 \left(b^{(r,i)}_j + \eta\,
c^{(r,i)}_j\right) \eta^{a_j} \, \frac{C_i(\Lambda)}{\Lambda^2}
\, R_r \langle \overline K \vert Q_1^{sd}\vert K \rangle \,.
\eeq
The magic numbers are~\cite{Ciuchini:1998ix}
\begin{equation}
\begin{array}{l l}
a_i=(0.29,-0.69,0.79,-1.1,0.14),& \\
& \\
b^{(11)}_i=(0.82,0,0,0,0),&
c^{(11)}_i=(-0.016,0,0,0,0),\\
b^{(22)}_i=(0,2.4,0.011,0,0),&
c^{(22)}_i=(0,-0.23,-0.002,0,0),\\
b^{(23)}_i=(0,-0.63,0.17,0,0),&
c^{(23)}_i=(0,-0.018,0.0049,0,0),\\
b^{(32)}_i=(0,-0.019,0.028,0,0),&
c^{(32)}_i=(0,0.0028,-0.0093,0,0),\\
b^{(33)}_i=(0,0.0049,0.43,0,0),&
c^{(33)}_i=(0,0.00021,0.023,0,0),\\
b^{(44)}_i=(0,0,0,4.4,0),&
c^{(44)}_i=(0,0,0,-0.68,0.0055),\\
b^{(45)}_i=(0,0,0,1.5,-0.17),&
c^{(45)}_i=(0,0,0,-0.35,-0.0062),\\
b^{(54)}_i=(0,0,0,0.18,0),&
c^{(54)}_i=(0,0,0,-0.026,-0.016),\\
b^{(55)}_i=(0,0,0,0.061,0.82),&
c^{(55)}_i=(0,0,0,-0.013,0.018).\\
\end{array}
\end{equation}
We use here only the first (SM) hadronic matrix element,
\beq
\langle \overline K \vert Q_{1}^{sd}  \vert K \rangle =
\frac{1}{3} m_K f_K^2 B_1^K\,,
\eeq
and the others are related to this one by the ratios $R_r$. The
other necessary inputs are thus~\cite{UTFit}
\beq
\begin{split}
m_K&=0.498 \textrm{ GeV} \,, \quad f_K=0.16 \textrm{ GeV} \,,
\quad B_1^K=0.6 \,, \\
R_1&=1 \,, \quad R_2=-12.9 \,, \quad R_3=3.98 \,, \quad
R_4=20.8 \,, \quad R_5=5.2 \,.
\end{split}
\eeq

%%%%%%%%%%%%%%%%%%%%%%%%%%%%%%%%%%%%%%%%%%%%%%%%
\subsubsection{Generic bounds from meson mixing}
%%%%%%%%%%%%%%%%%%%%%%%%%%%%%%%%%%%%%%%%%%%%%%%%
We now move to the actual derivation of bounds on new physics
from $\Delta F=2$ transitions. It is interesting to note that
only fairly recently has the data begun to disfavor models with
only LH currents, but with new sources of flavor and
CPV~\cite{Ligeti,NMFV1,NMFV2}, characterized by a CKM-like
suppression~\cite{Davidson:2007si,aps1, aps2}. In fact, this is
precisely the way that one can test the success of the
Kobayashi-Maskawa mechanism for flavor and CP
violation~\cite{NMFV1,NMFV2,Buras:2009us,UTFit,CKMFitter,
test3,test5,test1, test2,test4,test6,test7,test8}.

We start with the $B_d$ system, where the recent improvement in
measurements has been particularly dramatic, as an example. The
NP contributions to $B_d^0$ mixing can be expressed in terms of
two parameters, $h_d$ and $\sigma_d$, defined by
\beq
M_{12}^d=(1+h_d e^{2i\sigma_d}) M_{12}^{d,{\rm SM}}\,,
\eeq
where $M_{12}^{d,{\rm SM}}$ is the dispersive part of the
$B^0_d-\overline B^0_d$ mixing amplitude in the SM.

In order to constrain deviations from the SM in these
processes, one can use measurements which are directly
proportional to $M_{12}^d$ (magnitude and phase). The relevant
observables in this case are $\Delta m_{B_d}$ and the CPV in
decay with and without mixing in $B^0_d\to \psi K$, $S_{\psi
K}$. These processes are characterized by hard GIM suppression,
and proceed, within the SM, via one loop (see
Eqs.~\eqref{estimate} and~\eqref{fsup}). In the presence of NP,
they can be written as (see \eg~\cite{Nirrev1,
Nirrev2,Nirrev3}):
\beq \label{par}
\begin{split}
\Delta m_{B_d} &= \Delta m_{B_d}^{\rm SM}\, \big|1+h_d
e^{2i\sigma_d} \big| \,,\\ S_{\psi K} &= \sin \big[2\beta +
\arg\big(1+h_d e^{2i\sigma_d}\big)\big]\,.
\end{split}
\eeq
The fact that the SM contribution to these processes involve
the CKM elements which are not measured directly prevents one
from independently constraining the NP contributions. Yet the
situation was dramatically improved when BaBar and Belle
experiments managed to measure CPV processes which, within the
SM, are mediated via tree level amplitudes. The information
extracted from these CP asymmetries in $B^\pm\to DK^\pm$ and
$B\to\rho\rho$ is probably hardly affected by new physics. The
most recent bounds (ignoring $2\sigma$ anomaly in $B\to \tau
\nu$) are~\cite{CKMfitter091,CKMfitter092}
\beq \label{bd_bound}
h_d\lesssim0.3 \ \ \ {\rm and} \ \ \  \pi\lesssim2\sigma_d
\lesssim 2\pi\,.
\eeq

\begin{figure}[htb]
  \begin{center}
   \includegraphics[width=.75\textwidth]{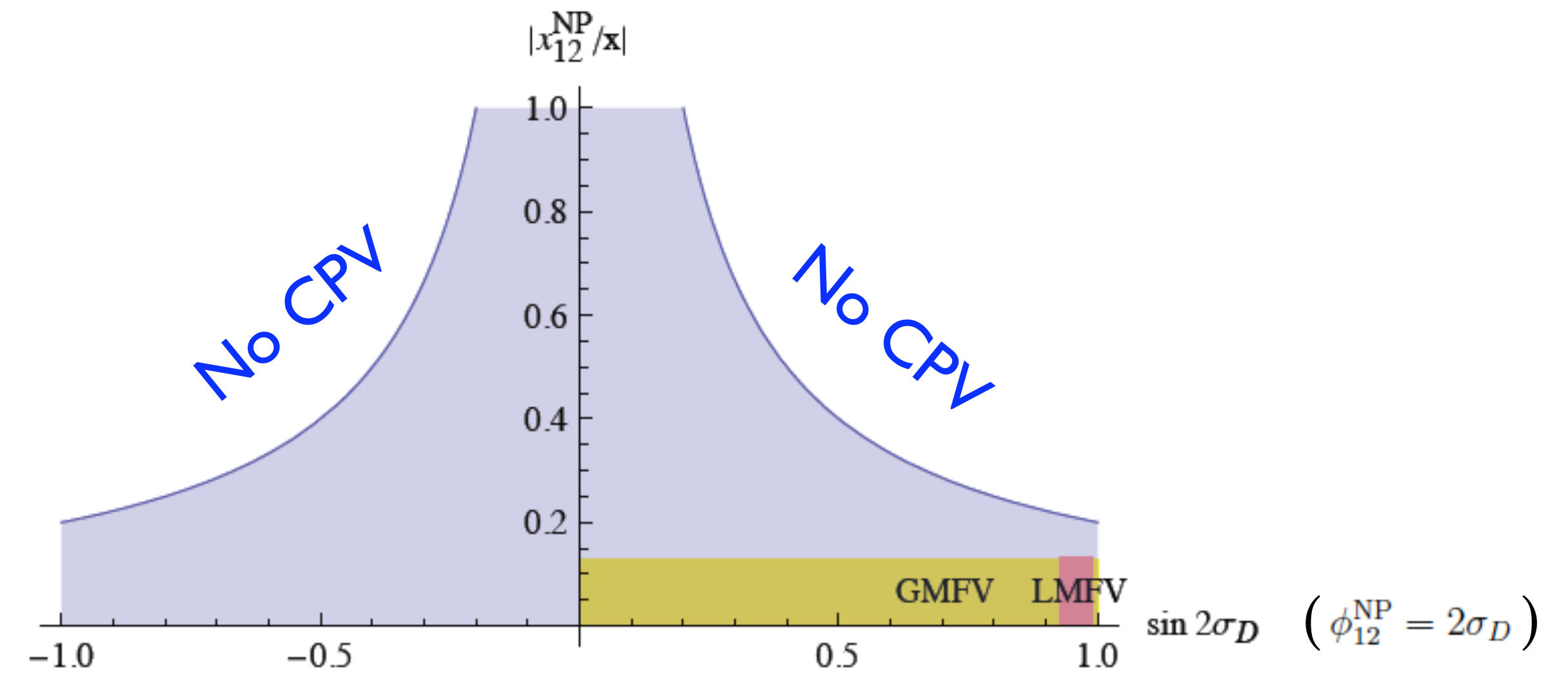}
    \caption{The allowed region, shown in grey, in the
$x_{12}^{\rm NP}/x_{12}-\sin\phi_{12}^{\rm NP}$ plane.
The pink and yellow  regions correspond to the ranges predicted by,
respectively, the linear MFV and general MFV classes of models~\cite{Gedalia:2009kh}.
}
    \label{hDsD_Dsystem}
  \end{center}
\end{figure}

Another example where recent progress has been achieved is in
measurements of CPV in $D^0-\overline D^0$ mixing, which led to
an important improvement of the NP constraints. However, in
this case the SM contributions are unknown~\cite{Dlong1,
Dlong2}, and the only robust SM prediction is the absence of
CPV~\cite{DCPV}. The three relevant physical quantities related
to the mixing can be defined as
\beq \label{thepar}
y_{12}\equiv|\Gamma_{12}|/\Gamma,\qquad x_{12}\equiv2|M_{12}|/
\Gamma,\qquad \phi_{12}\equiv\arg(M_{12}/\Gamma_{12})\,,
\eeq
where $M_{12},\Gamma_{12}$ are the total dispersive and
absorptive part of the $D^0-\overline D^0$ amplitude,
respectively. Fig.~\ref{hDsD_Dsystem} shows (in grey) the
allowed region in the $x_{12}^{\rm NP}/x-\sin\phi_{12}^{\rm
NP}$ plane. $x_{12}^{\rm NP}$ corresponds to the NP
contributions and $x\equiv(m_2-m_1)/\Gamma$, with $m_i,\Gamma$
being the neutral $D$ meson mass eigenstates and average width,
respectively. The pink and yellow regions correspond to the
ranges predicted by, respectively, the linear MFV and general
MFV classes of models~\cite{Gedalia:2009kh} (see Sec.~\ref{MFV}
for details). We see that the absence of observed CP violation
removes a sizable fraction of the possible NP parameter space,
in spite of the fact that the magnitude of the SM contributions
cannot be computed!

An updated analysis of $\Delta F=2$ constraints has been
presented in~\cite{UTFit}. The main conclusions drawn from this
analysis can be summarized as follows:

(i) In all the three accessible short distance amplitudes
($K^0$--$\Knotbar$, $B_d$--$\Bbar_d$, and $B_s$--$\Bbar_s$) the
magnitude of the NP amplitude cannot exceed the SM short
distance contribution. The latter is suppressed by both the GIM
mechanism and the hierarchical structure of the CKM matrix,
\beq
\cA_{\rm SM}^{\Delta F=2} \approx \frac{ G_F^2 m_t^2 }{16
\pi^2} \left[\left(\VKM_{ti}\right)^* \VKM_{tj} \right]^2
\times \langle \Mbar |  (\Qbar_{Li} \gamma^\mu Q_{Lj} )^2 | M
\rangle \times F\left(\frac{M_W^2}{m_t^2}\right),
\eeq
where $F$ is a loop function of $\cO(1)$. As a result, NP
models with TeV scale flavored degrees of freedom and
$\mathcal{O}(1)$ effective flavor mixing couplings are ruled
out. To set explicit bounds, let us consider for instance the
LH $\Delta F=2$ operator $Q_1$ from Eq.~\eqref{4q_operators},
and rewrite it as
\begin{equation}\label{eq:qlql}
\sum_{i\not=j} \frac{c_{ij}}{\Lambda^2} (\Qbar_{Li} \gamma^\mu
Q_{Lj} )^2~,
\end{equation}
where the $c_{ij}$ are dimensionless couplings. The condition
$|\mathcal{A}^{\Delta F=2}_{\rm NP}| < |\mathcal{A}^{\Delta
F=2}_{\rm SM} |$ implies~\cite{Isidori:2010kg}
\begin{eqnarray}
\Lambda > \frac{4.4~{\rm TeV} }{| \left(\VKM_{ti}\right)^* \VKM_{tj}|
/|c_{ij}|^{1/2}} \sim \left\{ \begin{array}{l}
1.3\times 10^4~{\rm TeV} \times |c_{sd}|^{1/2} \!\!\!\!\!\!\! \\
5.1\times 10^2~{\rm TeV} \times |c_{bd}|^{1/2} \!\!\!\!\!\!\! \\
1.1\times 10^2~{\rm TeV} \times |c_{bs}|^{1/2} \!\!\!\!\!\!\!
\end{array}
\right.
\label{eq:bound}
\end{eqnarray}
The strong bounds on $\Lambda$ for generic $c_{ij}$ of order 1
is a manifestation of what in many specific frameworks
(supersymmetry, technicolor, etc.)  goes by the name of the
{\em flavor problem}: if we insist that the new physics emerges
in the TeV region, then it must possess a highly non-generic
flavor structure.

(ii) In the case of $B_d$--$\Bbar_d$ and $K^0$--$\Knotbar$
mixing, where both CP conserving and CP violating observables
are measured with excellent accuracy, there is still room for a
sizable NP contribution (relative to the SM one), provided that
it is, to a good extent, aligned in phase with the SM amplitude
(${\cal O}\left(0.01\right)$ for the $K$ system and ${\cal
  O}\left(0.3\right)$ for the $B_d$ system). This is because the
theoretical errors in the observables used to constrain the
phases, $S_{B_d \to \psi K}$ and $\epsilon_K$, are smaller with
respect to the theoretical uncertainties in $\Delta m_{B_d}$
and $\Delta m_K$, which constrain the magnitude of the mixing
amplitudes.

(iii) In the case of $B_s$--$\Bbar_s$ mixing, the precise
determination of $\Delta m_{B_s}$ does not allow large
deviations in modulo with respect to the SM. The constraint is
particularly severe if we consider the ratio $\Delta
m_{B_d}/\Delta m_{B_s}$, where hadronic uncertainties cancel to
a large extent. However, the constraint on the CPV phase is
quite poor. Present data from CDF~\cite{Aaltonen:2007he} and
D0~\cite{d0:2008fj} indicate a large central value for this
phase, contrary to the SM expectation. The errors are, however,
still large, and the disagreement with the SM is at about the
$2 \sigma$ level.  If the disagreement persists, and becomes
statistically significant, this would not only signal the
presence of physics beyond the SM, but would also rule out a
whole subclass of MFV models (see Sec.~\ref{MFV}).

(iv) The resulting constraints in the $D$ system discussed
above are only second to those from $\epsilon_K$, and unlike
the case of $\epsilon_K$, they are controlled by experimental
statistics, and could possibly be significantly improved in the
near future.

\begin{table}[t]
\begin{center}
\begin{tabular}{c|c c|c c|c} \hline\hline
\rule{0pt}{1.2em}%
Operator &  \multicolumn{2}{c|}{Bounds on $\Lambda$~in~TeV~($c_{ij}=1$)} &
\multicolumn{2}{c|}{Bounds on
$c_{ij}$~($\Lambda=1$~TeV) }& Observables\cr
&   Re& Im & Re & Im \cr
 \hline $(\bar s_L \gamma^\mu d_L )^2$  &~$9.8 \times 10^{2}$& $1.6 \times 10^{4}$
&$9.0 \times 10^{-7}$& $3.4 \times 10^{-9}$ & $\Delta m_K$; $\epsilon_K$ \\
($\bar s_R\, d_L)(\bar s_L d_R$)   & $1.8 \times 10^{4}$& $3.2 \times 10^{5}$
&$6.9 \times 10^{-9}$& $2.6 \times 10^{-11}$ &  $\Delta m_K$; $\epsilon_K$ \\
 \hline $(\bar c_L \gamma^\mu u_L )^2$  &$1.2 \times 10^{3}$& $2.9 \times 10^{3}$
&$5.6 \times 10^{-7}$& $1.0 \times 10^{-7}$ & $\Delta m_D$; $|q/p|, \phi_D$ \\
($\bar c_R\, u_L)(\bar c_L u_R$)   & $6.2 \times 10^{3}$& $1.5 \times 10^{4}$
&$5.7 \times 10^{-8}$& $1.1 \times 10^{-8}$ &  $\Delta m_D$; $|q/p|, \phi_D$\\
\hline$(\bar b_L \gamma^\mu d_L )^2$    &  $5.1 \times 10^{2}$ & $9.3
\times 10^{2}$ &  $3.3 \times 10^{-6}$ &
$1.0 \times 10^{-6}$ & $\Delta m_{B_d}$; $S_{\psi K_S}$  \\
($\bar b_R\, d_L)(\bar b_L d_R)$  &   $1.9 \times 10^{3}$ & $3.6
\times 10^{3}$ &  $5.6 \times 10^{-7}$ &   $1.7 \times 10^{-7}$
&   $\Delta m_{B_d}$; $S_{\psi K_S}$ \\
\hline $(\bar b_L \gamma^\mu s_L )^2$    &  \multicolumn{2}{c|}{$1.1 \times 10^{2}$} &
 \multicolumn{2}{c|}{$7.6\times10^{-5}$}  & $\Delta m_{B_s}$ \\
($\bar b_R \,s_L)(\bar b_L s_R)$  &   \multicolumn{2}{c|}{$3.7 \times 10^{2}$}   &
 \multicolumn{2}{c|}{$1.3\times10^{-5}$} & $\Delta m_{B_s}$ \\
 \hline $(\bar t_L \gamma^\mu u_L )^2$ &  \multicolumn{2}{c|}{$12$} &
 \multicolumn{2}{c|}{$7.1\times10^{-3}$} & $pp \to tt$ \\ \hline\hline
\end{tabular}
\caption{\label{tab:DF2} Bounds on representative dimension six
$\Delta F=2$ operators (taken from~\cite{Isidori:2010kg}, and
the last line is from~\cite{Gedalia:2010zs,Gedalia:2010mf}).
Bounds on $\Lambda$ are quoted assuming an effective coupling
$1/\Lambda^2$, or, alternatively, the bounds on the respective
$c_{ij}$'s assuming $\Lambda=1$ TeV. Observables related to CPV
are separated from the CP conserving ones with semicolons. In
the $B_s$ system we only quote a bound on the modulo of the NP
amplitude derived from $\Delta m_{B_s}$ (see text). For the
definition of the CPV observables in the $D$ system see
Ref.~\cite{DDbarform2}.}
\end{center}
\end{table}

To summarize this discussion, a detailed list of constraints
derived from $\Delta F=2$ observables is shown in
Table~\ref{tab:DF2}, where we quote the bounds for two
representative sets of dimension six operators~-- the left-left
operators (present also in the SM) and operators with a
different chirality, which arise in specific SM extensions
($Q_1$ and $Q_4$ from Eq.~\eqref{4q_operators}, respectively).
The bounds on the latter are stronger, especially in the Kaon
case, because of the larger hadronic matrix elements and
enhanced renormalization group evolution (RGE) contributions.
The constraints related to CPV correspond to maximal phases,
and are subject to the requirement that the NP contributions
are smaller than $30\%$ ($60\%$) of the total
contributions~\cite{NMFV1,NMFV2} in the $B_d$ ($K$) system (see
Eq.~\eqref{bd_bound}). Since the experimental status of CP
violation in the $B_s$ system is not yet settled, we simply
require that the NP contributions would be smaller than the
observed value of $\Delta m_{B_s}$ (for less naive treatments
see {\it e.g.}~\cite{UTFit,Charles:2004jd}, and for a different
type of $\Delta F=2$ analysis see~\cite{Pirjol:2009vz}).

%%%%%%%%%%%%%%%%%%%%%%%%%%%%%%%%%%%%%%%%%%%%%%%%%%%
\subsection{Robust bounds immune to alignment mechanisms} \label{immune}
%%%%%%%%%%%%%%%%%%%%%%%%%%%%%%%%%%%%%%%%%%%%%%%%%%

There are two interesting features for models that can provide
flavor-related suppression factors: degeneracy and alignment.
The former means that the operators generated by the NP are
flavor-universal, that is diagonal in any basis, thus producing
no flavor violation. Alignment, on the other hand, occurs when
the NP contributions are diagonal in the corresponding quark
mass basis. In general, low energy measurements can only
constrain the product of these two factors. An interesting
exception occurs, however, for the left-left (LL) operators of
the type defined in Eq.~\eqref{o1}, where there is an
independent constraint on the level of
degeneracy~\cite{combine}. The crucial point is that operators
involving only quark doublets cannot be simultaneously aligned
with both the down and the up mass bases. For example, we can
take $X_Q$ from Eq.~\eqref{o1} to be proportional to $\ad$.
Then it would be diagonal in the down mass basis, but it would
induce flavor violation in the up sector. Hence, these types of
theories can still be constrained by measurements. The ``best''
alignment is obtained by choosing the NP contribution such that
it would minimize the bounds from both sectors. The strength of
the resulting constraint, which is the weakest possible one, is
that it is unavoidable in the context of theories with only one
set of quark doublets. Here we briefly discuss this issue, and
demonstrate how to obtain such bounds.

%%%%%%%%%%%%%%%%%%%%%%%%%%%%%%%%%%%%%%%%%%%%%%%
\subsubsection{Two generation $\Delta F=2$ transitions}
\label{sec:2g_df2}

As mentioned before, the strongest experimental constraints
involve transitions between the first two generations. When
studying NP effects, ignoring the third generation is often a
good approximation to the physics at hand. Indeed, even when
the third generation does play a role, a two generations
framework is applicable, as long as there are no strong
cancelations with contributions related to the third
generation. Hence, for this analysis we can use the formalism
of Sec.~\eqref{sec:2g}.

The operator defined in Eq.~\eqref{o1}, when restricted to the
first two generations, induces mixing in the $K$ and $D$
systems, and possibly also CP violation. We can use the
covariant bases defined in Eq.~\eqref{2g_basis} to parameterize
$X_Q$,
\beq \label{xq_2g}
X_Q=L \left(X^{u,d} \haud+X^J \hj+X^{J_{u,d}} \hjud \right) \,,
\eeq
and the two bases are related through
\beq \label{u_to_d_2g}
X^u=\cos 2\theta_{\rm C} X^d-\sin 2\theta_{\rm C} X^{J_d} \, ,
\quad X^{J_u}=-\sin 2\theta_{\rm C} X^d-\cos 2\theta_{\rm C}
X^{J_d} \, ,
\eeq
while $X^J$ remains invariant. We choose the $X^i$ coefficients
to be normalized,
\beq
\left(X^d\right)^2+ \left(X^J\right)^2+ \left(X^{J_d}\right)^2=
\left(X^u\right)^2+ \left(X^J\right)^2+
\left(X^{J_u}\right)^2=1 \,,
\eeq
such that $L$ signifies the ``length'' of $X_Q$ under the
definitions in Eq.~\eqref{definitions},
\beq \label{xq_length}
L= \left| X_Q \right|=\left( X_Q^2-X_Q^1 \right)/2 \,,
\eeq
where $X_Q^{1,2}$ are the eigenvalues of $X_Q$ before removing
the trace.

Plugging Eqs.~\eqref{xq_2g} and~\eqref{u_to_d_2g} into
Eq.~\eqref{2g_fv}, we obtain expressions for the contribution
of $X_Q$ to $\Delta m_K$ and $\Delta m_D$, without CPV,
\beq
\begin{split}
C_1^K&=L^2 \left[ \left(X^J\right)^2+ \left(X^{J_d}\right)^2
\right] \,, \\ C_1^D&=\frac{L^2}{2} \left[ 2 \left(X^J\right)^2
+ \left(X^d\right)^2+ \left(X^{J_d}\right)^2 +\left(
\left(X^{J_d}\right)^2- \left(X^d\right)^2 \right) \cos (4
\theta_{\rm C}) +2 X^d X^{J_d} \sin(4 \theta_{\rm C}) \right]
\,.
\end{split}
\eeq
In order to minimize both contributions, we first need to set
$X^J=0$. Next we define
\beq
\tan \alpha \equiv \frac{X^{J_d}}{X^d} \,, \qquad r_{KD} \equiv
\sqrt{\frac{\left(C_1^K\right)_{\rm
exp}}{\left(C_1^D\right)_{\rm exp}}} \,,
\eeq
where the experimental constraints $\left(C_1^K\right)_{\rm
exp}$ and $\left(C_1^D\right)_{\rm exp}$ can be extracted from
Table~\ref{tab:DF2}. Then the weakest bound is obtained for
\beq
\tan \alpha = \frac{r_{KD} \sin (2\theta_{\rm C})}{1+ r_{KD}
\cos (2\theta_{\rm C})} \,,
\eeq
and is given by
\beq
L \leq 3.8 \times 10^{-3} \ltev \,.
\eeq

\begin{figure}[hbt]
\centering
\includegraphics[width=3.8In]{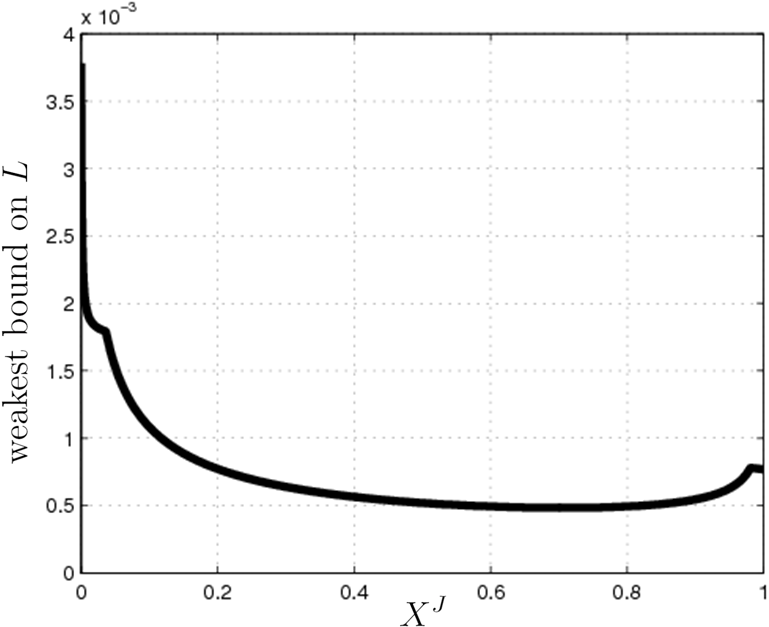}
\caption{The weakest upper bound on $L$ coming from flavor and CPV
 in the $K$ and $D$ systems, as a function of the CP violating
parameter $X^J$, assuming $\Lambda_{\rm NP}=1$ TeV. The figure is
taken from~\cite{combine}.}
\label{fig:2g_bounds}
\end{figure}

A similar process can be carried out for the CPV in $K$ and $D$
mixing, by plugging Eqs.~\eqref{xq_2g} and~\eqref{u_to_d_2g}
into Eq~\eqref{cpv_2g}. Now we do not set $X^J=0$, otherwise
there would be no CPV (since $X_Q$ would reside in the same
plane as $\ad$ and $\au$). Moreover, there are many types of
models in which we can tweak the alignment, but we do not
control the phase (we do not expect the NP to be CP-invariant),
hence they might give rise to CPV. The weakest bound in this
case, as a function of $X^J$, is given by
\beq \label{2g_bound}
L \leq \frac{3.4 \times 10^{-4}}{\left[ \left(X^J\right)^2-
\left(X^J\right)^4 \right]^{1/4}} \ltev \,.
\eeq
The combination of the above two bounds is presented in
Fig.~\ref{fig:2g_bounds}.

We should note that $L$ is simply the difference between the
eigenvalues of $X_Q$ (see Eq.~\eqref{xq_length}), thus the
bounds above put limits on the degeneracy of the NP
contribution.

%--------------------------------------------------------------
\subsubsection{Third generation $\Delta F=1$ transitions} \label{sec:app}
%--------------------------------------------------------------
Similar to the analysis of the previous subsection, we can use
other types of processes to obtain model independent constrains
on new physics. Here we consider flavor violating decays of
third generation quarks in both sectors, utilizing the three
generations framework discussed in Sec.~\ref{sec:3g}. Since the
existing bound on top decay is rather weak, we use the
projection for the LHC bound, assuming that no positive signal
is obtained.

We focus on the following operator
\beq \label{3g_operator}
O^h_{LL}= i \left[ \overline{Q}_i \gamma^\mu (X_Q)_{ij} Q_j
\right] \left[ H^\dagger \overleftrightarrow{D}_\mu H \right] +
\mathrm{h.c.} \, ,
\eeq
which contributes at tree level to both top and bottom
decays~\cite{Fox:2007in}. We omit an additional operator for
quark doublets, $O^u_{LL}= i\left[ {\overline Q}_3 {\tilde H}
\right] \left[ \big( D\!\!\!\!\slash {\tilde H} \big)^\dagger
Q_2 \right] - i\left[ {\overline Q}_3 \big( D\!\!\!\!\slash
{\tilde H} \big) \right] \left[ {\tilde H}^\dagger Q_2
\right]$, which induces bottom decays only at one loop, but in
principle it should be included in a more detailed analysis.

The experimental constraints we use
are~\cite{Aubert:2004it,Iwasaki:2005sy,Carvalho:2007yi}
\beq \label{tbdecay}
\begin{split}
\mathrm{Br}&(B \to X_s\ell^+ \ell^-)_{1 \textrm{
GeV}^2<q^2<6 \textrm{ GeV}^2}=(1.61 \pm 0.51) \times 10^{-6} \, , \\
\mathrm{Br}&(t\to (c,u)Z) <5.5 \times 10^{-5} \, ,
\end{split}
\eeq
where the latter corresponds to the prospect of the LHC bound
in the absence of signal for 100~fb$^{-1}$ at a center of mass
energy of 14 TeV. We adopt the weakest limits on the
coefficient of the operator in Eq.~\eqref{3g_operator},
$C^h_{LL}$, derived in~\cite{Fox:2007in}:
\beq \label{exp_constraints}
\begin{split}
\mathrm{Br}&(B \to X_s\ell^+ \ell^-) \longrightarrow \left|
C^h_{LL} \right|_b < 0.018 \ltev^2 \, , \\ \mathrm{Br}&(t\to
(c,u)Z) \longrightarrow \left| C^h_{LL} \right|_t < 0.18
\ltev^2 \, ,
\end{split}
\eeq
and define $r_{tb} \equiv \left| C^h_{LL} \right|_t/\left|
C^h_{LL} \right|_b\,$.

The NP contribution can be decomposed in the covariant bases as
\beq \label{xq_param}
X_Q =L \left( X'^{u,d} \haudp +X^J \hj+X^{J_{u,d}} \hjud
+X^{J_Q} \hjq + X^{\vec D} \hD \right) \,,
\eeq
where again the coefficients are normalized such that $L=\left|
X_Q \right|$. The contribution of $X_Q$ to third generation
decays is given by Eq.~\eqref{inclusive_decay}. The weakest
bound for a fixed $L$ is obtained, as before, by finding a
direction of $X_Q$ that minimizes the contributions to $\left|
C^h_{LL} \right|_t$ and $\left| C^h_{LL} \right|_b$, thus
constituting the ``best'' alignment. However, since $\hjq$
commutes with $\aud$, as discussed above, it does not
contribute to third generation decay in neither sectors. In
other words, $X_Q \propto \hjq$ is not constrained by such a
process. On the other hand, any component of $X_Q$ may also
generate flavor violation among the first two generations (when
their masses are switched back on), which is more strongly
constrained. Specifically, the bound that stems from the case
of $X_Q \propto \hjq$ is~\cite{Gedalia:2010mf}
\beq \label{3g2g_constraint}
L<0.59 \ltev^2; \quad  \Lambda_{\rm NP}>1.7 \, \mathrm{TeV} \,
,
\eeq
where the latter is for $L=1$. This is stronger than the limit
given below for other forms of $X_Q$, hence this does not
constitute optimal alignment. To conclude this issue, all
directions that contribute to first two generations flavor and
CPV at ${\cal O}(\lambda)$, that is $\hjq$, $\hD$ and $\haudp$,
are not favorable in terms of alignment~\cite{Gedalia:2010mf}.

The induced third generation flavor violation, after removing
these contributions, is then given by
\beq \label{3gfv_explicit}
\frac{4}{3} \left| X_Q\times \haud \right|^2 =
\left(X^J\right)^2+ \left(X^{J_{u,d}}\right)^2  \,,
\eeq
and in order to see this in a common basis, we express
$X^{J_u}$ as
\beq \label{xju}
X^{J_u}=\cos 2\theta \,X^{J_d}+ \sin 2\theta \,X'^d\,,
\eeq
with $\theta$ as defined in Eq.~\eqref{theta}. From this it is
clear that $X^J$ contributes the same to both the top and the
bottom decay rates, so it should be set to zero for optimal
alignment. Thus the best alignment is obtained by varying
$\alpha$, defined as before by
\beq \label{tanalpha}
\tan\alpha \equiv \frac{X^{J_{d}}}{X^d}\,.
\eeq
Here we use $X^d$, which is the coefficient of $\had$, instead
of $X'^d$, since the former does not produce flavor violation
among the first two generations to leading order (up to
$\mathcal{O}(\lambda^5)$).

\begin{figure}[hbt]
\centering
\includegraphics[width=4In]{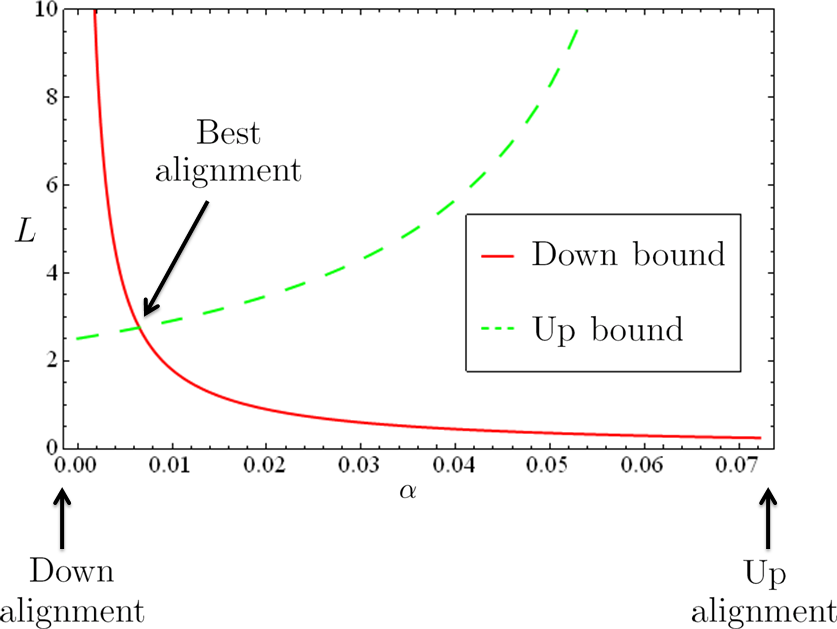}
\caption{Upper bounds on $L$ as a function of $\alpha$, coming from the measurements
of flavor violating decays of the bottom and the top quarks, assuming
$\Lambda_{\rm NP}=1$ TeV. The figure is taken from~\cite{Gedalia:2010zs,Gedalia:2010mf}}
\label{fig:3g_bounds}
\end{figure}

We now consider two possibilities: (i) complete alignment with
the down sector; (ii) the best alignment satisfying the bounds
of Eq.~\eqref{exp_constraints}, which gives the weakest
unavoidable limit. Note that we can also consider up alignment,
but it would give a stronger bound than down alignment, as a
result of the stronger experimental constraints in the down
sector. The bounds for these cases
are~\cite{Gedalia:2010zs,Gedalia:2010mf}
\beq \label{3g_bounds}
\begin{split}
\mathrm{(i)} & \quad \alpha=0 \, , \quad L<2.5 \ltev^2 ; \quad
\Lambda_{NP}>0.63 \,(7.9)\; \mathrm{TeV} \, ,
\\ \mathrm{(ii)} & \quad \alpha=\frac{\sqrt{3}\,\theta}{1+r_{tb}}
\, , \quad L<2.8 \ltev^2 ; \quad \Lambda_{NP}>0.6 \, (7.6) \;
\mathrm{TeV} \, ,
\end{split}
\eeq
as shown in Fig.~\ref{fig:3g_bounds}, where in parentheses we
give the strong coupling bound, in which the coefficient of the
operators in Eqs.~\eqref{o1} and~\eqref{3g_operator} is assumed
to be $16 \pi^2$. Note that these are weaker than the bound in
Eq.~\eqref{3g2g_constraint}.

It is important to mention that the optimized form of $X_Q$
generates also $c \to u$ decay at higher order in $\lambda$,
which might yield stronger constraints than the top decay.
However, the resulting bound from the former is actually much
weaker than the one from the top~\cite{Gedalia:2010mf}.
Therefore, the LHC is indeed expected to strengthen the model
independent constraints.

%--------------------------------------------------------------
\subsubsection{Third generation $\Delta F=2$ transitions} \label{sec:uutt}
%--------------------------------------------------------------
Finally, we analyze $\Delta F=2$ transitions involving the
bottom and the top. For simplicity, we only consider complete
alignment with the down sector
\beq \label{xq2}
X_Q=L \had \,,
\eeq
as the constraints from this sector are much stronger. This
generates in the up sector top flavor violation, and also
$D^0-\overline{D^0}$ mixing at higher order. Yet there is no
top meson, as the top quark decays too rapidly to hadronize.
Instead, we analyze the process $pp \to tt$ (related to mixing
by crossing symmetry), which is most appropriate for the LHC.
It should be emphasized, however, that in this case the parton
distribution functions of the proton strongly break the
approximate $U(2)_Q$ symmetry of the first two generations. The
simple covariant basis introduced in Sec.~\ref{sec:3g_u2},
which is based on this approximate symmetry, cannot be used as
a result. Furthermore, this LHC process is dominated by $uu \to
tt$, so we focus only on the operator involving up (and not
charm) quarks.

The bound that would stem from this process at the LHC was
evaluated in~\cite{Gedalia:2010zs,Gedalia:2010mf} to be
\beq
C_1^{tt}<7.1 \times 10^{-3} \ltev ^2 \, ,
\eeq
for 100 fb$^{-1}$ at 14 TeV. Since the form of $X_Q$ that we
consider also contributes to transitions between the first two
generations, we should additionally take into account the
experimental constraints in the $D$ system, given in
Table~\ref{tab:DF2} (we use the CPV observable).

The contribution of $X_Q$ to these processes is calculated by
applying a CKM rotation to Eq.~\eqref{xq2}. CPV in the $D$
system is then given by $\mathrm{Im}\left[ \left(X_Q
\right)^2_{12} \right]$, and $\left| \left( X_Q \right)_{13}
\right|^2$ describes $uu \to tt$. Note that we have
\beq
\begin{split}
\left(X_Q \right)_{12} &\cong -\sqrt3 \, L \VKM_{ub}
\left(\VKM_{cb}\right)^* \, ,
\\ \left( X_Q \right)_{13} &\cong -\sqrt3 \, L \VKM_{ub} \left(\VKM_{tb}\right)^*
\,.
\end{split}
\eeq
The resulting bounds are
\beq \label{uutt_bound}
L<12 \ltev\,; \quad \Lambda_{\rm NP}>0.08 \, (1.0)\,
\mathrm{TeV} \, ,
\eeq
for $uu \to tt$ and
\beq \label{dd_bound}
L<1.8 \ltev\,; \quad \Lambda_{\rm NP}>0.57 \, (7.2)\,
\mathrm{TeV} \, ,
\eeq
for $D$ mixing.

The limits in Eqs.~\eqref{uutt_bound} and~\eqref{dd_bound} can
be further weakened by optimizing the alignment between the
down and the up sectors, as in the previous subsection. Since
this would only yield a marginal improvement of about 10\%, we
do not analyze this case in detail.

To conclude, we learn that for $\Delta F=2$ processes, the
existing bound is stronger than the one which will be obtained
at the LHC for top quarks, as opposed to $\Delta F=1$ case
considered above.

%%%%%%%%%%%%%%%%%%%%%%%%%%%%%%%%%%%%%%%%%%%%%%%%%%%%%%%%%%%%%%%
\section{Minimal flavor violation}\label{MFV}
%%%%%%%%%%%%%%%%%%%%%%%%%%%%%%%%%%%%%%%%%%%%%%%%%%%%%%%%%%%%%%%

As we have seen above, SM extensions with general flavor
structure are strongly constrained by measurements. This is a
consequence of the fact that, within the SM, flavor conversion
and CP violation arise in a unique and suppressed manner. It is
therefore valuable to investigate beyond the SM theories, where
the breaking of the global flavor symmetries is induced by the
same source as in the SM. In such models, which go under the
name of minimal flavor violation (MFV), flavor violating
interactions are only generated by the SM Yukawa
couplings~~(see {\it e.g.~}\cite{MFVspurions1,MFV1,MFV2,MFV3,
MFV4,MFV5}). Although we only consider here the quark sector,
the notion of MFV can be extended also to the lepton sector.
However, there is no unique way to define the minimal sources
of flavor symmetry breaking, if one wants to keep track of
non-vanishing neutrino masses~\cite{Cirigliano:2005ck,
Davidson:2006bd,Gavela:2009cd}.

In addition to the suppression of FCNCs, there are two
important aspects of the MFV framework: First, low energy
flavor conversion processes can be described by a small set of
operators in an effective Lagrangian, without reference to a
specific model. Furthermore, MFV arises naturally as a low
energy limit of a sizable class of models, in which the flavor
hierarchy is generated at a scale much higher than other
dynamical scales. Examples of microscopic theories which flow
to MFV at the IR are supersymmetric models with gauge or
anomaly mediation~\cite{GMSB1,GMSB2,AMSB1,AMSB2} and a certain
class of warped extra dimension models~\cite{shining1,shining2,
shining3,shining4,shining5,shining6}.

The basic idea can be described in the language of effective
field theory, without the need of referring to a specific
framework. MFV models can have a very different microscopical
dynamics, yet by definition they all have a common origin of
flavor breaking~-- the SM Yukawa matrices. After integrating
out the NP degrees of freedom, we expect to obtain a low energy
effective theory which involves only the SM fields and a bunch
of higher dimension, Lorentz and gauge invariant, operators
suppressed by the NP scale $\LMFV$. Since flavor is broken only
via the SM Yukawas, we can study the flavor breaking of the MFV
framework by the simple following prescription: We should
construct the most general set of higher dimensional operators,
which in addition of being Lorentz and gauge invariant, they
are required also to be flavor invariant, using the spurion
analysis that we have introduced above. A simple example for
such an operator is the one from Eq.~\eqref{o1}, where the
matrix that mixes the generations is a combination of the
appropriate Yukawa matrices,
\beq
O_1^{\rm MFV}=\frac{1}{(\LMFV)^2}\left[\overline Q_i\left( a_u
A_{Q^u} +a_d A_{Q^d}+\ldots \right)_{ij} \gamma_\mu
Q_j\right]^2 \,.
\eeq
Here the dots represent higher order terms in $\aud$.

It is important to realize that quite often, in models which
exhibit MFV-like behavior, the Yukawa couplings are associated
with constant factors, such that they appear as $x_U Y_U,x_D
Y_D$. These factors might come from loop suppression, RGE etc.
In general they are not necessarily small, since for example
large Logs from RGE flow might compensate for the loop
suppression. We should thus consider these ``effective'' Yukawa
couplings $x_i Y_i$, rather than just $Y_i$, as operators would
usually involve $f\left( x_U Y_U,x_D Y_D \right)$.

To get a further insight on the structure of MFV models, it is
useful to classify the framework according to the strength of
breaking of the individual flavor group components. Since,
within the SM, and also as suggested by the data, the only
source of CP breaking is the CKM phase, it is also useful to
extend $\GSM$ of Eq.~\eqref{GSM} to include CP as a discrete
group:
\beq \label{GSMCP}
\GSMCP= U(3)_Q\times U(3)_U\times U(3)_D\times CP\,.
\eeq
The low energy phenomenology of MFV models can be divided as
follows (see below for details):
\begin{itemize}
\item[(i)] Small effective Yukawas~-- The SM flavor group,
    $\GSMCP$, is approximately preserved by the NP
    dynamics, in the sense that all the effective Yukawa
    couplings are small
\beq
\left| x_U y_U^i \right|, \left| x_D y_D^i \right| \ll1\,.
\eeq
\item[(ii)] Large effective top Yukawa~-- The effective
    down type Yukawas are still small, but the top coupling
    is $\mathcal{O}(1)$.
\item[(iii)] Large third generation Yukawas~-- Both the top
    and the bottom effective Yukawa couplings are large.
    This can happen for instance in two Higgs doublet
    models (see {\it e.g.~}\cite{2HDM1,2HDM2,2HDM3,2HDM4})
    with large $\tan \beta$, but also in theories with only
    one Higgs doublet, but a large $x_D$ factor. However,
    in this case CP is only broken by the up and down
    Yukawa matrices, hence no extra sources of flavor
    diagonal phases are present in the microscopic theory.
\item[(iv)] Large effective Yukawas and flavor diagonal
    phases~-- This is the most general case, where both the
    top and bottom Yukawa couplings are large and new
    flavor diagonal CP violating phases are present. It is
    thus denoted as general MFV (GMFV)~\cite{GMFV}.
\end{itemize}

Obviously, within the MFV framework, a built-in approximate
$U(2)$ symmetry for the first two generation is guaranteed for
the low energy phenomenology~\cite{GMFV}. As we shall see, the
models which belong to the MFV class, especially in the cases
of (i)-(iii), enjoy much of the protection against large flavor
violation that we have found to exist in the SM case, and
therefore tend to be consistent with current flavor precision
measurements (for reviews see
\eg~\cite{BurasMFV1,BurasMFV2,BurasMFV3} and refs.~therein).

%%%%%%%%%%%%%%%%%%%%%%%%%%%%%%%%%%%%%%%%%%%%
\subsection{MFV with small effective bottom Yukawa}
%%%%%%%%%%%%%%%%%%%%%%%%%%%%%%%%%%%%%%%%%%%%
Here we deal with the first two cases (i)+(ii), where $x_D y_b
\ll1$. In the following we absorb the $x$ factors into the
Yukawa for simplicity of notation.
%%%%%%%%%%%%%%%%%%%%%%%%%%%%%%%%%%%%%%%%%%%%
\subsubsection{Small effective top Yukawa}
%%%%%%%%%%%%%%%%%%%%%%%%%%%%%%%%%%%%%%%%%%%%
If we are interested in SM processes where the typical energy
scale is much lower than $\LMFV$ and the NP is not strongly
coupled, then we expect that the dominant non-SM flavor
violation would arise from the lowest order higher dimension
operators. For processes involving quark fields, the leading
operators are of dimension six. Consider, for instance, the
following $\Delta F=2$ MFV Lagrangian:
\beq \label{lmfv_df2}
\LMFVF=\frac{1}{(\LMFV)^2}\left[\overline Q_i\left( a_u A_{Q^u}
+a_d A_{Q^d}\right)_{ij} Q_j\right]^2+ \frac{1}{(\LMFV)^2}
\left[ \overline Q_i \left(b \au Y_D\right)_{ij} D_j\right]^2+
\ldots \,,
\eeq
where we write a LL operator and a LR operator for down quarks,
and we assume that they are both suppressed by the same MFV
scale\footnote{Strictly speaking, this does not have to be the
case, as these operators might be generated by different
processes in the underlying theory.}. We can immediately reach
two conclusions: First, the LR operator is subdominant, since
its lowest order flavor violating contribution contains three
Yukawa matrices, compared to two for the LL operator (note that
a term of the form $\overline Q_i (Y_D)_{ij} D_j$ does not
induce down flavor conversion). Next, we only need to take into
account the leading terms, as a result of the small effective
Yukawas. Therefore, this case can be named linear MFV
(LMFV)~\cite{GMFV}.

Let us, for instance, focus on flavor violation in the down
sector, which is more severely constrained. We want to estimate
what is the size of flavor violation which is mediated by
$\LMFVF$, restricting ourselves to the first operator for now.
The experimental information is obtained by looking at the
dynamics (masses, mass differences, decay, time evolution etc.)
of down type mesons, hence we can just look at the form that
$\LMFVF$ takes in the down quark mass basis. By definition
$A_{Q^d}$ is then diagonal, and does not mediate flavor
violation, but $A_{Q^u}$ is not diagonal and is given by
\beq \label{lmfv_exp}
\left(\au\right)_{\rm down}=\VKM \diag (0,0,y_t^2) \left(\VKM
\right)^\dagger- \frac{y_t^2}{3} \mathds{1}_3+
\Ord{\frac{m_c^2}{m_t^2}} \approx y_t^2 \VKM_{ti} \left(
\VKM_{tj} \right)^* \,,
\eeq
where we take advantage of the approximate $U(2)$ symmetry
discussed before. As expected, we find that within the MFV
framework, FCNC processes are suppressed by roughly the same
amount as the SM processes, and therefore are typically
consistent with present data, at least to leading order.

Within the LMFV framework, several of the constraints used to
determine the CKM matrix (and in particular the unitarity
triangle) are not affected by NP~\cite{MFV4,MFV5}. In this
context, NP effects are negligible not only in tree level
processes, but also in a few clean observables sensitive to
loop effects, such as the time dependent CP asymmetry in $B_d
\to \psi K_{L,S}$. Indeed the structure of the basic flavor
changing coupling which results from Eqs.~\eqref{lmfv_df2}
and~\eqref{lmfv_exp} implies that the weak CPV phase of the
operator $(\bar b d)^2$, related to $B_d$--$\Bbar_d$ mixing, is
arg$\left[\left(\VKM_{td} \left( \VKM_{tb} \right)^*\right)^2
\right]$, exactly as in the SM. This construction provides a
natural (\textit{a posteriori}) justification of why no NP
effects have been observed in the quark sector: By
construction, most of the clean observables measured at $B$
factories are insensitive to NP effects in the LMFV framework.

\begin{table}[t]
\begin{minipage}{\textwidth}
\begin{center}
\begin{tabular}{l|c|l}
Operator & ~Bound on $\LMFV$~  & ~Observables \\
\hline\hline
$H^\dagger \left( \Dbar_R Y_D^\dagger \au \sigma_{\mu\nu}
Q_L \right) (e F_{\mu\nu})$
& ~$6.1$~TeV & ~$B\to X_s \gamma$, $B\to X_s \ell^+ \ell^-$\\
$\frac{1}{2} (\Qbar_L  \au \gamma_{\mu} Q_L)^2
\phantom{X^{X^X}_{iii}}$
& ~$5.9$~TeV & ~$\epsilon_K$, $\Delta m_{B_d}$, $\Delta m_{B_s}$ \\
$H_D^\dagger \left( \Dbar_R  Y_D^\dagger \au \sigma_{\mu\nu}
T^a  Q_L \right) (g_s G^a_{\mu\nu})$
&~$3.4$~TeV & ~$B\to X_s \gamma$, $B\to X_s \ell^+ \ell^-$\\
$\left( \Qbar_L \au \gamma_\mu Q_L \right) (\Ebar_R \gamma_\mu E_R)$
& ~$2.7$~TeV & ~$B\to X_s \ell^+ \ell^-$, $B_s\to\mu^+\mu^-$ \\
$~i \left( \Qbar_L \au \gamma_\mu Q_L \right) H_U^\dagger D_\mu H_U$
&~$2.3$~TeV %\footnote{A discrete ambiguity is removed at $90\%$ probability}
&~$B\to X_s \ell^+ \ell^-$, $B_s\to\mu^+\mu^-$\\
$\left( \Qbar_L \au \gamma_\mu Q_L \right) (\Lbar_L \gamma_\mu L_L)$
&~$1.7$~TeV & ~$B\to X_s \ell^+ \ell^-$, $B_s\to\mu^+\mu^-$\\
$\left( \Qbar_L \au \gamma_\mu Q_L \right) (e D_\mu F_{\mu\nu})$
&~$1.5$~TeV & ~$B\to X_s \ell^+ \ell^-$\\
\end{tabular}
\end{center}
\end{minipage}
\caption{\label{tab:MFV} Bounds on the NP scale (at 95\% C.L.)
for some representative $\Delta F=1$~\cite{MFVspurions2} and
$\Delta F=2$~\cite{UTFit} MFV operators (assuming effective
coupling $\pm 1/\Lambda^2$), and corresponding observables used
to set the bounds.}
\end{table}

In Table~\ref{tab:MFV} we report a few representative examples
of bounds on higher dimensional operators in the LMFV
framework. For simplicity, only leading spurion dependence is
shown on the left handed column.  The built-in CKM suppression
leads to bounds on the effective NP scale not far from the TeV
region. These bounds are very similar to the bounds on flavor
conserving operators derived by precision electroweak tests.
This observation reinforces the conclusion that a deeper study
of rare decays is definitely needed in order to clarify the
flavor problem: The experimental precision on the clean FCNC
observables required to obtain bounds more stringent than those
derived from precision electroweak tests (and possibly discover
new physics) is typically in the $1-10\%$ range.
Table~\ref{tab:MFVbounds} demonstrates that discriminating
between the SM and a theory with LMFV behavior is expected to
be a difficult task.

\begin{table}[t]
\begin{minipage}{\textwidth}
\begin{center}
\begin{tabular}{l|l|l|l}
 Observable & ~Experiment  & ~{LMFV prediction}~ & ~{SM prediction} \\
  \hline
  \hline
  $ \beta_s$~from~$\cA_{\rm CP}(B_s \to \psi \phi)$ & ~[0.10, 1.44] @ 95\% CL~
  &  ~$0.04(5)$  & ~$0.04(2)$   \\  \hline
  $\cA_{\rm CP}(B \to X_s \gamma)$ &  ~$< 6\%$ @ 95\% CL~
  &  ~$<0.02$  & ~$<0.01$   \\  \hline
  $\cB(B_d \to \mu^+ \mu^-)$ & ~$<1.8 \times 10^{-8}$
  \qquad    & ~$<1.2 \times 10^{-9}$ & ~$1.3(3)\times 10^{-10}$ \\
  \hline
  $\cB(B \to X_s \tau^+ \tau^-)$ & ~ -- ~  &
  ~$< 5 \times 10^{-7}$ & ~$1.6(5)\times 10^{-7}$ \\
 \hline
  $\cB(K_L \to \pi^0 \nu \bar \nu)$ & ~$<2.6 \times 10^{-8}$ @ 90\% CL
  & ~$<2.9\times 10^{-10}$ & ~$2.9(5)\times 10^{-11}$ \\
\end{tabular}
\end{center}
\end{minipage}
\caption{Some predictions derived in the LMFV framework,
compared to the SM~\cite{Isidori:2010kg}. \label{tab:MFVbounds}
}
\end{table}

%%%%%%%%%%%%%%%%%%%%%%%%%%%%%%%%%%%%%%%%%%%%
\subsubsection{Large effective top Yukawa}
\label{sec:large_top}
%%%%%%%%%%%%%%%%%%%%%%%%%%%%%%%%%%%%%%%%%%%%
The consequence of a large effective top Yukawa, $x_U y_t
\gtrsim 1$, is the need to take into account higher order terms
in the up Yukawa matrix, and resum all these terms to a single
effective contribution. However, the results derived for the
LMFV case are in principle still valid for a large effective
top Yukawa.

Yet, one subtlety does arise in this case: Contributions to
$1\to2$ transitions which proceed through the charm and the top
are correlated within LMFV, but are independent in the current
case (see Sec.~\ref{GMFV} below, and specifically the
discussion around Eqs.~\eqref{compareLMFV} and~\eqref{cbctcc}).
Distinguishing between these cases can be achieved by comparing
$K^+\to \pi^+ \nu\bar\nu$ and the CPV decay $K_L\to \pi^0
\nu\bar\nu$, or via $\epsilon_K$. This needs to be accomplished
both theoretically and experimentally to the level of ${\cal
O}(m_c^2/m_t^2)$. Unfortunately, the smallness of this
difference prevents tests of the first in the near future,
while the second is masked by long distance contributions at
the level of a few percents~\cite{Buras:2010pz}. Nevertheless,
the ability to discriminate between these two cases is of high
theoretical importance, since it yields information about short
distance physics (such as the mediation scale of supersymmetry
breaking via the Logs' size or anomalous dimensions) well
beyond the direct reach of near future experiments.

%%%%%%%%%%%%%%%%%%%%%%%%%%%%%%%%%%%%%%%%%%%%
\subsection{Large bottom Yukawa}
%%%%%%%%%%%%%%%%%%%%%%%%%%%%%%%%%%%%%%%%%%%%

The effects of a large effective bottom Yukawa usually appear
in two Higgs doublet models (such as supersymmetry), but they
can also be found in other NP frameworks without an extended
Higgs sector, where $x_D y_b$ is of order one due to a large
value of $x_D$. In any case, we can still assume that the
Yukawa couplings are the only irreducible breaking sources of
the flavor group.

For concreteness, we analyze the case of a two Higgs doublet
model, which is described by the Lagrangian in
Eq.~\eqref{Lflavor} (focusing only on the quark sector) with
independent $H_U$ and $H_D$. This Lagrangian is invariant under
an extra $U(1)$ symmetry with respect to the one Higgs case~--
a symmetry under which the only charged fields are $D$ (charge
$+1$) and $H_D$ (charge $-1$). This symmetry, denoted
$U(1)_{\rm PQ}$, prevents tree level FCNCs, and implies that
$Y_{U,D}$ are the only sources of flavor breaking appearing in
the Yukawa interaction (similar to the one Higgs doublet
scenario). By assumption, this also holds for all the low
energy effective operators. This is sufficient to ensure that
flavor mixing is still governed by the CKM matrix, and
naturally guarantees a good agreement with present data in the
$\Delta F=2$ sector. However, the extra symmetry of the Yukawa
interaction allows us to change the overall normalization of
$Y^{U,D}$ with interesting phenomenological consequences in
specific rare modes.

The normalization of the Yukawa couplings is controlled by the
ratio of the vacuum expectation values of the two Higgs fields,
or by the parameter
\beq
\tan\beta = \langle H_U\rangle/\langle H_D\rangle \,.
\eeq
For $\tan\beta\gg1 $, the smallness of the $b$ quark (and
$\tau$ lepton) mass can be attributed to the smallness of
$1/\tan\beta$, rather than to the corresponding Yukawa
coupling. As a result, for $\tan\beta\gg1$ we cannot anymore
neglect the down type Yukawa coupling. Moreover, the $U(1)_{\rm
PQ}$ symmetry cannot be exact~-- it has to be broken at least
in the scalar potential in order to avoid the presence of a
massless pseudoscalar Higgs. Even if the breaking of $U(1)_{\rm
PQ}$ and $\GSM$ are decoupled, the presence of $U(1)_{\rm PQ}$
breaking sources can have important implications on the
structure of the Yukawa interaction, especially if $\tan\beta$
is large~\cite{MFVspurions1, Hall:1993gn,Blazek:1995nv,
Isidori:2001fv}.

Since the $b$ quark Yukawa coupling becomes $\mathcal{O}(1)$,
the large $\tan\beta$ regime is particularly interesting for
helicity-suppressed observables in $B$ physics. One of the
clearest phenomenological consequences is a suppression
(typically in the $10-50\%$ range) of the $B \to \ell \nu$
decay rate with respect to its SM expectation~\cite{Hou:1992sy,
Akeroyd:2003zr,Isidori:2006pk}. Potentially measurable effects
in the $10-30\%$ range are expected also in $B\to X_s
\gamma$~\cite{Carena:2000uj,Degrassi:2000qf,Carena:1999py} and
$\Delta M_{B_s}$~\cite{Buras:2002vd,Buras:2001mb}. Given the
present measurements of $B \to \ell \nu$, $B\to X_s \gamma$ and
$\Delta M_{B_s}$, none of these effects seems to be favored by
data. However, present errors are still sizable compared to the
estimated NP effects.

The most striking signature could arise from the rare decays
$B_{s,d}\to \ell^+\ell^-$, whose rates could be enhanced over
the SM expectations by more than one order of
magnitude~\cite{Hamzaoui:1998nu,Choudhury:1998ze, Babu:1999hn}.
An enhancement of both $B_{s}\to \ell^+\ell^-$ and $B_{d}\to
\ell^+\ell^-$ respecting the MFV relation $\Gamma(B_{s}\to
\ell^+\ell^-)/\Gamma(B_{d}\to \ell^+\ell^-) \approx
|\VKM_{ts}/\VKM_{td}|^2$ would be an unambiguous signature of
MFV at large $\tan\beta$~\cite{MFVspurions2}.

Dramatic effects are also possible in the up sector. The
leading contribution of the LL operator to $D-\overline D$
mixing is given by
\beq
C_1^{cu}\propto \left[y_s^2\left(\VKM_{cs}\right)^* \VKM_{us}
+(1+r_{\rm GMFV}) y_b^2\left(\VKM_{cb}\right)^* \VKM_{ub}
\right]^2 \sim 3\times10^{-8}\zeta_1 \,,
\eeq
for $\tan \beta \sim m_t/m_b$, where $r_{\rm GMFV}$ accounts
for the necessary resummation of the down Yukawa, and is
expected to be an order one number. In such a case, the simple
relation between the contribution from the strange and bottom
quarks does not apply~\cite{GMFV}. We thus have
\beq
\begin{split}
\zeta_1&= e^{2i\gamma} + 2r_{sb}
e^{i\gamma}+r_{sb}^2\sim1.7i+r_{\rm GMFV}
\left[2.4 i-1-0.7 \,r_{\rm GMFV}\left(1+i\right)\right]\,,\\
r_{sb}&\equiv \frac{y_s^2}{y_b^2} \left|
\frac{\VKM_{us}\VKM_{cs}}{\VKM_{ub}\VKM_{cb}} \right|\sim0.5
\,,
\end{split}
\eeq
where $\gamma\approx67^o$ is the relevant phase of the
unitarity triangle. We thus learn that MFV models with two
Higgs doublets can contribute to $D-\overline D$ mixing up to
${\cal O}(0.1)$ for very large $\tan\beta$, assuming a TeV NP
scale. Moreover, the CPV part of these contributions is not
suppressed compared to the CP conserving part, and can provide
a measurable signal. In Fig.~\ref{hDsD_Dsystem} we show in pink
(yellow) the range predicted by the LMFV (GMFV) class of
models. The GMFV yellow band is obtained by scanning the range
$r_{\rm GMFV}\in(-1,+1)$ (but keeping the magnitude of
$C_1^{cu}$ fixed for simplicity).

Sizeable contributions to top FCNC can also emerge for large
$\tan\beta$. For a MFV scale of~$\sim1$~TeV, this can lead to
$Br(t \to cX) \sim \mathcal{O}(10^{-5})$~\cite{GMFV}, which may
be within the reach of the LHC.

%%%%%%%%%%%%%%%%%%%%%%%%%%%%%%%
\subsection{General MFV}\label{GMFV}
%%%%%%%%%%%%%%%%%%%%%%%%%%%%%%%

The breaking of the $\GSM$ flavor group and the breaking of the
discrete CP symmetry are not necessarily related, and we can
add flavor diagonal CPV phases to generic MFV
models~\cite{CPdiag1,CPdiag4,Ellis:2007kb}. Because of the
experimental constraints on electric dipole moments (EDMs),
which are generally sensitive to such flavor diagonal
phases~\cite{CPdiag4}, in this more general case the bounds on
the NP scale are substantially higher with respect to the
``minimal'' case, where the Yukawa couplings are assumed to be
the only breaking sources of both
symmetries~\cite{MFVspurions1}.

If $\tan\beta$ is large, the inclusion of flavor diagonal
phases has interesting effects also in flavor changing
processes~\cite{CPdiag3,CPdiag5,CPdiag6}. The main
consequences, derived in a model independent manner, can be
summarized as follows~\cite{GMFV}: (i) extra CPV can only arise
from flavor diagonal CPV sources in the UV theory; (ii) the
extra CP phases in $B_s-\overline B_s$ mixing provide an upper
bound on the amount of CPV in $B_d-\overline B_d$ mixing; (iii)
if operators containing RH light quarks are subdominant, then
the extra CPV is equal in the two systems, and is negligible in
$2\to1$ transitions. Conversely, these operators can break the
correlation between CPV in the $B_s$ and $B_d$ systems, and can
induce significant new CPV in $\epsilon_K$.

We now analyze in detail this general MFV case, where both top
and bottom effective Yukawas are large and flavor diagonal
phases are present, to prove the above conclusions. We
emphasize the differences between the LMFV case and the
non-linear MFV (NLMFV) one. It is shown below that even in the
general scenario, there is a systematic expansion in small
quantities, $\VKM_{td},\VKM_{ts}$, and light quark masses,
while resumming in $y_t$ and $y_b$. This is achieved via a
parametrization borrowed from non-linear
$\sigma$-models\footnote{Another non-linear parameterization of
MFV was presented in~\cite{Feldmann:2008ja}.}. Namely, in the
limit of vanishing weak gauge coupling (or $m_W\to \infty$),
$U(3)_Q$ is enhanced to $U(3)_{Q^u}\times U(3)_{Q^d}$, as
discussed in Sec.~\ref{spurion}. The two groups are broken down
to $U(2)\times U(1)$ by large third generation eigenvalues in
$\aud$, so that the low energy theory is described by a
$[U(3)/U(2)\times U(1)]^2$ non-linear $\sigma$-model. Flavor
violation arises due to the misalignment of $Y_U$ and $Y_D$,
given by $\VKM_{td}$ and $\VKM_{ts}$, once the weak interaction
is turned on. It should be stressed that while below we
implicitly assume a two Higgs doublet model to allow for a
large bottom Yukawa coupling, this assumption is not necessary,
and the analysis is essentially model independent.

As discussed in Sec.~\ref{sec:appsym}, the breaking of the
flavor group is dominated by the top and bottom Yukawa
couplings. Yet here we also assume that the relevant
off-diagonal elements of $\VKM$ are small, so the residual
approximate symmetry is $\H=U(2)_{Q}\times U(2)_{U}\times
U(2)_{D} \times U(1)_3$ ($U(1)_Q$ is enhanced to $U(2)_Q$, and
there is also a $U(1)_3$ symmetry for the third generation).
The broken symmetry generators live in $\G/\H$ cosets. It is
useful to factor them out of the Yukawa matrices, so we
parameterize
\beq \label{eq:3}
Y_{U,D}=e^{i \hat \rho_Q} e^{\pm i \hat \chi /2} \tilde Y_{U,D}
e^{-i\hat \rho_{U,D}},
\eeq
where the reduced Yukawa spurions, $ \tilde Y_{U,D} $, are
\beq \label{eq:4}
\tilde Y_{U,D}= \begin{pmatrix} \phi_{U,D} & 0\\ 0 & y_{t,b}
\end{pmatrix}.
\eeq
Here $\phi_{U,D}$ are $2\times 2$ complex spurions, while
$\hat\chi$ and $\hat \rho_i$, $i=Q,U,D$, are the $3\times 3$
matrices spanned by the broken generators. Explicitly,
\beq \label{eq:5}
\hat \chi= \begin{pmatrix} 0 & \chi\\ \chi^\dagger & 0
\end{pmatrix}, \qquad \hat \rho_{i}= \begin{pmatrix} 0 &
\rho_{i}\\ \rho_{i}^\dagger & \theta_{i} \end{pmatrix}, \qquad
i=Q,U,D,
\eeq
where $\chi$ and $\rho_i$ are two dimensional vectors. The
$\hat\rho_i$ shift under the broken generators, and therefore
play the role of spurion ``Goldstone bosons''. Thus the
$\rho_i$ have no physical significance. On the other hand,
$\chi$ parameterizes the misalignment of the up and down Yukawa
couplings, and therefore corresponds to $\VKM_{td}$ and
$\VKM_{ts}$ in the low energy effective theory (see
Eq.~(\ref{eq:chi})).

Under the flavor group, the above spurions transform as,
\beq  \label{eq:9}
e^{i\hat\rho_{i}'} = V_ie^{i\hat \rho_{i}}U_i^\dagger, \ \
e^{i\hat\chi'} = U_Q e^{i\hat\chi} U_Q^\dagger, \ \ \tilde
Y_{i}^\prime = U_Q \tilde Y_{i} U_{i}^\dagger.
\eeq
Here $U_{i}=U_{i}(V_i,\hat \rho_i)$ are (reducible) unitary
representations of the unbroken flavor subgroup $U(2)_i\times
U(1)_3$,
\beq  \label{eq:6}
U_{i}= \begin{pmatrix} U_{i}^{2\times2} & 0\\ 0 &
e^{i\varphi_Q} \end{pmatrix}, ~~i=Q,U,D.
\eeq
For $V_i\in\H$, $U_i=V_i$. Otherwise the $U_i$ depend on the
broken generators and $\hat\rho_i$. They form a nonlinear
realization of the full flavor group. In particular,
Eq.~\eqref{eq:9} defines $U_{i}(V_i,\hat \rho_i)$ by requiring
that $\hat \rho_i'$ is of the same form as $\hat \rho_i$,
Eq.~(\ref{eq:5}). Consequently $\hat\rho_i$ is shifted under
$\G/\H$, and can be set to a convenient value as discussed
below.  Under $\H$, $\chi$\,[$\rho_i$] are fundamentals of
$U(2)_Q$\,[$U(2)_i$] carrying  charge $-1$ under the $U(1)_3$,
while $\phi_{U,D}$ are bi-fundamentals of $U(2)_Q\times
U(2)_{U,D}$.

As a final step we also redefine the quark fields by moding out
the ``Goldstone spurions'',
\begin{eqnarray} \label{def-tilded-q}
\tilde u_L= e^{-i \hat \chi /2} e^{-i \hat \rho_Q}  u_L, &\quad \tilde d_L=
e^{ i \hat \chi /2} e^{- i \hat \rho_Q} d_L, \\ \label{eq:7}
\tilde u_R= e^{-i \hat \rho_u} u_R, \qquad  & \tilde d_R= e^{-i \hat \rho_d} d_R.
\end{eqnarray}
The latter form reducible representations of $\H$.
Concentrating here and below on the down sector, we therefore
define $\tilde d_{L,R}=(\d_{L,R},0)+(0,\b_{L,R})$. Under flavor
transformations $\d_{L}{}'=U_{Q}^{2\times2} \d_{L}$ and
$\b_L{}'=\exp(i \varphi_Q)\b_L$. A similar definition can be
made for the up quarks.

With the redefinitions above, invariance under the full flavor
group is captured by the invariance under the unbroken flavor
subgroup $\H$~(see {\it e.g.}~\cite{Weinberg:1996kr}). Thus,
\emph{GMFV can be described without loss of generality as a
formally $\H$--invariant expansion in $\phi_{U,D}, \chi$.} This
is a straightforward generalization of the known effective
field theory description of spontaneous symmetry
breaking~\cite{Weinberg:1996kr}.  The only difference in our
case is that $Y_{U,D}$ are not aligned, as manifested by
$\chi\neq0$. Since the background field values of the relevant
spurions are small, we can expand in them.

We are now in a position to write down the flavor structure of
quark bilinears from which low energy flavor observables can be
constructed. We work to leading order in the spurions that
break $\H$, but to all orders in the top and bottom Yukawa
couplings. Beginning with the LL bilinears, to second order in
$\chi$ and $\phi_{U,D}$, one finds (omitting gauge and Lorentz
indices)
\begin{eqnarray} \label{eq:8}
& \bbar_L \b_L, \quad \dbarL\d_L, \quad \dbarL\phi_U
\phi_U^\dagger \d_L, \\ \label{eq:10} &
\dbarL\chi \b_L,\quad \bbar_L\chi^\dagger \chi \b_L,\quad
\dbarL\chi \chi^\dagger \d_L.
\end{eqnarray}
The first two bilinears in Eq.~\eqref{eq:8} are diagonal in the
down quark mass basis, and do not induce flavor violation. In
this basis, the Yukawa couplings take the form
\beq
Y_U=\left(\VKM \right)^\dagger \diag(m_u,m_c,m_t)\,, \quad
Y_D=\diag(m_d,m_s,m_b)\,.
\eeq
This corresponds to spurions taking the background values
$\rho_Q = \chi /2$, $\hat \rho_{U,D}=0$ and $\phi_D=\diag(m_d,
m_s)/m_b$, while flavor violation is induced via
\beq \label{eq:chi}
\chi^\dagger= i(\VKM_{td},\VKM_{ts})\,,\qquad \phi_U=\left(
\VKM_{(2)}\right)^\dagger \,\diag\left( \frac{m_u}{m_t},
\frac{m_c}{m_t}\right)\,.
\eeq
$\VKM_{(2)}$ stands for a two generation CKM matrix. In terms
of the Wolfenstein parameter $\lambda$, the flavor violating
spurions scale as $\chi \sim (\lambda^3, \lambda^2)$,
$(\phi_U)_{12} \sim \lambda^5$. Note that the redefined down
quark fields, Eqs.~(\ref{def-tilded-q},\ref{eq:7}), coincide
with the mass eigenstate basis, $\tilde d_{L,R}=d_{L,R}$, for
the above choice of spurion background values.

The LR and RR bilinears which contribute to flavor mixing are
in turn (at leading order in $\chi$ and $\phi_{U,D}$ spurions),
\begin{eqnarray} \label{eq:13}
& \dbarL \chi \b_R,  \quad  \dbarL\chi\chi^\dagger
\phi_D \d_R, \quad  \bbar_L \chi^\dagger \phi_D \d_R,  \\
\label{eq:13a}
& \dbarR \phi_D^\dagger \chi  \b_R, \quad \dbarR\phi_D^\dagger
\chi \chi^\dagger \phi_D \d_R.
\end{eqnarray}

To make contact with the more familiar MFV notation, consider
down quark flavor violation from LL bilinears. We can then
expand in the Yukawa couplings,
\beq \label{eq:a1}
\overline Q \left[a_1 Y_U Y_U^\dagger +a_2 \left(Y_U
Y_U^\dagger \right)^2 \right]Q+ \left[ b_2\, \overline Q Y_U
Y_U^\dagger Y_D Y_D^\dagger Q +{\rm h.c.}\right]+\ldots,
\eeq
with $a_{1,2}=\mathcal{O}(x_U^{2,4})$, $b_2= \mathcal{O}(x_U^2
x_D^2 )$. Note that the LMFV limit corresponds to $a_1\gg
a_2,b_2$, and the NLMFV limit to $a_1\sim a_2\sim b_2$. While
$a_{1,2}$ are real, the third operator in Eq.~\eqref{eq:a1} is
not Hermitian and $b_2$ can be complex~\cite{CPdiag1},
introducing a new CP violating phase beyond the SM phase. The
leading flavor violating terms in Eq.~\eqref{eq:a1} for the
down quarks are
\beqa \label{compareLMFV}
&&\hspace*{-.8cm} \bar d_L^i \left[\left(a_1+a_2
y_t^2\right)\xi^t_{ij}+a_1\xi^c_{ij}\right] d_L^j + \left[ b_2
y_b^2 \, \bar d_L^i \xi^t_{ib} b_L+ {\rm h.c.}\right]
=\nonumber\\
&&\hspace*{-.8cm} c_b \left(\dbarL \chi \b_L+{\rm h.c.}\right)+
c_t \dbarL\chi \chi^\dagger \d_L+ c_c\dbarL\phi_U
\phi_U^\dagger \d_L\,,
\eeqa
where $\xi^k_{ij}=y_k^2 \left(\VKM_{ki}\right)^* \VKM_{kj}$
with $i\neq j$. On the right hand side we have used the general
parameterization in Eqs.~(\ref{eq:8},\ref{eq:10}) with
\beq \label{cbctcc}
c_b\simeq(a_1 y_t^2+a_2 y_t^4+b_2 y_b^2),\quad c_t\simeq a_1
y_t^2+a_2 y_t^4 \quad \mathrm{and} \quad c_c\simeq a_1\,,
\eeq
to leading order. The contribution of the $c_c$ bilinear in
flavor changing transitions is $\mathcal{O}(1\%)$, compared to
the $c_t$ bilinear, and can thus be neglected in practice.

A novel feature of NLMFV is the potential for observable CPV
from RH currents, to which we return below. Other important
distinctions can be readily understood from
Eq.~\eqref{compareLMFV}. In NLMFV (with large $\tan\beta$) the
extra flavor diagonal CPV phase $\Im(c_b)$ can be large,
leading to observable deviations in the $B_{d,s}-\overline
B_{d,s}$ mixing phases, but none in LMFV. Another example is
$b\to s \nu \bar{\nu}$ and $s \to d \nu \bar{\nu}$ transitions,
which receive contributions only from a single operator in
Eq.~\eqref{compareLMFV} multiplied by the neutrino currents.
Thus, new contributions to $B \to X_s \nu \bar{\nu}$, $B \to K
\nu \bar{\nu}$ vs.~$K_L \to \pi^0 \nu\bar{\nu}$, $K^+ \to \pi^+
\nu \bar{\nu}$ are correlated in LMFV ($c_b \simeq c_t $), see
{\it e.g.}~\cite{MFVspurions2,Bergmann:2001pm,Bobeth:2005ck},
but are independent in NLMFV with large $\tan \beta$.
$\mathcal{O}(1)$ effects in the rates would correspond to an
effective scale $\Lambda_{\rm MFV} \sim 3$ TeV in the four
fermion operators, with smaller effects scaling like
$1/\Lambda_{\rm MFV}$ due to interference with the SM
contributions. Other interesting NLMFV effects involving the
third generation, such as large deviations in ${\rm
Br}(B_{d,s}\rightarrow \mu^+\mu^-)$ and $b\rightarrow s\gamma$,
arise in the minimal supersymmetric standard model (MSSM) at
large $\tan\beta$, where resummation is
required~\cite{Carena:2000uj, Degrassi:2000qf, Buras:2002vd,
Bobeth:2002ch}.

Assuming MFV, new CPV effects can be significant if and only if
the UV theory contains new flavor diagonal CP sources. The
proof is as follows. If no flavor diagonal phases are present,
CPV only arises from the CKM phase. In the exact $U(2)_Q$
limit, the CKM phase can be removed and the theory becomes CP
invariant (at all scales). The only spurions that break the
$U(2)_Q$ flavor symmetry are $\phi_{U,D}$ and $\chi$. CPV in
operators linear in $\chi$ is directly proportional to the CKM
phase (see Eq.~\eqref{compareLMFV}). Any additional
contributions are suppressed by at least $[\phi_U^\dagger
\phi_U, \phi_D^\dagger \phi_D]\sim (m_s/m_b)^2 (m_c/m_t)^2
\sin\theta_C\sim 10^{-9}$, and are therefore negligible.

Flavor diagonal weak phases in NLMFV can lead to new CPV
effects in $3\to 1$ and $3\to 2$ decays. An example is $\Delta
B=1$ electromagnetic and chromomagnetic dipole operators
constructed from the first bilinear in Eq.~\eqref{eq:13}. The
operators are not Hermitian, hence their Wilson coefficients
can contain new CPV phases. Without new phases, the untagged
direct CP asymmetry in $B \to X_{d,s} \gamma$ would essentially
vanish due to the residual $U(2)$ symmetry, as in the
SM~\cite{Soares:1991te}, and the $B \to X_s \gamma$ asymmetry
would be less than a percent. However, in the NLMFV limit
(large $y_b$), non-vanishing phases can yield significant CPV
in untagged and $B \to X_s \gamma$ decays, and the new CPV in
$B \to X_s \gamma$ and $B \to X_d \gamma$ would be strongly
correlated. Supersymmetric examples of this kind were studied
in~\cite{CPdiag2,Buras:2002wq,Hurth:2003dk}, where new phases
were discussed.

Next, consider the NLMFV $\Delta b=2$ effective operators. They
are not Hermitian, hence  their Wilson coefficients $C_i
/\Lgmfv^2$ can also contain new CP violating phases. The
operators can be divided into two classes:  class-1, which does
not contain light RH quarks [$(\overline{\d_L} \chi\b_{L,R})^2
$,\ldots]; and class-2, which does [$(\overline{\d_R}
\phi_D^\dagger \chi\b_{L})\,(\overline{\d_L} \chi\b_{R}
)$,\ldots]. Class-2 only contributes to $B_{s}-\overline B_{s}$
mixing, up to $m_d/m_s$ corrections. Taking into account that
$SU(3)_F$ (approximate $u$-$d$-$s$ flavor symmetry of the
strong interaction) breaking in the bag parameters of the
$B_s-\overline B_s$ vs.~$B_d-\overline B_d$ mixing matrix
elements is only at the few percent level in lattice
QCD~\cite{Becirevic:2001xt, Gamiz:2009ku}, we conclude that
class-1 yields the \emph{same weak phase shift in
$B_{d}-\overline B_{d}$ and $B_s-\overline B_s$ mixing relative
to the SM}. The class-1 contribution would dominate if $\Lgmfv$
is comparable for all the operators. For example, in the limit
of equal Wilson coefficients $C_i /\Lgmfv^2$, the class-2
contribution to $B_s-\overline B_s$ mixing would be $\approx
5\%$ of class-1. The maximal allowed magnitude of CPV in the
$B_d$ system is smaller than roughly 20\%. Quantitatively, for
$\Im(C_i) \approx 1$, this corresponds to $\Lgmfv \approx 18$
TeV for the leading class-1 operator, which applies to the
$B_s$ system as well. Thus, sizable CPV in the $B_s$ system
would require class-2 contributions, with $\mathcal{O}(1)$ CPV
corresponding to $\Lgmfv \approx 1.5$ TeV for the leading
class-2 operator. Conversely, barring cancelations,
\emph{within NLFMV models NP CPV in $B_s-\overline B_s$ mixing
provides an upper bound on NP CPV in $B_d-\overline B_d$
mixing.}

For $2\to1$ transitions, the new CPV phases come suppressed by
powers of $m_{d,s}/m_b$. All the $2\to 1$ bilinears in
Eqs.~\eqref{eq:8}, \eqref{eq:10}, \eqref{eq:13}
and~\eqref{eq:13a} are Hermitian, with the exception of
$\dbarL\chi\chi^\dagger \phi_D \d_R$. This provides the leading
contribution to $\epsilon_K$ from a non-SM phase, coming from
the operator $O_{LR} = ( \dbarL \chi\chi^\dagger \phi_D
\d_R)^2$. Its contribution is $\approx 2\% $ of the SM operator
$O_{LL} =(\dbarL\chi \chi^\dagger \d_L)^2$ for comparable
Wilson coefficients $C_{LR\,,LL}/\Lgmfv^2$. For $C_{LL}
\,,\Im(C_{LR}) \approx 1$, a new contribution to $\epsilon_K$
that is 50\% of the measured value would correspond to $\Lgmfv
\approx 5$ TeV for $O_{LL}$ and $\Lgmfv \approx 0.8$ TeV for
$O_{LR}$.

Note that the above new CPV effects can only be sizable in the
large $\tan \beta$ limit. They arise from non-Hermitian
operators (such as the second operator in \eqref{eq:a1}), and
are therefore of higher order in the $Y_D$ expansion. Whereas
we have been working in the large $\tan \beta$ limit, it is
straightforward to incorporate the small $\tan \beta$ limit
(discussed above in Sec.~\ref{sec:large_top}) into our
formalism. In that case the flavor group is broken down to
$U(2)_Q\times U(2)_U\times U(1)_t\times U(3)_D\,$, and the
expansion in Eq.~\eqref{eq:3} no longer holds. In particular,
resummation over $y_b$ is not required. Flavor violation is
described by linearly expanding in the down type Yukawa
couplings, from which it follows that contributions
proportional to the bottom Yukawa are further suppressed beyond
the SM CKM suppression.

It should also be pointed out that NLMFV differs from the
next-to-MFV framework~\cite{NMFV1,NMFV2}, since the latter
exhibits additional spurions at low energy.

%%%%%%%%%%%%%%%%%%%%%%%%%%%%%%%%%%%%%%%%%%%%%%%%%%%%
\subsection{MFV in covariant language}
%%%%%%%%%%%%%%%%%%%%%%%%%%%%%%%%%%%%%%%%%%%%%%%%%%%%%%

The covariant formalism described in Sec.~\ref{covdes} enables
us to offer further insight on the MFV framework. In the LMFV
case, the NP source $X_Q$ from Eq.~\eqref{o1} or
Eq.~\eqref{3g_operator} is a linear combination of the $\ad$
and $\au$ ``vectors'', naturally with $\mathcal{O}(1)$
coefficients at most. Hence we can immediately infer that no
new CPV sources exist, as all vectors are on the same plane,
and that the induced flavor violation is small (recall that the
angle between $\au$ and $\ad$ is small~-- $\mathcal{O}
(\lambda^2)$). These conclusions are of course already known,
but they emerge naturally when using the covariant language.

In the GMFV scenario, $X_Q$ is a general function of $\au$ and
$\ad$. We can alternatively express it in terms of the
covariant basis introduced in Sec.~\ref{sec:3g_full}, since
this basis is constructed using only $\au$ and $\ad$. Then, it
is easy to see that an arbitrary function of the Yukawa
matrices could produce any kind of flavor and CP
violation~\cite{CPdiag1,CPdiag4, Ellis:2009di}. However, the
directions denoted by $\hD$ require higher powers of the
Yukawas, so their contribution is generically much smaller
(in~\cite{CPdiag1} it was noticed that some directions, which
we identify as $\hD$, are not generated via RGE flow).
Therefore, the induced flavor and CP violation tend to be
restricted to the submanifold which corresponds to the $U(2)_Q$
limit (that is, the directions denoted by $\haud$, $\hj$,
$\hjud$ and $\hat C_{u,d}$).

%%%%%%%%%%%%%%%%%%%%%%%%%%%%%%%%%%%%%%%%%%%%%%%%%%%%%%%%%
%%%%%%%%%%%%%%%%%%%%%%%
\section{Supersymmetry}
\label{sec:susy}
Supersymmetric models provide, in general, new sources of
flavor violation, for both the quark and the lepton sectors.
The main new sources are the supersymmetry breaking soft mass
terms for squarks and sleptons and the trilinear couplings of a
Higgs field with a squark-antisquark or slepton-antislepton
pairs. Let us focus on the squark sector. The new sources of
flavor violation are most commonly analyzed in the basis in
which the corresponding (down or up) quark mass matrix and the
neutral gaugino vertices are diagonal. In this basis, the
squark masses are not necessarily flavor-diagonal, and have the
form
\beq
\tilde q_{Mi}^*(M_{\tilde q}^2)^{MN}_{ij}\tilde q_{Nj}= (\tilde
q_{Li}^*\ \tilde q_{Rk}^*)\left(\begin{array}{cc} (M^2_{\tilde
q})_{Lij} & A^q_{il}v_q \cr A^q_{jk}v_q & (M^2_{\tilde
q})_{Rkl} \cr \end{array}\right) \left(\begin{array}{c} \tilde
q_{Lj} \cr \tilde q_{Rl} \cr \end{array}\right),
\eeq
where $M,N=L,R$ label chirality, and $i,j,k,l=1,2,3$ are
generation indices. $(M^2_{\tilde q})_L$ and $(M^2_{\tilde
q})_R$ are the supersymmetry breaking squark masses-squared.
The $A^q$ parameters enter in the trilinear scalar couplings
$A^q_{ij}H_q\widetilde q_{Li}\widetilde q_{Rj}^*$, where $H_q$
$(q=u,d)$ is the $q$-type Higgs boson and $v_q=\langle
H_q\rangle$.

In this basis, flavor violation takes place through one or more
squark mass insertion. Each mass insertion brings with it a
factor of $(\delta_{ij}^q)_{MN}\equiv(M^2_{\tilde
q})^{MN}_{ij}/\tilde m_q^2$, where $\tilde m^2_q$ is a
representative $q$-squark mass scale. Physical processes
therefore constrain
\beq
[(\delta^q_{ij})_{MN}]_{\rm eff}\sim{\rm max}
[(\delta^q_{ij})_{MN}, (\delta^q_{ik})_{MP}
(\delta^q_{kj})_{PN},\ldots,(i\leftrightarrow j)].
\eeq
For example,
\beq
[(\delta^d_{12})_{LR}]_{\rm eff}\sim{\rm max}[A^d_{12}v_d/
\tilde m_d^2, \, (M^2_{\tilde d})_{L1k}A^d_{k2}v_d/\tilde
m_d^4,\, A^d_{1k}v_d(M^2_{\tilde d})_{Rk2}/\tilde
m_d^4,\ldots,(1\leftrightarrow2)].
\eeq
Note that the contributions with two or more insertions may be
less suppressed than those with only one.

In terms of mass basis parameters, the $(\delta^q_{ij})_{MM}$'s
stand for a combination of mass splittings and mixing angles:
\beq
(\delta^q_{ij})_{MM}=\frac{1}{\tilde m_{q}^2}\sum_\alpha
(K^q_M)_{i\alpha}(K^q_M)_{j\alpha}^*\Delta\tilde m^2_{q_\alpha},
\eeq
where $K^q_M$ is the mixing matrix in the coupling of the
gluino (and similarly for the bino and neutral wino) to
$q_{Li}-\tilde q_{M\alpha}$; $\tilde
m^2_q=\frac13\sum_{\alpha=1}^3 m_{\tilde q_{M\alpha}}^2$ is the
average squark mass-squared, and $\Delta\tilde
m^2_{q_\alpha}=m^2_{\tilde q_\alpha}-\tilde m^2_q$. Things
simplify considerably when the two following conditions are
satisfied \cite{Hiller:2008sv,Hiller:2010dv}, which means that
a two generation effective framework can be used (for
simplicity, we omit here the chirality index):
\beq
|K_{ik}K_{jk}^*|\ll|K_{ij}K_{jj}^*|,\ \ \
|K_{ik}K_{jk}^*\Delta\tilde
m^2_{q_kq_i}|\ll|K_{ij}K_{jj}^*\Delta\tilde m^2_{q_jq_i}|,
\eeq
where there is no summation over $i,j,k$ and where $\Delta
\tilde m^2_{q_jq_i}= m^2_{\tilde q_j}- m^2_{\tilde q_i}$. Then,
the contribution of the intermediate $\tilde q_k$ can be
neglected, and furthermore, to a good approximation,
$K_{ii}K_{ji}^*+K_{ij}K_{jj}^*=0$. For these cases, we obtain a
simpler expression for the mass insertion term
\beq\label{eq:delmass}
(\delta^q_{ij})_{MM}=\frac{\Delta\tilde m^2_{q_jq_i}}{ \tilde
m_{q}^2} (K^q_M)_{ij}(K^q_M)_{jj}^*\,,
\eeq
In the non-degenerate case, in particular relevant for
alignment models, it is useful to take instead of $\tilde m_q$
the mass scale $\tilde m^q_{ij}=\frac12(m_{\tilde q_i}+
m_{\tilde q_j})$~\cite{Raz:2002zx}, which better approximates
the full expression. We also define
\beq
\langle\delta^q_{ij}\rangle=\sqrt{(\delta_{ij}^{q})_{LL}
(\delta^{q}_{ij})_{RR}}\,.
\eeq

The new sources of flavor and CP violation contribute to FCNC
processes via loop diagrams involving squarks and gluinos (or
electroweak gauginos, or higgsinos). If the scale of the soft
supersymmetry breaking is below TeV, and if the new flavor
violation is of order one, and/or if the phases are of order
one, then these contributions could be orders of magnitude
above the experimental bounds. Imposing that the supersymmetric
contributions do not exceed the phenomenological constraints
leads to constraints of the form $(\delta^q_{ij})_{MM}\ll1$.
Such constraints imply that either quasi-degeneracy
($\Delta\tilde m^2_{q_jq_i}\ll (\tilde m^{q}_{ij})^2$) or
alignment ($|K^q_{ij}|\ll1$) or a combination of the two
mechanisms is at work.

Table~\ref{tab:exp} presents the constraints obtained in
Refs.~\cite{Ciuchini:2007cw, Gedalia:2009kh,Masiero:2005ua,
Buchalla:2008jp} as appear in~\cite{Hiller:2008sv}. Wherever
relevant, a phase suppression of order 0.3 in the mixing
amplitude is allowed, namely we quote the stronger between the
bounds on $\mathrm{Re}(\delta^q_{ij})$ and $3\mathrm{Im}
(\delta^q_{ij})$. The dependence of these bounds on the average
squark mass $\tilde m_q$, the ratio $x\equiv m_{\tilde g}^2
/\tilde m_q^2$ as well as the effect of arbitrary strong CP
violating phases can be found in~\cite{Hiller:2008sv}.

\begin{table}[t]
\begin{center}
\begin{tabular}{cc|cc} \hline\hline
\rule{0pt}{1.2em}%
$q$\ & $ij\ $\ &  $(\delta^{q}_{ij})_{MM}$ &
$\langle\delta^q_{ij}\rangle$ \cr \hline
$d$ & $12$\ & $\ 0.03\ $ & $\ 0.002\ $ \cr
$d$ & $13$\ & $\ 0.2\ $ & $\ 0.07\ $ \cr
$d$ & $23$\ & $\ 0.6\ $ & $\ 0.2\ $ \cr
$u$ & $12$\ & $\ 0.1\ $ & $\ 0.008\ $ \cr
\hline\hline
\end{tabular}
\caption{The phenomenological upper bounds on
$(\delta_{ij}^{q})_{MM}$ and on $\langle\delta^q_{ij}\rangle$,
where $q=u,d$ and $M=L,R$. The constraints are given for
$\tilde m_q=1$ TeV and $x\equiv m_{\tilde g}^2/\tilde m_q^2=1$.
We assume that the phases could suppress the imaginary parts by
a factor $\sim0.3$. The bound on $(\delta^{d}_{23})_{RR}$ is
about 3 times weaker than that on $(\delta^{d}_{23})_{LL}$
(given in table). The constraints on
$(\delta^{d}_{12,13})_{MM}$, $(\delta^{u}_{12})_{MM}$ and
$(\delta^{d}_{23})_{MM}$ are based on, respectively,
Refs.~\cite{Masiero:2005ua}, \cite{Ciuchini:2007cw}
and~\cite{Buchalla:2008jp}. \label{tab:exp}}
\end{center}
\end{table}

For large $\tan\beta$, some constraints are modified from those
in Table~\ref{tab:exp}. For instance, the effects of neutral
Higgs exchange in $B_s$ and $B_d$ mixing give, for $\tan \beta
=30$ and $x=1$ (see~\cite{Hiller:2008sv,Isidori:2002qe,
Foster:2006ze} and refs.~therein for details):
\beq \label{eq:bmixbounds}
\langle \delta^d_{13}\rangle < 0.01 \left( \frac{M_{A^0}}{200
\, \mbox{GeV}} \right) , ~~~~~ \langle \delta^d_{23} \rangle <
0.04 \left( \frac{M_{A^0}}{200 \, \mbox{GeV}} \right) ,
\eeq
where $M_{A^0}$ denotes the pseudoscalar Higgs mass, and the
above bounds scale roughly as $(30/\tan \beta)^2$.

The experimental constraints on the $(\delta^q_{ij})_{LR}$
parameters in the quark-squark sector are presented in
Table~\ref{tab:expLRme}. The bounds are the same for
$(\delta^q_{ij})_{LR}$ and $(\delta^q_{ij})_{RL}$, except for
$(\delta^d_{12})_{MN}$, where the bound for $MN=LR$ is 10 times
weaker.  Very strong constraints apply for the phase of
$(\delta^q_{11})_{LR}$ from EDMs. For $x=4$ and a phase smaller
than 0.1, the EDM constraints on
$(\delta^{u,d,\ell}_{11})_{LR}$ are weakened by a factor
$\sim6$.

\begin{table}[t]
\begin{center}
\begin{tabular}{cc|c} \hline\hline
\rule{0pt}{1.2em}
$q$\ & $ij$\ & $(\delta^{q}_{ij})_{LR}$\cr \hline
$d$ & $12$\  &    $\ 2\times10^{-4} \ $ \cr
$d$ & $13$\  &  $\ 0.08 \ $  \cr
$d$ & $23$\  &  $\ 0.01 \ $ \cr
$d$ & $11$\  & $4.7\times10^{-6}$ \cr
$u$ & $11$\  &  $9.3\times 10^{-6}$ \cr
$u$ & $12$\  & $\ 0.02 \ $\cr
\hline\hline
\end{tabular}
\caption{The phenomenological upper bounds on chirality-mixing
$(\delta_{ij}^{q})_{LR}$, where $q=u,d$. The constraints are
given for $\tilde m_q=1$ TeV and $x\equiv m_{\tilde g}^2/\tilde
m_q^2=1$.  The constraints on $\delta^{d}_{12,13}$,
$\delta^{u}_{12}$, $\delta^{d}_{23}$ and $\delta^{q}_{ii}$ are
based on, respectively, Refs.~\cite{Masiero:2005ua},
\cite{Ciuchini:2007cw}, \cite{Buchalla:2008jp}
and~\cite{Raidal:2008jk} (with the relation between the neutron
and quark EDMs as in~\cite{Gabbiani:1996hi}).
\label{tab:expLRme}}
\end{center}
\end{table}

While, in general, the low energy flavor measurements constrain
only the combinations of the suppression factors from
degeneracy and from alignment, such as Eq.~\eqref{eq:delmass},
an interesting exception occurs when combining the measurements
of $K^0$--$\Knotbar$ and $D^0$--$\Dnotbar$ mixing to test the
first two generation squark doublets (based on the analysis in
Sec.~\ref{sec:2g_df2}). Here, for masses below the TeV scale,
some level of degeneracy is unavoidable \cite{combine}:
\beq
\frac{m_{\widetilde Q_2}-m_{\widetilde Q_1}}{m_{\widetilde
Q_2}+m_{\widetilde Q_1}}\leq\begin{cases} 0.034 & {\rm maximal\
phases} \cr 0.27 & {\rm vanishing\ phases}\cr \end{cases}
\eeq
Similarly, using $\Delta F=1$ processes involving the third
generation (Sec.~\ref{sec:app}), the following bound is
obtained~\cite{Gedalia:2010mf}
\beq
\frac{\left|m_{\tilde Q_2}^2- m_{\tilde Q_3}^2\right|}{\left(2
m_{\tilde Q_2}+m_{\tilde Q_3}\right)^2}
 <20\left(\frac{\tilde m_{Q}}{100\mbox{\,GeV}}\right)^2\,,
\eeq
which is rather weak and insignificant in practice. The bound
that stems from $\Delta F=2$ third generation processes
(Sec.~\ref{sec:uutt}) is~\cite{Gedalia:2010zs, Gedalia:2010mf}
\beq
\frac{\left|m_{\tilde Q_1}^2- m_{\tilde Q_3}^2\right|}{\left(2
m_{\tilde Q_1}+m_{\tilde Q_3}\right)^2}
 <0.45\left(\frac{\tilde m_{Q}}{100\mbox{\,GeV}}\right)^2\,.
\eeq
Note that the latter limit is actually determined by CPV in $D$
mixing (see discussion in Sec.~\ref{sec:uutt}). It should be
mentioned that by carefully tuning the squark and gluino
masses, one finds a region in parameter space where the above
bounds can be ameliorated~\cite{Crivellin:2010ys}.

The strong constraints in Tables~\ref{tab:exp}
and~\ref{tab:expLRme} can be satisfied if the mediation of
supersymmetry breaking to the MSSM is MFV. In particular, if at
the scale of mediation, the supersymmetry breaking squark
masses are universal, and the $A$-terms (couplings of squarks
to the Higgs bosons) vanish or are proportional to the Yukawa
couplings, then the model is phenomenologically safe. Indeed,
there are several known mechanisms of mediation that are MFV
(see, {\it e.g.}~\cite{Shadmi:1999jy}). In particular,
gauge-mediation~\cite{GMSB1, GMSB2, Dine:1993yw,Meade:2008wd},
anomaly-mediation~\cite{AMSB1,AMSB2}, and
gaugino-mediation~\cite{Chacko:1999mi} are such mechanisms.
(The renormalization group flow in the MSSM with generic MFV
soft-breaking terms at some high scale has recently been
discussed in Refs.~\cite{CPdiag1,Paradisi:2008qh}.) On the
other hand, we do not expect gravity-mediation to be MFV, and
it could provide subdominant, yet observable flavor and CP
violating effects~\cite{Feng:2007ke}.

%%%%%%%%%%%%%%%%%%%%%%%%%%%%%%%%%%%%%%%%%%%%%%%%%%%%%%%%%%%%%
\section{Extra Dimensions} \label{sec:exdim}
%%%%%%%%%%%%%%%%%%%%%%%%%%%%%%%%%%%%%%%%%%%%%%%%%%%%%%%%%%%%
Models of extra dimensions come in a large variety, and the
corresponding phenomenology, including the implications for
flavor physics, changes from one extra dimension framework to
another.  Yet, as in the supersymmetric case, one can classify
the new sources of flavor violation which generically arise:

{\bf Bulk masses}~-- If the SM fields propagate in the bulk of
the extra dimensions, they can have bulk vector-like masses.
These mass terms are of particular importance to flavor
physics, since they induce fermion localization which may yield
hierarchies in the low energy effective couplings. Furthermore,
the bulk masses, which define the extra dimension interaction
basis, do not need to commute with the Yukawa matrices, and
hence might induce contributions to FCNC processes, similarly
to the squark soft masses-squared in supersymmetry.

{\bf Cutoff, UV physics}~-- Since, generically, higher
dimensional field theories are non-renormalizable, they rely on
unspecified microscopic dynamics to provide UV completion of
the models. Hence, they can be viewed as effective field
theories, and the impact of the UV physics is expected to be
captured by a set of operators suppressed by the framework
dependent cutoff scale.  Without precise knowledge of the short
distance dynamics, the additional operators are expected to
carry generic flavor structure and contribute to FCNC
processes. This is somewhat similar to ``gravity mediated''
contributions to supersymmetry breaking soft terms, which are
generically expected to have an anarchic flavor structure, and
are suppressed by the Planck scale.

{\bf ``Brane''-localized terms}~-- The extra dimensions have to
be compact, and typically contain defects and boundaries of
smaller dimensions [in order, for example, to yield a chiral
low energy four dimension (4D) theory].  These special points
might contain different microscopical degrees of freedom.
Therefore, generically, one expects that a different and
independent class of higher dimension operators may be
localized to this singular region in the extra dimension
manifold. (These are commonly denoted `brane terms', even
though, in most cases, they have very little to do with string
theory). The brane-localized terms can, in principle, be of
anarchic flavor structure, thus providing new flavor and CP
violating sources.  One important class of such operators are
brane kinetic terms: their impact is somewhat similar to that
of non-canonical kinetic terms, which generically arise in
supersymmetric flavor models.

We focus on flavor physics of five dimension (5D) models, with
bulk SM fields, since most of the literature focuses on this
class. Furthermore, the new flavor structure that arises in 5D
models captures most of the known effects of extra dimension
flavor models. Assuming a flat extra dimension, the energy
range, $\Lambda_{\rm 5D} R$ (where $\Lambda_{\rm 5D}$ is the 5D
effective cutoff scale and $R$ is the extra dimension radius
with the extra dimension coordinate $y\in(0,\pi R)$), for which
the 5D effective field theory holds, can be estimated as
follows. Since gauge couplings in extra dimensional theories
are dimensionful, {\it i.e.}~$\alpha_{\rm 5D}$ has mass
dimension $-1$, a rough guess (which is confirmed, up to order
one corrections, by various naive dimensional analysis methods)
is~\cite{Kribs:2006mq} $\Lambda_{\rm 5D} \sim 4
\pi/{\alpha_{\rm 5D}} \,.$ Matching this 5D gauge coupling to a
4D coupling of the SM at leading order, ${1}/{g^2} = \pi
R/{g_{\rm 5D}^2}\,,$ we obtain
\begin{equation}
\Lambda_{\rm 5D} R \sim \frac{4}{\alpha} \sim 30\,.
\end{equation}
Generically, the mass of the lightest Kaluza-Klein (KK) states,
$\Mkk$, is of ${\cal O}\big(R^{-1}\big)$.  If the extra
dimension theory is linked to the solution of the hierarchy
problem and/or directly accessible to near future experiments,
then $R^{-1}= {\cal O}\big(\rm TeV\big)$.  This implies an
upper bound on the 5D cutoff:
\begin{equation}
\Lambda_{\rm 5D}\lesssim 10^2\,{\rm TeV}\ll \Lambda_K\sim 2 \times
10^5\,\rm TeV \,,
\end{equation}
where $ \Lambda_K$ is the scale required to suppress the generic
contributions to $\epsilon_K$, discussed above (see
Table~\ref{tab:DF2}).

The above discussion ignores the possibility of splitting the
fermions in the extra dimension. In split fermion models,
different bulk masses are assigned to different generations,
which gives rise to different localizations of the fermions in
the extra dimension. Consequently, they have different
couplings to the Higgs, in a manner which may successfully
address the SM flavor puzzle~\cite{ArkaniHamed:1999dc}.
Separation in the extra dimension may suppress the
contributions to $\epsilon_K$ from the higher dimension
cutoff-induced operators.  As shown in Table~\ref{tab:DF2}, the
most dangerous operator is
\beq
{O^4_K} = \frac{1}{\Lambda_{\rm 5D}^2}  \left(\bar s_L
\,d_R\right) \left(\bar s_R\, d_L\right)\,. \label{eq:O4K}
\eeq
This operator contains $s$ and $d$ fields of both chiralities.
As a result, in a large class of split fermion models, the
overlap suppression would be similar to that accounting for the
smallness of the down and strange 4D Yukawa couplings. The
integration over the 5D profiles of the four quarks may yield a
suppression factor of ${\cal O}\big(m_d
m_s/v^2\big)\sim10^{-9}$. Together with the naive scale
suppression, $1/\Lambda_{\rm 5D}^2$, the coefficient of
${O^4_K}$ can be sufficiently suppressed to be consistent with
the experimental bound.

In the absence of large brane kinetic terms, however, fermion
localization generates order one non-universal couplings to the
gauge KK fields~\cite{Delgado:1999sv} (the case with large
brane kinetic terms is similar to the warped scenario discussed
below). The fact that the bulk masses are, generically, not
aligned with the 5D Yukawa couplings implies that KK gluon
exchange processes induce, among others, the following operator
in the low energy theory:
$\left[(D_L)_{12}^2/(6\Mkks)\right]\left(\bar s_L\,
d_L\right)^2$, where $\left(D_{L}\right)_{12}\sim \lambda$ is
the LH down quark rotation matrix from the 5D interaction basis
to the mass basis. This structure provides only a mild
suppression to the resulting operator. It implies that to
satisfy the $\epsilon_K$ constraint, the KK and the inverse
compactification scales have to be above $10^3$\,TeV, beyond
the direct reach of near future experiments, and too high to be
linked to a solution of the hierarchy problem. This problem can
be solved by tuning the 5D flavor parameters, and imposing
appropriate 5D flavor symmetries to make the tuning stable.
Once the 5D bulk masses are aligned with the 5D Yukawa
matrices, the KK gauge contributions vanish, and the
configuration becomes radiatively stable.

The warped extra dimension [Randall Sundrum (RS)]
framework~\cite{Randall:1999ee} provides a solution to the
hierarchy problem. Moreover, with SM fermions propagating in
the bulk, both the SM and the NP flavor puzzles can be
addressed.  The light fermions can be localized away from the
TeV brane~\cite{Grossman:1999ra}, where the Higgs is localized.
Such a configuration can generate the observed Yukawa
hierarchy, and at the same time ensure that higher dimensional
operators are suppressed by a high cutoff scale, associated
with the location of the light fermions in the extra
dimension~\cite{Gherghetta:2000qt,Huber:2000ie}. Furthermore,
since the KK states are localized near the TeV brane, the
couplings between the SM quarks and the gauge KK fields exhibit
the hierarchical structure associated with SM masses and CKM
mixings.  This hierarchy in the couplings provides an extra
protection against non-standard flavor violating
effects~\cite{Huber:2003tu}, denoted as RS-GIM
mechanism~\cite{aps1,aps2} (see also~\cite{Burdman:2002gr,
Burdman:2003nt}). It is interesting to note that an analogous
mechanism is at work in models with strong dynamics at the TeV
scale, with large anomalous dimension and partial
compositeness~\cite{Kaplan:1983fs,Georgi:1984ef,
Georgi:1984af}. The link with strongly interacting models is
indeed motivated by the AdS/CFT
correspondence~\cite{Maldacena:1997re, Witten:1998qj}, which
implies that the above 5D framework is a dual description of 4D
composite Higgs models~\cite{shining1, ArkaniHamed:2000ds}.

Concerning the quark zero modes, the flavor structure of the
above models as well as the phenomenology can be captured by
using the following simple rules~\cite{aps1,aps2,
Contino:2006nn,Gedalia:2009ws}. In the 5D interaction basis,
where the bulk masses $k\, C^{ij}_{x}$ are diagonal ($x=Q,U,D$;
$i,j=1,2,3$; $k$ is the AdS curvature), the value $f_{x_i}$ of
the profile of the quark zero modes is given by \be \label{fs}
f_{x_i}^2=(1-2c_{x_i})/( 1-\epsilon^{ 1-2c_{x_i} })\,.
 \ee
Here $c_{x_i}$ are the eigenvalues of the $C_x$ matrices,
$\epsilon=\exp[-\xi]$, $\xi=\log[M_{\rm \overline{Pl}}/{\rm
TeV}]$, and $M_{\rm \overline{Pl}}$ is the reduced Planck mass.
If $c_{x_i}<1/2$, then $f_{x_i}$ is exponentially suppressed.
Hence, order one variations in the 5D masses yield large
hierarchies in the 4D flavor parameters.  We consider the cases
where the Higgs VEV either propagates in the bulk or is
localized on the IR brane.  For a bulk Higgs case, the profile
is given by $\tilde{v}( \beta, z) \simeq v\sqrt{k(1+\beta)}
\bar z^{2+\beta}/ \epsilon $, where $\bar z\in(\epsilon,1)$
($\bar z=1$ on the IR brane), and $\beta\geq0$. The $\beta=0$
case describes a Higgs maximally-spread into the bulk
(saturating the AdS stability
bound~\cite{Breitenlohner:1982bm}).  The relevant part of the
effective 4D Lagrangian, which involves the zero modes and the
first KK gauge states ($G^1$), can be approximated
by~\cite{aps1, aps2,Gedalia:2009ws}
\beq \label{lagrangian}
\hspace*{-.15cm}\mathcal{L}^{4D} \supset  (Y_{U,D}^{\rm
5D})_{ij} H_{U,D} \, \bar Q_i f_{Q_i} \left(U,D\right)_j
f_{U_j,D_j} r^\phi_{00} (\beta,c_{Q_i},c_{U_j,D_j}) +g_*  G^1
x^\dagger_i x_i \left[f_{x_i}^2 \rg(c_{x_i}) -{1}/{\xi} \right]
,
\eeq
where $g_*$ stands for a generic effective gauge coupling and
summation over $i,j$ is implied. The corrections for the
couplings relative to the case of fully IR-localized Higgs and
KK states are given by the functions
$r^\phi_{00}$~\cite{Gedalia:2009ws} and
$\rg$~\cite{Csaki:2008zd,Csaki:2009bb}, respectively: \be
\label{r00} r^\phi_{00}(\beta,c_L,c_R) \approx
\frac{\sqrt{2(1+\beta)}}{2+\beta-c_L-c_R} \,, \ \  \ \rg(c)
\approx \frac{\sqrt{2}}{J_1(x_1)} \frac{0.7}{6-4c} \left(
1+e^{c/2} \right) \,, \ee where $r^\phi_{00}(\beta,c_L,c_R)=1$
for brane-localized Higgs and $x_1 \approx 2.4$ is the first
root of the Bessel function, $J_0(x_1)=0$.

In Table~\ref{fstab} we present an example of a set of
$f_{x_i}$ values that, starting from anarchical 5D Yukawa
couplings, reproduce the correct hierarchy of the flavor
parameters.  We assume for simplicity an IR-localized Higgs.
The values depend on two input parameters: $f_{U_3}$, which has
been determined assuming a maximally IR-localized $t_R$
($c_{U_3}=-0.5$), and $\yfd$, the overall scale of the 5D
Yukawa couplings in units of $k$, which has been fixed to its
maximal value assuming three KK states. On general grounds, the
value of $\yfd$ is bounded from above, as a function of the
number of KK levels, $\Nkk$, by the requirement that Yukawa
interactions are perturbative below the cutoff of the theory,
$\Lambda_{\rm 5D}$. In addition, it is bounded from below in
order to account for the large top mass. Hence the following
range for $\yfd$ is obtained (see {\it
e.g.}~\cite{shining6,Agashe:2008uz}):
\begin{equation}
\frac{1}{ 2}\lesssim \yfd \lesssim \frac{2\pi}{ \Nkk} {\rm \  \ for \
  brane \ Higgs\,;} \ \ \ \ \
\frac{1}{ 2}\lesssim \yfd \lesssim \frac{4\pi}{ \sqrt\Nkk}  {\rm \  \
  for \ bulk \ Higgs\,,}
\label{lam5D}\end{equation} where we use the rescaling $y_{\rm
5D}\to y_{\rm 5D}\, \sqrt{1+\beta}$, which produces the correct
$\beta\to \infty$ limit~\cite{Azatov:2009na} and avoids
subtleties in the $\beta=0$ case.

{\small
\begin{table}[t]\begin{center}
 \begin{tabular}{c|c|c|c}
    \hline\hline
    { Flavor}& { $f_Q$} & { $f_U$} & { $f_D$}\cr
    \hline
    1 &$ {{A \lambda^{3}} { f_{Q_3}}}\sim 3 \times
    10^{-3}$&
    $\frac{m_u}{m_t}\, \frac{ f_{U_3} }{ A \lambda^3} \sim1\times10^{-3}$&
     $\frac{m_d}{m_b}\, \frac{ f_{D_3} }{ A \lambda^3} \sim 2\times 10^{-3}$
    \cr
    2&$ {{ A \lambda^{2}} { f_{Q_3}}}\sim
    1\times 10^{-2}$&
    $\frac{ m_c }{ m_t} \, \frac{ f_{U_3} }{  A \lambda^2} \sim 0.1$&
      $\frac{ m_s}{ m_b}\, \frac{ f_{D_3} }{ A \lambda^2} \sim 1\times 10^{-2}$
    \cr
    3 &$ \frac{ m_t}{ v \yfd f_{U_3}}\sim 0.3$ &$\sqrt2$&
    $\frac{ m_b }{ m_t} \, { f_{U_3}}\sim 2\times 10^{-2}$
    \vspace*{.05cm}\cr
\hline\hline\end{tabular} \caption{{\small Values of the
$f_{x_i}$ parameters (Eq.~(\ref{fs})) which reproduce the
observed quark masses and CKM mixing angles starting from
anarchical 5D Yukawa couplings. We fix $f_{U_3}=\sqrt2$ and
$\yfd=2$ (see text).}}\label{fstab}
  \end{center}
\end{table}
}

With anarchical 5D Yukawa matrices, an RS residual little CP
problem remains~\cite{shining6}: Too large contributions to the
neutron EDM~\cite{aps1,aps2} and sizable chirally enhanced
contributions to
$\epsilon_K$~\cite{UTFit,Davidson:2007si,Csaki:2008zd,
Casagrande:2008hr,Blanke:2008zb} are predicted. The RS leading
contribution to $\epsilon_K$ is generated by a tree level KK
gluon exchange, which leads to an effective coupling for the
chirality-flipping operator in Eq.~\eqref{eq:O4K} of the
type~\cite{Davidson:2007si,Gedalia:2009ws,Csaki:2008zd,
Casagrande:2008hr, Blanke:2008zb}
\beq \label{eq:C4K}
C_4^K \simeq\frac{g_{s*}^2}{M_{KK}^2} f_{Q_2} f_{Q_1} f_{d_2}
f_{d_1} \rg(c_{Q_2}) \rg(c_{d_2}) \sim
\frac{g_{s*}^2}{M_{KK}^2} \frac{2 m_d m_s }{(v \yfd)^2}
\frac{\rg(c_{Q_2})\rg(c_{d_2})}{r^H_{00}(\beta,c_{Q_1},c_{d_1})
r^H_{00}(\beta,c_{Q_2},c_{d_2})}\,.
\eeq
The final expression is independent of the $f_{x_i}$, so the
bound in Table~\ref{tab:DF2} can be translated into constraints
in the $\yfd-\Mkk$ plane. The analogous effects in the $D$ and
$B$ systems yield numerically weaker bounds. Another class of
contributions, which involves only LH quarks, is also important
to constrain the $f_Q-\Mkk$ parameter space.

\begin{table}[t]
\begin{center}
\begin{tabular}{c|c c|c c} \hline\hline
\rule{0pt}{1.2em}%
Observable &  \multicolumn{2}{c}{$M_G^{\rm min}[\mathrm{TeV}]$} &
\multicolumn{2}{|c}{$\yfdmin\ {\rm or}\ f_{Q_3}^{\rm max}$} \cr
&   IR Higgs& $\beta=0$ & IR Higgs & $\beta=0$ \cr
 \hline CPV-$B_d^{LLLL}$ &  $12 f_{Q_3}^2$ &$12 f_{Q_3}^2$&
 $f_{Q_3}^{\rm max}=0.5$ &$f_{Q_3}^{\rm max}=0.5$\cr
CPV-$B_d^{LLRR}$  & ${4.2/ \yfd}$ &${2.4/ \yfd}$&  $\yfdmin=1.4$ &$\yfdmin=0.82$ \cr
CPV-$D^{LLLL}$ &  $0.73 f_{Q_3}^2$ &$0.73 f_{Q_3}^2$& no bound & no bound\cr
CPV-$D^{LLRR}$   & ${4.9/ \yfd}$ &${2.4/ \yfd}$& $\yfdmin=1.6$ &$\yfdmin=0.8$ \cr
$\epsilon_K^{LLLL}$ &  $7.9 f_{Q_3}^2$& $7.9 f_{Q_3}^2$ &
$f_{Q_3}^{\rm max}=0.62$ &$f_{Q_3}^{\rm max}=0.62$\cr
$\epsilon_K^{LLRR}$  & ${49/ \yfd}$ &$ {24/ \yfd}$& above
(\ref{lam5D}) & $\yfdmin=8 $ \cr
 \hline
 \hline
\end{tabular}
\caption{Most significant flavor constraints in the RS
framework (taken from~\cite{Isidori:2010kg}). The values of
$\yfdmin$ and $ f_{Q_3}^{\rm max}$ correspond to $\Mkk=3$ TeV.
The bounds are obtained assuming maximal CPV phases and
$g_{s*}=3$. Entries marked `above (\ref{lam5D})' imply that for
$\Mkk=3$ TeV, $\yfd$ is outside the perturbative range.
\label{tab:rszi}}
\end{center}
\end{table}

In Table~\ref{tab:rszi} we summarize the resulting constraints.
For the purpose of a quantitative analysis we set $g_{s*}=3$,
as obtained by matching to the 4D coupling at
one-loop~\cite{Agashe:2008uz} (for the impact of a smaller RS
volume see~\cite{Davoudiasl:2008hx}).  The constraints related
to CPV correspond to maximal phases, and are subject to the
requirement that the RS contributions are smaller than $30\%$
($60\%$) of the SM contributions~\cite{NMFV1,NMFV2} in the
$B_d$ ($K$) system.  The analytical expressions in the table
have roughly a 10\% accuracy over the relevant range of
parameters. Contributions from scalar exchange, either
Higgs~\cite{Azatov:2009na,Agashe:2009di} or
radion~\cite{Azatov:2008vm}, are not included, since these are
more model dependent and known to be
weaker~\cite{Duling:2009pj} in the IR-localized Higgs case.

Constraints from $\epsilon'/\epsilon_K$ have a different
parameter dependence than the $\epsilon_K$ constraints.
Explicitly, for $\beta=0$, the $\epsilon'/\epsilon_K$ bound
reads $M_G^{\rm min}=1.2\yfd$ TeV. When combined with the
$\epsilon_K$ constraint, we find $M_G^{\rm min}=5.5$ TeV with a
corresponding $\yfdmin=4.5$~\cite{Gedalia:2009ws}.

The constraints summarized in Table~\ref{tab:rszi} and the
contributions to the neutron EDM which generically require
$\Mkk> {\cal O}\left(10\,\rm TeV\right)$~\cite{aps1,aps2} are a
clear manifestation of the RS little CP problem. The problem
can be amended by various alignment
mechanisms~\cite{shining3,shining5,shining6, Csaki:2009bb,
Santiago:2008vq}. In this case, the bounds from the up sector,
especially from CPV in the $D$ system~\cite{Gedalia:2009kh,
combine}, become important. Constraints from $\Delta F=1$
processes (in either the down
sector~\cite{aps1,aps2,Blanke:2008yr,Buras:2009ka,
Bauer:2009cf} or $t\to c Z$~\cite{Agashe:2006wa}) are not
included here, since they are weaker in general, and
furthermore, these contributions can be suppressed
(see~\cite{Blanke:2008yr, Buras:2009ka, Bauer:2009cf}) due to
incorporation of a custodial symmetry~\cite{Agashe:2006at}.

It is interesting to combine measurements from the down and the
up sector in order to obtain general bounds (as done for
supersymmetry above). Using $K$ and $D$ mixing,
Eq.~\eqref{2g_bound}, the constraint on the RS framework
is~\cite{combine}
\beq
m_{\rm KK}>2.1 f^2_{Q_3} \; \mathrm{TeV} \,,
\eeq
for a maximal phase, where $f_{Q_3}$ is typically in the range
of 0.4-$\sqrt{2}$. We thus learn that the case where the third
generation doublet is maximally localized on the IR brane
(fully composite) is excluded, if we insist on $m_{\rm
KK}=3$~TeV, as allowed by electroweak precision tests (see {\it
e.g.}~\cite{Davoudiasl:2009cd}). The bounds derived from
$\Delta F=1$ and $\Delta F=2$ processes involving the third
generation are~\cite{Gedalia:2010zs, Gedalia:2010mf}
\beq
\begin{split}
m_{\rm KK}&>0.33 f^2_{Q_3} \; \mathrm{TeV} \,, \\ m_{\rm
KK}&>0.4 f^2_{Q_3} \; \mathrm{TeV} \, ,
\end{split}
\eeq
respectively.

%%%%%%%%%%%%%%%%%%%%%%%%%%%%%%%%%%%%%%%%%%%%%%%%%%%%
%%%%%%%%%%%%%%%%%%%%%%%%%%%%%%%%%%%%%%%%%%%%%%%%%%%%
\section{High $p_T$ Flavor Physics Beyond the SM} \label{sec:highpt}
%%%%%%%%%%%%%%%%%%%%%%%%%%%%%%%%%%%%%%%%%%%%%%%%%%%%
%%%%%%%%%%%%%%%%%%%%%%%%%%%%%%%%%%%%%%%%%%%%%%%%%%%%
So far we have mostly focused on information that can be
gathered from observables related to flavor conversion and in
particular to low energy experiments, the exception being top
flavor violation, which will be studied in great detail at the
LHC. However, much insight can be obtained on short distance
flavor dynamics, if one is to observe new degrees of freedom
which couple to the SM flavor sector. This is why high $p_T$
collider analyses are also useful for flavor physics (see {\it
e.g.}~\cite{Feng:2007ke,Grossman:2007bd,Dittmaier:2007uw,
delAguila:2008iz,Hiller:2008wp,Kribs:2009zy, Hurth:2009ke,
Bartl:2009au, Hiller:2009ii, Hurth:2009hp}). Below we discuss
implications of measurements related to both flavor diagonal
information and flavor conversion transitions.

Most of the analysis discussed in the following is rather
challenging to be done at the LHC for the quark sector, due to
the difficulty in distinguishing between jets originated from
first and second generation quarks. However, it is certainly
possible to distinguish the third generation quarks from the
other ones. Furthermore, even though not discussed in this
review, the charged lepton sector, which possesses a similar
approximate symmetry structure, allows for rather
straightforward flavor tagging. Therefore, some of the analysis
discussed below can be applied more directly to the lepton
sector (see {\it e.g.}~\cite{Bartl:2005yy,Bartl:2007ua,
Feng:2009bd,Buras:2009sg, Gross:2010ce}). For the quark sector,
future progress in the frontier of charm tagging\footnote{Some
progress has been recently achieved at the Tevatron in this
direction~\cite{cdfcharmtag}, and one might expect that the LHC
would perform at least as well, given that its detectors are
better (we thank Gustaaf Brooijmans for bringing this point to
our attention).} may play a crucial role in extracting further
information regarding the breaking of the SM approximate
symmetries.

In general, one may say that not much work has been done on the
issues discussed below, and that there are many issues, both
theoretical and experimental, to study on how to improve the
treatment related to high $p_T$ flavor physics at the LHC era.
While we do not attempt here to give a complete or even in
depth description of the subject of flavor at the LHC, we at
least try to touch upon a few of the relevant ingredients which
may help the reader to understand the potential richness and
importance of this topic.

%%%%%%%%%%%%%%%%%%%%%%%%%%%%%%%%%%%%%%%%%%%%%%%%%%%%
\subsection{Flavor diagonal information} \label{sec:flav_diag}
%%%%%%%%%%%%%%%%%%%%%%%%%%%%%%%%%%%%%%%%%%%%%%%%%%%%
Naively, one might think that flavor physics is related to
flavor converting amplitudes, say when the sum of the flavor
charges of the incoming particles is different from that of the
outgoing particles. However, this is not entirely true, since
(as we have discussed in detail in Sec.~\ref{covdes}) any form
of non-universality, if not aligned with the quark mass basis,
would induce some form of flavor conversion. Furthermore,
non-universal terms involving new states, which transform
non-trivially under the $SU(2)_L$ gauge group and are gauge
invariant (such as LH squark square masses), unavoidably induce
flavor conversion at some level, since these cannot be
simultaneously diagonalized in the up and down mass bases (see
discussion in Sec.~\ref{immune}).

The information that can be extracted is most usefully
expressed in terms of the manner that the SM flavor symmetry,
$\GSM$, is broken by the NP flavor diagonal sources. Of
particular importance is whether the approximate $U(2)$
symmetry, which acts on the light quarks, is broken, since in
this case the data implies that a strong mechanism of alignment
must be at work. Even if the $U(2)$ symmetry is respected by
the new degrees of freedom, any non-universal information,
related to the breaking of $\GSM$, would be also extremely
useful. In general, this kind of experimental insight is linked
to the microscopic nature of the new dynamics. Such knowledge
is invaluable, and is typically related to scales well beyond
the direct reach of near future experiments. As an example of
flavor diagonal information that can be extracted at the LHC
era, we discuss the spectrum of new degrees of freedom which
transform under the SM flavor group and the coupling of a
flavor singlet state to the SM quarks.

%%%%%%%%%%%%%%%%%%%%%%%%%%%%%%%%%%%%%%%%%%%%%%%%%%%%%%%
\subsubsection{Spectrum}
%%%%%%%%%%%%%%%%%%%%%%%%%%%%%%%%%%%%%%%%%%%%%%%%%%%%%%%%

Among the first parameters that can be extracted once new
degrees of freedom are found are their masses. The
phenomenology changes quantitatively based on the
representation of the new particles under the flavor group.
However, the interesting experimental information that one
would wish to extract is similar in essence. Suppose, for
instance, that we have the new states, discovered at the LHC,
transforming as an irreducible representation of the $U(3)_U$
SM flavor group (this is a reasonable assumption, given that
the top couplings yield the most severe hierarchy problem). If
the masses of all new states are identical or universal, then
not much flavor information could be extracted. Otherwise, it
is useful to break the states according to their representation
of the approximate $U(2)_U$ symmetry, obtained by setting the
up and charm masses to zero. The simplest non-trivial case,
which we now consider, is when the new states transform in the
fundamental representation of the flavor group. The most
celebrated example of this case is the up type squarks, but
also the KK partner of the up type quarks in universal/warped
extra dimension. Under the $U(2)_U$ approximate flavor group,
the fundamental states would transform as a doublet and
singlet. Thus, we can think of the following three
possibilities listed by order of significance (regarding flavor
physics):
\begin{itemize}
\item[(i)] The spectrum is universal, and the $U(2)_U$
    doublet and singlet are of identical masses. This
    implies a flavor blind underlying dynamics.
\item[(ii)] The spectrum exhibits an approximate $2+1$
    structure, {\it i.e.}~the doublet and singlet differ in
    mass. This spectrum is expected in a wide class of
    models, where the NP flavor dynamics preserve the SM
    approximate symmetry structure. Examples of this class
    are the MFV and next-to-MFV~\cite{NMFV1} frameworks,
    which contain various classes of supersymmetry models,
    warped extra dimension models etc. There is highly
    non-trivial physical content in this case, since the
    $U(3)_U\to U(2)_U$ breaking of the new physics cannot
    be generic: New physics with such breaking, if not
    aligned with the SM up type Yukawa, induces top flavor
    violation (as we have discussed in Sec.~\ref{immune} to
    be constrained at the LHC) and more importantly $c\to
    u$ transition contributing to $D-\overline D$ mixing.
    Furthermore, hints on the origin of the flavor puzzle
    and flavor mediation scale could be extracted.
\item[(iii)] The spectrum is anarchic, {\it i.e.}~there is
    no approximate degeneracy between the new particles'
    masses. This case is the most exciting in terms of
    flavor physics, since it suggests that some form of
    alignment mechanism is at work, to prevent too large
    contributions to various flavor violating processes.
    Thus, there is a potential that when combining the
    spectrum information with high $p_T$ and low energy
    measurements, information on the origin of the flavor
    hierarchies and flavor mediation scale could be
    extracted.
\end{itemize}
Let us also consider another case: Suppose that the newly
discovered particles are in the adjoint representation of the
$U(3)_{Q,U,D}$ flavor group. An example of this case is the KK
excitation of a flavor gauge boson of extra dimension
models~\cite{shining1,shining3,shining4,shining6}. As discussed
in Sec.~\ref{GMFV}, under the approximate $U(2)_{Q,U,D}$ flavor
group, an adjoint consists of a doublet (which corresponds to
the four broken generators), a triplet and a singlet (both
correspond to the unbroken generators). Once again, there are
three possible cases: A universal spectrum, an approximate
$3+2+1$ structure or an anarchic spectrum with alignment. The
case of a bi-fundamental representation has been recently
discussed in~\cite{Albrecht:2010xh}.

%%%%%%%%%%%%%%%%%%%%%%%%%%%%%%%%%%%%%%%%%%%%%%%%%%%%%%%%%%%%%%%%%%%%%%
\subsubsection{Couplings}
%%%%%%%%%%%%%%%%%%%%%%%%%%%%%%%%%%%%%%%%%%%%%%%%%%%%%%%%%%%%%%%%%%%%%%%%
Another source of precious flavor diagonal information, which
has not been widely studied, is the coupling of a flavor
singlet object. Celebrated examples would be in the form of
non-oblique and non-universal corrections to the coupling of
the $Z$ to the bottom due to the top Yukawa, or just the
predicted Higgs branching ratio into quarks, which favors third
generation final states. A more exotic example is the quark
coupling of a new gauge boson, such as the $Z'$ variety,
supersymmetric gauginos\footnote{In the case of softly broken
supersymmetry, it is most likely that the gauginos' coupling
will be characterized by a unitary matrix~-- a remnant of
supersymmetric gauge invariance. In such a case, unless large
flavor violation in the gauginos' couplings is present, they
are expected to exhibit universality.} or KK gauge bosons in
extra dimension models. In these cases, we can view the
coupling as a spurion which either transforms under the
fundamental representation of the flavor group (the Higgs case)
or as an adjoint (the other cases). The approach would be
therefore to characterize the flavor information according to
the three items listed above. If the couplings are flavor
universal, then there is not much to learn. If, however, the
couplings obey the $2+1$ rule, it already tells us that the new
interactions do not only follow the SM approximate symmetry
structure, but are also quasi-aligned with the SM third
generation direction. The case where the couplings are
anarchical is the most exciting one, as it requires a strong
alignment mechanism, and may lead to a new insight on the SM
flavor puzzle.

As an example for the case of a $2+1$ structure, let us imagine
that a color octet resonance\footnote{A recent proposal to
distinguish between a color octet resonance an singlet one can
be found in~\cite{Sung:2009iq}.} is discovered at the LHC in
the $t\bar t$ channel~\cite{Agashe:2006hk,Lillie:2007yh}. One
may suggest that this is an observation of a KK gluon state,
yet other options are clearly possible as well (assuming that
the particle's spin is consistent with one). It would be a
particularly convincing argument in favor of the anarchic
warped extra dimension framework if one is to prove
experimentally that the decay channels into the light quarks
are much smaller than the $t\bar t$ one. The challenge in this
measurement would be to compete against the continuous di-jet
background. The ability to have charm tagging is obviously a
major advantage in such a scenario. Not only that it would help
to suppress the background, but also a bound on the deviation
from universality could be translated to a bound on the warped
extra dimension volume, and thus hint for the amount of
hierarchy produced by the warping~\cite{Davoudiasl:2008hx,
Davoudiasl:2009jk}.

\begin{figure}[hbt]
\centering
\includegraphics[width=3.8In]{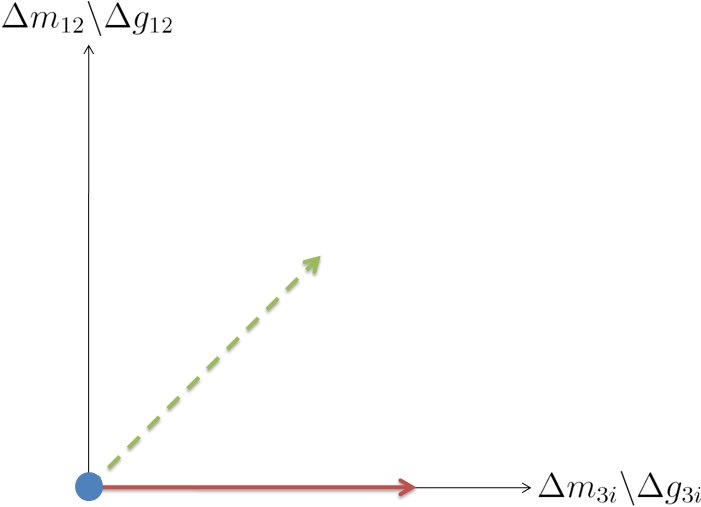}
\caption{A schematic representation of some possible spectra or
coupling structure of new degrees of freedom. The $x$ axis symbols
the difference in mass/coupling between the third generation and the
first two, and the $y$ axis is for the difference between the first
two generations. The red solid arrow represents a $2+1$ structure of
the spectrum/coupling, the dashed green arrow stands for an anarchic
structure (generally excluded) and the blue circle at the origin
signifies complete degeneracy.}
\label{fig:spec_meas}
\end{figure}

\begin{figure}[hptb]
\centering
\includegraphics[width=3.2In]{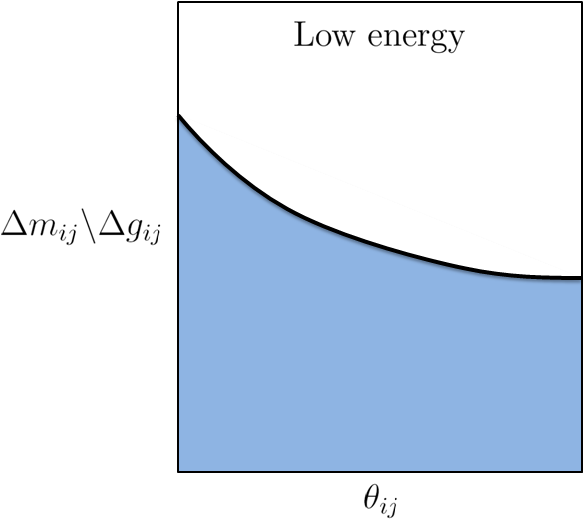} \hspace{0.1In}
\includegraphics[width=3.2In]{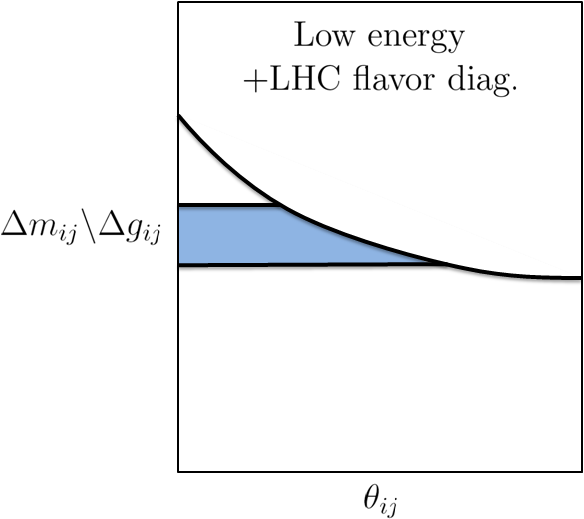}
\caption{A schematic representation of bounds on the new physics
parameter space, given by the mixing between two generations
$\theta_{ij}$ and the difference in mass/coupling. Left: A typical
present constraint arising from not observing deviations from the
SM predictions (the allowed region is colored). Right: Adding a
possible measurement of a mass/coupling difference at the LHC.
This figure is inspired by a plot from~\cite{Grossman:2009dw}.}
\label{fig:split_bound}
\end{figure}

To conclude the subject of flavor diagonal information, we
schematically show possible consequences in
Figs.~\ref{fig:spec_meas} and~\ref{fig:split_bound}. The former
presents different structures of the spectrum or coupling of
newly discovered degrees of freedom, and the latter
demonstrates how such a measurement at the LHC affects the NP
parameter space, in addition to existing low energy bounds.

\begin{figure}[pbh]
\centering
\includegraphics[width=3.2In]{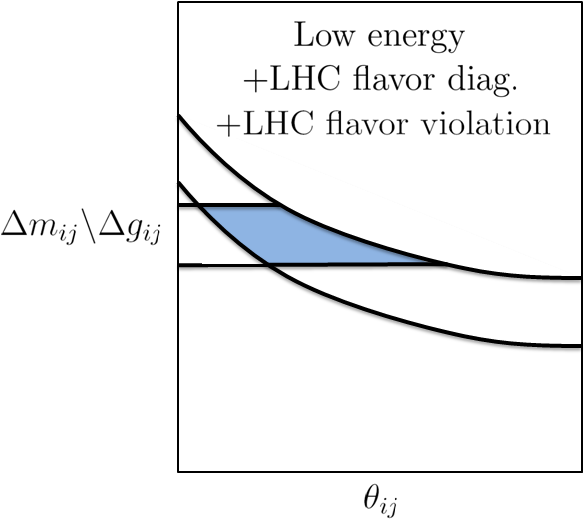}
\caption{A schematic representation of bounds on the new physics
parameter space. Here we include, in addition to the low energy
data and the mass/coupling difference measurement in
Fig.~\ref{fig:split_bound}, a positive signal of flavor violation
at the LHC.}
\label{fig:split_fv_bound}
\end{figure}

%%%%%%%%%%%%%%%%%%%%%%%%%%%%%%%%%%%%%%%%%%%%%%%%%%%%
\subsection{Flavor non-diagonal information}
%%%%%%%%%%%%%%%%%%%%%%%%%%%%%%%%%%%%%%%%%%%%%%%%%%%%

So far we have mostly considered flavor conversion at low
energies. In the following we briefly mention possible signals
in which new degrees of freedom are involved in flavor
converting processes, hopefully to be discovered soon at the
LHC. Clearly, more direct information regarding flavor physics
would be obtained in case the new states induce some form of
flavor breaking beyond non-universality. For concreteness, let
us give a few examples for such a possibility:
\begin{itemize}
\item A sfermion, say squark, which decays to a gaugino and
    either of two different quark flavors, both with
    considerable rate~\cite{Kribs:2009zy}.
\item A gluino which decays to quark and squark of a
    different flavor with a sizable
    rate~\cite{Bartl:2009au}.
\item A lifetime measurement of a long lived
    stop~\cite{Hiller:2008wp,Hiller:2009ii}.
\item A single stop production from the charm sea content
    due to large scharm-stop mixing.
\item A $Z'$ state or a KK gauge boson which decay into two
    quarks of a different flavor.
\item A charged higgs particle which decays to a top and a
    strange~\cite{Dittmaier:2007uw}.
\end{itemize}

As in the above, we separate the discussion to the case where
the approximate $U(2)$ flavor symmetry is respected by the new
dynamics and the one in which it is badly broken.
\begin{itemize}
\item[(i)] $U(2)$ preserving~-- flavor conversion occurs
    between the third generation and a light one. The
    corresponding processes then contain an odd number of
    third generation quarks. Since ATLAS and CMS have a top
    and bottom tagging capability, this class of processes
    can be observed with a reasonable efficiency. In the
    absence of charm tagging, there is no practical way to
    differentiate between the first two generation (thus,
    the information that can be extracted is well described
    by the covariant formalism presented in
    Sec.~\ref{sec:3g_u2}). Recall that in the exact
    massless $U(2)$ limit, the first two generations are
    divided into an active state and a sterile
    non-interacting one. In the absence of CP violating
    observables at the LHC, the measurement of flavor
    conversion is directly translated to determination of
    the amount of the third-active transition strength, or
    the corresponding mediating generator denoted
    as~$\hj_u$ in Sec.~\ref{sec:3g_u2}.
\item[(ii)] In order to go beyond case (i), charm tagging
    is required, which would enable to observe flavor
    violation that differentiate between the first two
    generations at high $p_T$. Almost no work has been
    performed on this case, but the corresponding
    measurement would be equivalent to probing the
    ``small'' CP conserving generators denoted by $\hat
    D_{1,4}$ in Sec.~\ref{sec:3g_full}.
\end{itemize}

Fig.~\ref{fig:split_fv_bound} demonstrates how detecting a
clear signal of flavor violation at the LHC affects the NP
parameter space, in addition to flavor diagonal information
(Fig.~\ref{fig:split_bound}).

%%%%%%%%%%%%%%%%%%%%%%%%%%%%%%%%%%%%%%%%%%%%%%%%%%%%
%%%%%%%%%%%%%%%%%%%%%%%%%%%%%%%%%%%%%%%%%%%%%%%%%%%%
\section{Conclusions}
%%%%%%%%%%%%%%%%%%%%%%%%%%%%%%%%%%%%%%%%%%%%%%%%%%%%
%%%%%%%%%%%%%%%%%%%%%%%%%%%%%%%%%%%%%%%%%%%%%%%%%%%%
The field of flavor physics is now approaching a new era marked
by the conclusion of the B-factories and the rise of the LHC
experiments. In the last decade or so, huge progress has been
achieved in precision flavor measurements. As of today, no
evidence for deviation from the standard model (SM) predictions
has been observed, and in particular it is established that the
SM is the dominant source of CP violation phenomena in quark
flavor conversion. Furthermore, strong bounds related to CP
violation in the up sector were recently obtained, which
provide another non-trivial test for the SM Kobayashi-Maskawa
mechanism.

The unique way of the SM to induce flavor violation implies
that the recent data is translated to stringent bounds on new
microscopical dynamics. To put it differently, any new physics
at the TeV scale, motivated by the hierarchy problem, cannot
have a general flavor structure. As we have discussed in detail
in these lectures, it is very likely that for a SM extension to
be phenomenologically viable, it has to possess the SM
approximate symmetry structure, characterized by the smallness
of the first two generation masses and their mixing with third
generation quarks.

In the LHC epoch, while continuous progress is expected in the
low energy precision tests frontier, dramatic progress is
foreseen in measurements related to top flavor changing neutral
processes. Moreover, in the event of new physics discovery, a
new arena for flavor physics tests would open, if the new
degrees of freedom carry flavor quantum numbers. At the LHC
high energy experiments, extraction of flavor information is
somewhat limited by its hadronic nature. In particular,
distinguishing between the first two generation quarks is
extremely challenging. Nevertheless, the power of this
information is in probing physics at scales well beyond the
direct reach of near future experiments. Thus, we expect flavor
physics to continue playing an important role in our
understanding of nature at short distances.

%%%%%%%%%%%%%%%%%%%%%%%%%%%%%%%%%%%%%%%%%%%%%%%%
%%%%%%%%%%%%%%%%%%%%%%%%%%%%%%%%%%%%%%%%%%%%%%%%

\section*{Acknowledgements} GP thanks the organizers of TASI09
for the successful school and great hospitality. GP is the
Shlomo and Michla Tomarin career development chair. The work of
GP is supported by the Israel Science Foundation (grant
\#1087/09), EU-FP7 Marie Curie, IRG fellowship and the Peter \&
Patricia Gruber Award.

\bibliographystyle{utcaps}
\bibliography{flavorbib}

\end{document}